\documentclass[11pt]{article}

\usepackage{amssymb}
\usepackage{amsmath}
\usepackage{graphicx}
\usepackage{latexsym}
\usepackage{subfigure}

\begin{document}

\begin{titlepage}
\begin{center}

\hfill STUPP-14-218 \\
\hfill \today

\vspace{0.5cm}
{\large\bf A way to crosscheck $\mu$-$e$ conversion in the case of no signals of 
$\mu \to e \gamma$ and $\mu \to 3e$}
\vspace{1cm}

{\bf Joe Sato}$^{a,\,}
$\footnote{joe@phy.saitama-u.ac.jp},
and 
{\bf Masato Yamanaka}$^{b,\,}
$\footnote{yamanaka@eken.phys.nagoya-u.ac.jp} \\

\vskip 0.15in

{\it
$^a${Department of Physics, Saitama University, 
Shimo-okubo, Sakura-ku, Saitama, 338-8570, Japan} \\
$^b${Department of Physics, Nagoya University, Nagoya 464-8602, Japan}
}

\vskip 0.4in

\abstract{We consider the case that $\mu$-$e$ conversion 
signal is discovered but other charged lepton flavor violating 
(cLFV) processes will never be found. In such a case, we need 
other approaches to confirm the $\mu$-$e$ conversion and 
its underlying physics without conventional cLFV searches. 
We study R-parity violating (RPV) SUSY models as a benchmark. 
We briefly review that our interesting case is realized in 
RPV SUSY models with reasonable settings according to current 
theoretical/experimental status. 
We focus on the exotic collider signatures at the LHC ($pp \to 
\mu^- e^+$ and $pp \to jj$) as the other approaches. We 
show the correlations between the branching ratio of $\mu$-$e$ 
conversion process and cross sections of these processes. 
It is first time that the correlations are graphically shown. 
We exhibit the RPV parameter dependence of the branching 
ratio and the cross sections, and discuss the feasibility to 
determine the parameters. 
}

\end{center}
\end{titlepage}

\tableofcontents

\section{Introduction}  \label{Sec:Intro} 

Lepton flavor violation (LFV) is the clearest signal for physics beyond
the Standard Model (SM) as it conserves lepton flavor exactly
\cite{Kuno:1999jp}. Therefore extensive searches for LFV have been made
since the muon was found.  There have been searches for $\mu\rightarrow
e\gamma$ \cite{Brooks:1999pu,Adam:2013mnn} , $\mu-e$ conversion
\cite{Bertl:2006up} and $\mu \rightarrow 3e$ \cite{Bellgardt:1987du}
. In all of these processes both muon and electron number are
violated. There are also LFV searches with the tau lepton
\cite{Aubert:2003pc,Abe:2003sx,Hayasaka:2005xw,Enari:2005gc} . Though a
lot of efforts have been made, we have not found any LFV signals with
charged leptons.  LFV had, however, been found in neutrino oscillation
\cite{Fukuda:1998mi,Abe:2013hdq} and it indeed requires us to extend the
SM so that physics beyond the SM must include LFV.  This fact also gives
us a strong motivation to search for charged lepton flavor violation
(cLFV).  Indeed the MEG collaboration has tried to observe the process
$\mu \rightarrow e \gamma$ and gave a significant upper bound on its
branching ratio \cite{Adam:2013mnn} .  Another effort at the LHC gave
some of upper limits on tau number violation \cite{Aaij:2013fia} though
at this moment more stringent limits are given by Belle collaboration.

Along this line new experiments to search for cLFV will start
soon. COMET \cite{Cui:2009zz,Kuno:2013mha} and DeeMe
\cite{Natori:2014yba} will launch within a few years and search $\mu-e$
conversion.  In these experiments, first, muons are trapped by target
nucleus (carbon, aluminum, titanium, and so on), then, if cLFV exists,
it converts into an electron.

If COMET and DeeMe observe the conversion process, then with what
kind of new physics should we interpret it?  Now it is worth 
considering again since we are in-between two kinds of cLFV experiments
with muon.

\begin{figure}[t!]
\begin{center}
\includegraphics[width=150mm]{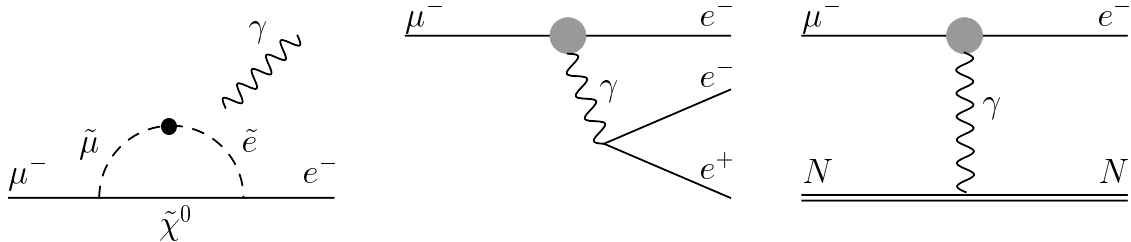}
\caption{cLFV processes in supersymmetric models with R-parity 
conservation. }
\label{fig:Diagrams}
\end{center}
\end{figure}

For these several decades, theories with supersymmetric extension have
been most studied. These theories include a source of LFV.  It is
realized by the fact that the scalar partner of the charged leptons can
have a different flavor basis from that of the charged leptons.  In
addition, R-parity is often imposed on this class of the
theory\cite{Hisano:1995cp,Sato:2000ff}. With it, $\mu\rightarrow
e\gamma$ process has the largest branching ratio among the three cLFV
processes. This occurs through the dipole process depicted in
Fig.~\ref{fig:Diagrams} and the other two, $\mu-e$ conversion and
$\mu\rightarrow 3e$ are realized by attaching a quark line and an
electron line at the end of the photon line respectively, giving an
O($\alpha$) suppression.  Those branching ratios must be smaller than
that of $\mu\rightarrow e\gamma$.  At this moment, however, the upper
bounds for those branching ratios are almost same each other. It means
if COMET and DeeMe observe a cLFV, that is the $\mu-e$ conversion
process, we have to discard this scenario.

It is, however, possible to find a theory easily in which COMET/DeeMe 
find cLFV first. 
To see this we first note that the $\mu\rightarrow e \gamma$ process 
occurs only at loop level due to the gauge invariance,
while other two can occur as a tree process. Therefore in this case we
have to consider a theory in which the $\mu-e$ conversion process occurs
as tree process.
In other words we have to assume a particle which violate
muon and electron number. Since $\mu - e$ conversion occurs in a nucleus,
it also couples with quarks with flavor conservation. 
Furthermore it is better to assume that it does not couple with two electrons 
as we have not observed $\mu\rightarrow 3e$.

In this paper we consider the case that COMET/DeeMe indeed observe 
the cLFV process, while all the other experiments will not observe 
anything new at that time.  With this situation, we need to understand 
how to confirm the cLFV in other experiments. It is dependent on a 
theory considered.
Unfortunately in this case other new physics signals are expected to be
quite few, since the magnitude of the cLFV interaction is so small due
to its tiny branching ratio. Therefore it is very important to simulate
now how to confirm the COMET signal and the new physics. As a benchmark
case we study a supersymmetric standard model without R
parity~\cite{deGouvea:2000cf} .  In this kind of theory the scalar lepton
mediates $\mu \leftrightarrow e$ flavor violation. It is important to
emphasize that the R parity violating theory is strongly motivated by
also the fact that we have not observed any typical SUSY signals.

The paper is organized as follows. First, in Sec.~\ref{Sec:interaction} 
we briefly review a theory with R parity violation and show our setup.
Next, in Sec.~\ref{Sec:Obs} we discuss what processes can be the signal of the
theory. Then in Sec.~\ref{Sec:result} we give the result and discuss how
to confirm the scenario here depending on the parameters.
Finally we summarize our work in Sec.~\ref{Sec:summary}.

\section{RPV interaction and our scenario}  \label{Sec:interaction} 

In general the supersymmetric gauge invariant superpotential contains 
the R-parity violating terms~\cite{Weinberg:1981wj, Sakai:1981pk, 
Hall:1983id}, 
\begin{equation}
\begin{split}
   \mathcal{W}_\text{RPV} = \lambda_{ijk} L_i L_j E_k^c 
   + \lambda'_{ijk} L_i Q_j D_k^c 
   + \lambda''_{ijk} U_i^c D_j^c D_k^c, 
\label{Eq:RPV_SP}   
\end{split}      
\end{equation}
where $E_i^c$, $U_i^c$ and $D_i^c$ are $SU(2)_L$ singlet superfields, 
and $L_i$ and $Q_i$ are $SU(2)_L$ doublet superfields. Indices $i$, $j$, 
and $k$ represent the generations. 
We take $\lambda_{ijk} = - \lambda_{jik}$ and $\lambda''_{ijk} 
= - \lambda''_{ikj}$. First two terms include lepton number violation, and 
the last term includes baryon number violation. Since some combinations 
of them accelerate proton decay, we omit the last term. 
Thus the RPV processes are described by following Lagrangian, 
\begin{equation}
\begin{split}
   &
   \mathcal{L}_\text{RPV} 
   = \mathcal{L}_{\lambda} 
   + \mathcal{L}_{\lambda'}, 
   \\&
   \mathcal{L}_\lambda 
   = \lambda_{ijk}  \bigl[ 
   \tilde \nu_{iL} \overline{e}_{kR} e_{jL} 
   + \tilde e_{jL} \overline{e}_{kR} \nu_{iL} 
   + \tilde e_{kR}^* \overline{(\nu_{iL})^c} e_{jL}
   - (i \leftrightarrow j) \bigr] + \text{h.c.}, 
   \\& 
   \mathcal{L}_{\lambda'}
   = \lambda'_{ijk}  \bigl[ 
   \tilde \nu_{iL} \overline{d}_{kR} d_{jL} 
   + \tilde d_{jL} \overline{d}_{kR} \nu_{iL} 
   + \tilde d_{kR}^* \overline{(\nu_{iL})^c} d_{jL} 
   \\& \hspace{10mm} - 
   \tilde e_{iL} \overline{d}_{kR} u_{jL} 
   - \tilde u_{jL} \overline{d}_{kR} e_{jL} 
   - \tilde d_{kR}^* \overline{(e_{iL})^c} u_{jL}
   \bigr] + \text{h.c.}. 
\label{Eq:RPV_L1}   
\end{split}      
\end{equation}

Our interesting situation is that only $\mu$-$e$ conversion is discovered,
and other cLFV processes will never be observed. The situation is realized 
under the following 3 setting on the RPV interaction: 
\begin{enumerate}
\item 
only the third generation slepton contributes to the RPV interactions 
\item
for quarks, flavor diagonal components are much larger than that 
of off-diagonal components, i.e., CKM-like matrix, $\lambda'_{ijj} 
\gg \lambda'_{ijk} (j \neq k)$
\item 
the generation between left-handed and right-handed leptons are 
different, $\lambda_{ijk} (i \neq k \text{ and } j \neq k)$. 
\end{enumerate}
The setting-1 is naturally realized by the RG evolved SUSY spectrum with 
universal soft masses at the GUT scale. For the simplicity, we decouple 
other SUSY particles except for the third generation sleptons. 
The setting-2 is also obtained in most cases unless we introduce additional 
sources of flavor violations. 
The setting-3 is artificially introduced to realize the interesting situation 
in this work, that the COMET find the cLFV process, while all the other 
experiments will not observe anything new at that time (see Introduction).
Under the settings, the general Lagrangian~\eqref{Eq:RPV_L1} is reduced 
as follows, 
\begin{equation}
\begin{split}
   &
   \mathcal{L}_\text{RPV} 
   = \mathcal{L}_{\lambda} 
   + \mathcal{L}_{\lambda'}, 
   \\&
   \mathcal{L}_\lambda = 2 \bigl[ 
   \lambda_{312} \tilde \nu_{\tau L} \overline{\mu} P_L e 
   + \lambda_{321} \tilde \nu_{\tau L} \overline{e} P_L \mu 
   + \lambda_{132} \tilde \tau_L \overline{\mu} P_L \nu_e 
   + \lambda_{231} \tilde \tau_L \overline{e} P_L \nu_\mu
   \\& \hspace{10mm}  
   + \lambda_{123} \tilde \tau_R^* \overline{(\nu_{eL})^c} P_L \mu 
   + \lambda_{213} \tilde \tau_R^* \overline{(\nu_{\mu L})^c} P_L e
   \bigr] + \text{h.c.}, 
   \\& 
   \mathcal{L}_{\lambda'} 
   = \bigl[ 
   \lambda'_{311} \bigl( \tilde \nu_{\tau L} \overline{d} P_L d 
   - \tilde \tau_L \overline{d} P_L u \bigr) 
   + \lambda'_{322} \bigl( \tilde \nu_{\tau L} \overline{s} P_L s 
   - \tilde \tau_L \overline{s} P_L c \bigr)
   \bigr] + \text{h.c.}.  
\label{Eq:RPV_L2}   
\end{split}      
\end{equation}

Some kind of processes described by the Lagrangian \eqref{Eq:RPV_L2} 
strongly depend on the values of $\lambda'_{311}$ and $\lambda'_{322}$. 
In this work, to clarify the dependence and to discuss the discrimination 
of each other, we study three cases: 
\begin{table}[htb]
\begin{tabular}{ll}
~~ case-I & $\lambda'_{311} \neq 0$ and $\lambda'_{322} = 0$ 
\\
~~ case-I\hspace{-1pt}I & $\lambda'_{311} = 0$ and $\lambda'_{322} \neq 0$ 
\\
~~ case-I\hspace{-1pt}I\hspace{-1pt}I & $\lambda'_{311} \neq 0$ 
and $\lambda'_{322} \neq 0$ 
\end{tabular}
\end{table}

\section{Exotic processes in our scenario}  \label{Sec:Obs} 

In our scenario we may have five types of exotic processes: 
$\mu$-$e$ conversion in a nucleus, $pp \to \mu^- e^+$, 
$pp \to jj$,  non-standard interaction (NSI) of neutrinos, and 
muonium conversion $\mu^+e^-\leftrightarrow \mu^-e^+$.
We formulate each reaction rate in our scenario. 

\begin{figure}[t!]
\begin{center}
\includegraphics[width=50mm]{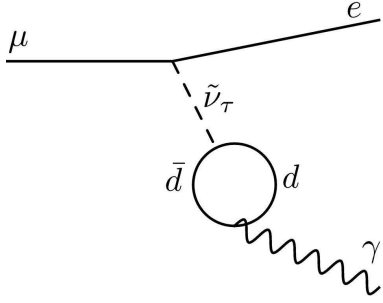}
\caption{Possible one-loop diagram for $\mu\rightarrow e\gamma$.
It is , however, proportional to $q^2q^\mu$ and hence vanish with
 on-shell photon ($q^2=0$) and with $\bar e \gamma_\mu e$ attached
due to gauge symmetry.}
\label{fig:mu-egamma}
\end{center}
\end{figure}

Note that in our scenario other muon cLFV processes 
($\mu \to e \gamma$, $\mu \to 3e$, $\mu^- e^- \to e^- e^-$ 
in muonic atom~\cite{Koike:2010xr}, and so on) occur at two-loop level.  
At one glance the tau sneutrino can connect with the photon via 
d-quark loop shown in Fig.~\ref{fig:mu-egamma}. 
The contribution of the loop of the diagram is
\begin{eqnarray}
 \lambda'\left(-\frac{1}{3}\right) e \frac{m_d q_\mu}{8\pi^2}
\int_0^1 dx(1-2x)\log(m_d^2-(x-x^2)q^2)\ \propto q^2q^\mu,
\end{eqnarray}
where $q$ is the four-momentum of the photon. The contribution to cLFV
is, therefore vanish with
 on-shell photon ($q^2=0$) for $\mu\rightarrow e\gamma$
and with $\bar e \gamma_\mu e$ attached for $\mu\rightarrow 3e$
due to gauge symmetry($q^\mu\bar e\gamma_\mu e=0$).

Thus these processes occur at two-loop level. Furthermore these loop 
processes are extremely suppressed further by higher order couplings, 
gauge invariance, and so on. Therefore we do not study these 
processes here.

\subsection{$\mu$-$e$ conversion}  \label{Sec:mue_conv} 

We briefly review the formulation of the branching ratio of $\mu$-$e$ 
conversion process based on Refs.~\cite{Kim:1997rr, Kitano:2002mt}. 
The $\mu$-$e$ conversion process via the tau sneutrino exchange is 
described by the effective interaction Lagrangian 
\begin{equation}
\begin{split}
   \mathcal{L}_{int} = - \frac{G_F}{\sqrt{2}} \sum_{q=d,s} 
   \Bigl\{ \left( 
   g_{LS(q)} \bar e P_R \mu + g_{RS(q)} \bar e P_L \mu
   \right) \bar q q \Bigr\} + \text{h.c.}, 
\label{Eq:Lint_mueconv}   
\end{split}     
\end{equation}
where $G_F$ is the Fermi coupling constant. 
The coefficients $g_{LS(q)}$ and $g_{RS(q)}$ are derived from the 
RPV interaction Lagrangian [Eq.~\eqref{Eq:RPV_L2}], 
\begin{equation}
\begin{split}
   g_{LS(d)} = \frac{\sqrt{2}}{G_F} \frac{2}{m_{\tilde \nu_\tau}^2} 
   \lambda'_{311} \lambda_{312}^{*}, 
\end{split}     
\end{equation}
\begin{equation}
\begin{split}
   g_{RS(d)} = \frac{\sqrt{2}}{G_F} \frac{2}{m_{\tilde \nu_\tau}^2} 
   \lambda_{311}'^{*} \lambda_{321}, 
\end{split}     
\end{equation}
\begin{equation}
\begin{split}
   g_{LS(s)} = \frac{\sqrt{2}}{G_F} \frac{2}{m_{\tilde \nu_\tau}^2} 
   \lambda'_{322} \lambda_{312}^{*}, 
\end{split}     
\end{equation}
\begin{equation}
\begin{split}
   g_{RS(s)} = \frac{\sqrt{2}}{G_F} \frac{2}{m_{\tilde \nu_\tau}^2} 
   \lambda_{322}'^{*} \lambda_{321}. 
\end{split}     
\end{equation}
The amplitude for the $\mu$-$e$ conversion process is calculated by the 
overlap of wave functions of the initial state muon $\psi_{1S}^{(\mu)}$, 
the final state electron $\psi_{\kappa, W}^{\mu (e)}$ with the eigenvalues 
of the orbital angular momentum $-\kappa$ and of the $z$-component 
angular momentum $\mu$, and the initial and final state nucleus as follows 
\begin{equation}
\begin{split}
   \mathcal{M} = \frac{G_F}{\sqrt{2}} \sum_{q=d,s}  
   \int \hspace{-0.8mm} d^3 \boldsymbol{x} 
   \bigl( 
   g_{LS(q)} \bar \psi_{\kappa, W}^{\mu (e)} P_R \psi_{1S}^{(\mu)} 
   + g_{RS(q)} \bar \psi_{\kappa, W}^{\mu (e)} P_L \psi_{1S}^{(\mu)} 
   \bigr) 
   \hspace{0.5mm} \langle N | \bar q q | N \rangle. 
\end{split}     
\end{equation}
Here we omitted the incoherent conversion process, because its fraction 
is much smaller than the coherent one. The matrix element 
$\langle N | \bar q q | N \rangle$ is given by the atomic number $Z$, 
the mass number $A$, and the proton (neutron) density in nucleus 
$\rho^{(p)}$ ($\rho^{(n)}$), 
\begin{equation}
\begin{split}
   \langle N |\bar q q| N \rangle = Z G_S^{(q,p)} \rho^{(p)} 
   + (A-Z) G_S^{(q,n)} \rho^{(n)}. 
\end{split}     
\end{equation}
The coefficients for scalar operators are evaluated in Ref.~\cite{Kosmas:2001mv}: 
$G_S^{(d,n)} = 5.1$, $G_S^{(d,p)} = 4.3$, and $G_S^{(s,p)} = 
G_S^{(s,n)} = 2.5$. This calculation assumes that the proton and the neutron 
densities are in spherical distribution and normalized as $\int \hspace{-0.8mm} 
dr 4\pi r^2 \rho^{(p,n)}=1$.

The reaction rate of the $\mu$-$e$ conversion is 
\begin{equation}
\begin{split}
   \omega_{conv} = 
   2G_F^2 \left| \tilde g_{LS}^{(p)} S^{(p)} 
   + \tilde g_{LS}^{(n)} S^{(n)} \right|^2 
   + 2G_F^2 \left| \tilde g_{RS}^{(p)} S^{(p)} 
   + \tilde g_{RS}^{(n)} S^{(n)} \right|^2. 
\end{split}     
\end{equation}
The overlap integral of wave functions of muon, electron, and protons 
(neutrons) gives $S^{(p)}$ ($S^{(n)}$) (explicit formulae and details 
of the calculation are explained in Ref.~\cite{Kitano:2002mt}).  We 
list $S^{(p)}$ and $S^{(n)}$ for relevant nuclei of SINDRUM-I\hspace{-1pt}I 
(Au), DeeMe (C and Si), COMET (Al and Ti), Mu2e (Al and Ti), and PRISM 
(Al and Ti) in Table~\ref{Tab:SpSn_cap}. The coefficients $\tilde g_{LS, 
RS}^{(p)}$ and $\tilde g_{LS, RS}^{(n)}$ are 
\begin{equation}
\begin{split}
   \tilde g_{LS, RS}^{(p)} 
   = \sum_{q} G_S^{q, p} g_{LS, RS(q)} 
   = G_S^{d,p} g_{LS, RS(d)} + G_S^{s,p} g_{LS, RS(s)}, 
\end{split}     
\end{equation}
\begin{equation}
\begin{split}
   \tilde g_{LS, RS}^{(n)} 
   = \sum_{q} G_S^{q, n} g_{LS, RS(q)} 
   = G_S^{d,n} g_{LS, RS(d)} + G_S^{s,n} g_{LS, RS(s)}. 
\end{split}     
\end{equation}
Thus the reaction rate of $\mu$-$e$ conversion via the $\tilde \nu_\tau$ 
exchange is obtained as follows, 
\begin{equation}
\begin{split}
   \omega_{conv} 
   &= 
   \frac{16}{m_{\tilde \nu_\tau}^4} 
   \bigl| 
   (4.3 S^{(p)} + 5.1 S^{(n)}) \lambda'_{311} \lambda_{312}^* 
   + 2.5 (S^{(p)} + S^{(n)}) \lambda'_{322} \lambda_{312}^*
   \bigr|^2 
   \\&~ + 
   \frac{16}{m_{\tilde \nu_\tau}^4} 
   \bigl| 
   (4.3 S^{(p)} + 5.1 S^{(n)}) \lambda_{311}'^{*} \lambda_{321} 
   + 2.5 (S^{(p)} + S^{(n)}) \lambda_{322}'^{*} \lambda_{321} 
   \bigr|^2. 
%
\label{Eq:omega_conv_2}
\end{split}     
\end{equation}
The branching ratio of $\mu$-$e$ conversion process is defined by
\begin{equation}
\begin{split}
   \text{BR}(\mu^- N \to e^- N) = \omega_{conv}/\omega_{capt}, 
\end{split}     
\end{equation}
where $\omega_{capt}$ is the muon capture rate of nucleus. We list the 
values of $\omega_{capt}$ in Table~\ref{Tab:SpSn_cap}. 
Assuming $\lambda_{311}'$ and $\lambda_{322}'$ are real and 
$\lambda_{312}^{*} = \lambda_{321} \equiv \lambda$, the branching 
ratio for $N = \text{C}$ is given by
\begin{equation}
\begin{split}
   \text{BR}(\mu^- \text{C} \to e^- \text{C}) 
   &= 
   1.383 \times 10^{-15} 
   \left( \frac{1\text{TeV}}{m_{\tilde \nu_\tau}} \right)^4  
   \left( \frac{\lambda_{311}' \lambda}{10^{-8}} \right)^2 
   \biggl| 
   1 + 0.532 \left( \frac{\lambda_{322}'}{\lambda_{311}'} \right) 
   \biggr|^2 
   \\&= 
   3.913 \times 10^{-16} 
   \left( \frac{1\text{TeV}}{m_{\tilde \nu_\tau}} \right)^4  
   \left( \frac{\lambda_{322}' \lambda}{10^{-8}} \right)^2 
   \biggl| 
   1 + 1.880 \left( \frac{\lambda_{311}'}{\lambda_{322}'} \right) 
   \biggr|^2, 
\label{Eq:BR_C}   
\end{split}     
\end{equation}
for $N = \text{Al}$, 
\begin{equation}
\begin{split}
   \text{BR}(\mu^- \text{Al} \to e^- \text{Al}) 
   &= 
   2.092 \times 10^{-15} 
   \left( \frac{1\text{TeV}}{m_{\tilde \nu_\tau}} \right)^4  
   \left( \frac{\lambda_{311}' \lambda}{10^{-8}} \right)^2 
   \biggl| 
   1 + 0.530 \left( \frac{\lambda_{322}'}{\lambda_{311}'} \right) 
   \biggr|^2 
   \\&= 
   5.881 \times 10^{-16} 
   \left( \frac{1\text{TeV}}{m_{\tilde \nu_\tau}} \right)^4  
   \left( \frac{\lambda_{322}' \lambda}{10^{-8}} \right)^2 
   \biggl| 
   1 + 1.886 \left( \frac{\lambda_{311}'}{\lambda_{322}'} \right) 
   \biggr|^2, 
\label{Eq:BR_Al}   
\end{split}     
\end{equation}
for $N = \text{Si}$, 
\begin{equation}
\begin{split}
   \text{BR}(\mu^- \text{Si} \to e^- \text{Si}) 
   &= 
   2.080 \times 10^{-15} 
   \left( \frac{1\text{TeV}}{m_{\tilde \nu_\tau}} \right)^4  
   \left( \frac{\lambda_{311}' \lambda}{10^{-8}} \right)^2 
   \biggl| 
   1 + 0.532 \left( \frac{\lambda_{322}'}{\lambda_{311}'} \right) 
   \biggr|^2 
   \\&= 
   5.886 \times 10^{-16} 
   \left( \frac{1\text{TeV}}{m_{\tilde \nu_\tau}} \right)^4  
   \left( \frac{\lambda_{322}' \lambda}{10^{-8}} \right)^2 
   \biggl| 
   1 + 1.880 \left( \frac{\lambda_{311}'}{\lambda_{322}'} \right) 
   \biggr|^2, 
\label{Eq:BR_Si}      
\end{split}     
\end{equation}
and for $N = \text{Ti}$, 
\begin{equation}
\begin{split}
   \text{BR}(\mu^- \text{Ti} \to e^- \text{Ti}) 
   &= 
   3.571 \times 10^{-15} 
   \left( \frac{1\text{TeV}}{m_{\tilde \nu_\tau}} \right)^4  
   \left( \frac{\lambda_{311}' \lambda}{10^{-8}} \right)^2 
   \biggl| 
   1 + 0.528 \left( \frac{\lambda_{322}'}{\lambda_{311}'} \right) 
   \biggr|^2 
   \\&= 
   9.962 \times 10^{-16} 
   \left( \frac{1\text{TeV}}{m_{\tilde \nu_\tau}} \right)^4  
   \left( \frac{\lambda_{322}' \lambda}{10^{-8}} \right)^2 
   \biggl| 
   1 + 1.893 \left( \frac{\lambda_{311}'}{\lambda_{322}'} \right) 
   \biggr|^2. 
\label{Eq:BR_Ti}   
\end{split}     
\end{equation}

\begin{table}[h]
\begin{center}
\caption{The overlap factor of wave functions (explicit formulae and details 
of the calculation are explained in Ref.~\cite{Kitano:2002mt}) and the 
muon capture rate  $\omega_{capt}$ for each nucleus. Here $m_\mu$ is 
muon mass. }
\vspace{2mm}
\begin{tabular}{llllll}
\hline
Nucleus 
& $S^{(p)}$
& $S^{(n)}$
& $\omega_{capt}(s^{-1})$
\\ \hline \hline
C
& $0.00308 m_\mu^{5/2}$
& $0.00308 m_\mu^{5/2}$
& $0.388 \times 10^{5}$
\\[0.5mm]
Si
& $0.0179 m_\mu^{5/2}$
& $ 0.0179m_\mu^{5/2}$
& $8.712 \times 10^{5}$
\\[0.5mm]
Al
& $0.0155 m_\mu^{5/2}$
& $ 0.0167m_\mu^{5/2}$
& $7.054 \times 10^{5}$
\\[0.5mm]
Ti
& $ 0.0368m_\mu^{5/2}$
& $ 0.0435m_\mu^{5/2}$
& $ 2.590 \times 10^{6}$
\\[0.5mm]
Au
& $0.0614 m_\mu^{5/2}$
& $ 0.0918m_\mu^{5/2}$
& $ 1.307\times 10^{7}$
\\ \hline
\label{Tab:SpSn_cap}
\end{tabular} 
\end{center}
\end{table}

\subsection{$pp \to \mu^- e^+$ and $pp \to jj$}  \label{Sec:collider} 

We formulate the cross sections of $pp \to \mu^- e^+$ and $pp 
\to jj$ in the RPV scenario. In the scenario, these processes are 
dominated by $s$-channel exchange resonance, and hence the 
cross sections are well approximated by the Breit-Wigner formula. 
The cross section for a final state $f_1 f_2$ is decomposed with 
$\gamma_{\tilde \nu_\tau} = 
\Gamma_{\tilde \nu_\tau}/m_{\tilde \nu_{\tau}}$ as follows
\begin{equation}
\begin{split}
   \sigma(pp \to f_1 f_2) 
   &= 
   F(\sqrt{s}, m_{\tilde \nu_{\tau}}, q_1, q_2) 
   \times \Gamma_{\tilde \nu_\tau}  
   \text{BR}(\tilde \nu_\tau \to q_1 q_2) 
   \text{BR}(\tilde \nu_\tau \to f_1 f_2) 
   \\&= 
   F(\sqrt{s}, m_{\tilde \nu_{\tau}}, q_1, q_2) 
   m_{\tilde \nu_{\tau}} 
   \times \gamma_{\tilde \nu_\tau} 
   \text{BR}(\tilde \nu_\tau \to q_1 q_2) 
   \text{BR}(\tilde \nu_\tau \to f_1 f_2) 
\label{Eq:cross_general}   
\end{split}      
\end{equation}
The front part, $F(\sqrt{s}, m_{\tilde \nu_{\tau}}, q_1, q_2) 
m_{\tilde \nu_{\tau}}$, is determined by the kinematics of each 
process, and is a function of collision energy $\sqrt{s}$, mediator 
mass $m_{\tilde \nu_{\tau}}$, and the flavors of initial quarks 
$(q_1 \text{ and } q_2)$. 
The decay width $\Gamma_{\tilde \nu_{\tau L}}$ is calculated by 
the Lagrangian [Eq.~\eqref{Eq:RPV_L2}], 
\begin{equation}
\begin{split}
   &
   \Gamma_{\tilde \nu_{\tau L}} 
   = \frac{m_{\tilde \nu_{\tau L}}}{16 \pi} 
   \bigl( 3\lambda_{311}'^2 + 3\lambda_{322}'^2
   + 4 \lambda_{312}^2 + 4 \lambda_{321}^2 \bigr). 
\label{Eq:width}   
\end{split}      
\end{equation}
The remaining part, $\gamma_{\tilde \nu_\tau} \text{BR}(\tilde \nu_\tau 
\to q_1 q_2) \text{BR} (\tilde \nu_\tau \to f_1 f_2)$, depends only on 
the coupling constants of RPV interactions.

\begin{figure}[h!]
\begin{center}
\includegraphics[width=90mm]{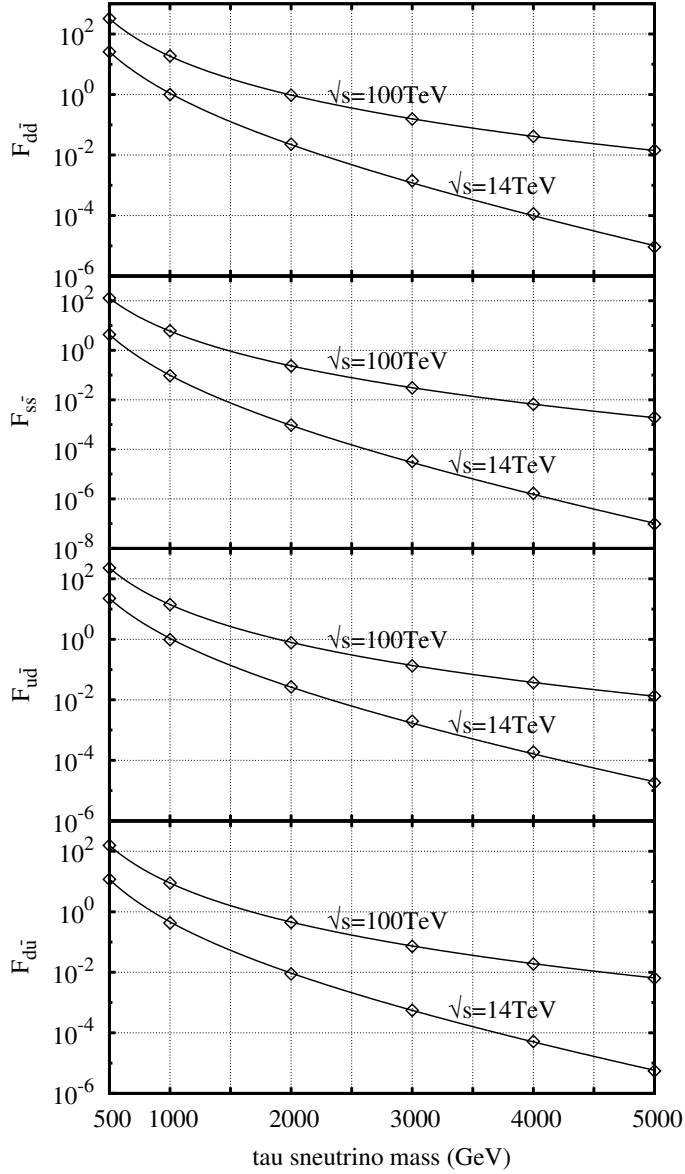}
\caption{Fit functions and numerical results of $F(\sqrt{s}, 
m_{\tilde \nu_{\tau}}, q_1, q_2)$ for collision energy $\sqrt{s} 
= 14\text{TeV}$ and $\sqrt{s} = 100\text{TeV}$. Rotated squares 
are numerical results calculated from Eq.~\eqref{Eq:F}, and lines are 
fit functions. }
\label{fig:Fqqbar}
\end{center}
\end{figure}

First we formulate $F(\sqrt{s}, m_{\tilde \nu_{\tau}}, q_1, q_2)$. 
With regardless of final state, once $\sqrt{s}$, $m_{\tilde \nu_{\tau}}$, 
and an initial state are fixed, $F(\sqrt{s}, m_{\tilde \nu_{\tau}}, q_1, 
q_2)$ is uniquely determined. It is really important and useful for analyzing 
the RPV coupling dependence on the cross sections to derive the explicit 
formula of $F(\sqrt{s}, m_{\tilde \nu_{\tau}}, q_1, q_2)$. 
The expression of $F(\sqrt{s}, m_{\tilde \nu_{\tau}}, q_1, q_2)$ is 
given from Eq.~\eqref{Eq:cross_general}, 
\begin{equation}
\begin{split}
   F(\sqrt{s}, m_{\tilde \nu_{\tau}}, q_1, q_2) 
   = 
   \frac{\sigma(pp \to f_1 f_2)}
  {m_{\tilde \nu_{\tau}} 
   \gamma_{\tilde \nu_\tau} 
   \text{BR}(\tilde \nu_\tau \to q_1 q_2) 
   \text{BR}(\tilde \nu_\tau \to f_1 f_2)}. 
\label{Eq:F}   
\end{split}      
\end{equation}
Numerical results from Eq.~\eqref{Eq:F} are shown by rotated squares in 
Fig.~\ref{fig:Fqqbar}. In Fig.~\ref{fig:Fqqbar}, we use an abbreviation 
$F_{q_1 q_2}$ as $F(\sqrt{s}, m_{\tilde \nu_{\tau}}, q_1, q_2)$. 
For each set of $\sqrt{s}$ and initial state quarks, we can parameterize 
$F(\sqrt{s}, m_{\tilde \nu_{\tau}}, q_1, q_2)$ as a function of 
$m_{\tilde \nu_{\tau}}$ as follows, 
\begin{equation}
\begin{split}
   F(\sqrt{s}, m_{\tilde \nu_{\tau}}, q_1, q_2) 
   = 
   \alpha \times 10^{-\beta m_{\tilde \nu_{\tau}}}
   m_{\tilde \nu_{\tau}}^{-\gamma} 
   \hspace{1.5mm} [\text{pb} \cdot \text{GeV}^{-1}] 
   \hspace{1mm},  
\label{Eq:Ffit}   
\end{split}      
\end{equation}
where coefficients $\alpha$, $\beta$, and $\gamma$ are calculated from 
numerical calculations of $F(\sqrt{s}, m_{\tilde \nu_{\tau}}, q_1, q_2)$, 
and we list the coefficients in table~\ref{Tab:F}. 
The fitted function of $F(\sqrt{s}, m_{\tilde \nu_{\tau}}, q_1, q_2)$ 
for collision energy $\sqrt{s} = 14\text{TeV}$ and $\sqrt{s} = 
100\text{TeV}$ are shown by lines in Fig.~\ref{fig:Fqqbar}.

\begin{table}[h]
\begin{center}
\caption{The coefficients for fit function of $F(\sqrt{s}, 
m_{\tilde \nu_\tau}, q_1, q_2)$ (see Eq.~\eqref{Eq:F}) for each 
set of the collision energy $\sqrt{s}$ and initial state quarks. 
We use CTEQ6L parton distribution function \cite{Pumplin:2002vw} 
for the evaluation.}
\vspace{2mm}
\begin{tabular}{llllll}
\hline
 ($\sqrt{s}$, $q_1$, $q_2$)
& $\alpha \hspace{0.9mm} [\text{pb} \cdot \text{GeV}^{\gamma-1}]$
& $\beta \hspace{0.9mm} [\text{GeV}^{-1}]$
& $\gamma$
\\ \hline \hline
$(14\text{TeV}, d, \bar d)$
& $ 1.352 \times 10^{11}$
& $ 6.500 \times 10^{-4}$
& $ 3.480$
\\[0.5mm]
$(14\text{TeV}, u, \bar d)$
& $ 6.652 \times 10^{10}$
& $ 5.900 \times 10^{-4}$
& $ 3.400$
\\[0.5mm]
$(14\text{TeV}, d, \bar u)$
& $ 2.233 \times 10^{11}$
& $ 5.800 \times 10^{-4}$
& $ 3.700$
\\[0.5mm]
$(14\text{TeV}, s, \bar s)$
& $ 2.248 \times 10^{12}$
& $ 7.600 \times 10^{-4}$
& $ 4.200$
\\[0.5mm]
$(100\text{TeV}, d, \bar d)$
& $ 2.220 \times 10^{13}$
& $ 8.000 \times 10^{-5}$
& $ 4.000$
\\[0.5mm]
$(100\text{TeV}, u, \bar d)$
& $ 8.385 \times 10^{12}$
& $ 7.500 \times 10^{-5}$
& $ 3.900$
\\[0.5mm]
$(100\text{TeV}, d, \bar u)$
& $ 1.084 \times 10^{13}$
& $ 8.500 \times 10^{-5}$
& $ 4.000$
\\[0.5mm]
$(100\text{TeV}, s, \bar s)$
& $ 3.265\times 10^{13}$
& $ 1.400 \times 10^{-4}$
& $ 4.200$
\\ \hline
\label{Tab:F}
\end{tabular} 
\end{center}
\end{table}

From Eq.~\eqref{Eq:cross_general}, the cross section of $pp \to \mu^- 
e^+$ is analytically calculated with the decay rate [Eq.~\eqref{Eq:width}] 
and the fit function of $F(\sqrt{s}, m_{\tilde \nu_\tau}, q_1, q_2)$ 
[Eq.~\eqref{Eq:Ffit}] as follows, 
\begin{equation}
\begin{split}
   \sigma(pp \to \mu^- e^+) 
   &= 
   \sum_{i=1,2}  \Bigl\{
   F(\sqrt{s}, m_{\tilde \nu_{\tau}}, d_i, \bar d_i) 
   m_{\tilde \nu_{\tau}} \times 
   \frac{1}{16 \pi} 
   \bigl( 3\lambda_{311}'^2 + 3\lambda_{322}'^2
   + 4 \lambda_{312}^2 + 4 \lambda_{321}^2 \bigr) 
   \\& ~~ \times 
   \frac{3\lambda_{3ii}'^2}
   {3\lambda_{311}'^2 + 3\lambda_{322}'^2
   + 4 \lambda_{312}^2 + 4 \lambda_{321}^2} 
   \cdot
   \frac{4 \lambda_{312}^2}
   {3\lambda_{311}'^2 + 3\lambda_{322}'^2
   + 4 \lambda_{312}^2 + 4 \lambda_{321}^2} \Bigr\}. 
\label{Eq:sigma_muebar}   
\end{split}      
\end{equation}
Here $d_1 = d$ and $d_2 = s$. 
The cross section of dijet production, $\sigma(pp \to jj)$, is similarly
calculated as follows\footnote{Both the $s$-channel and $t$-channel 
$\tilde \nu_{\tau_L}$ ($\tilde \tau_L$) exchange processes 
contribute the dijet production in our scenario. Since the $s$-channel 
processes are highly dominant, we can formulate $\sigma(pp \to jj)$ 
with the Breit-Wigner formula.}, 
\begin{equation}
\begin{split}
   \sigma(pp \to jj) 
   &= 
   \frac{9}{16\pi}  
   \Bigl\{F_{d \bar d} + F_{u \bar d} + F_{\bar u d} \Bigr\} 
   m_{\tilde \nu_{\tau}}
   \times 
   \frac{\lambda_{311}'^4}
   {3\lambda_{311}'^2 + 3\lambda_{322}'^2
   + 4 \lambda_{312}^2 + 4 \lambda_{321}^2} 
   \\& ~ + 
   \frac{9}{16\pi}  
   \Bigl\{F_{d \bar d} + F_{u \bar d} 
   + F_{\bar u d} + F_{s \bar s} \Bigr\} 
   m_{\tilde \nu_{\tau}}
   \times
   \frac{\lambda_{311}'^2 \lambda_{322}'^2}
   {3\lambda_{311}'^2 + 3\lambda_{322}'^2
   + 4 \lambda_{312}^2 + 4 \lambda_{321}^2} 
   \\& ~ + 
   \frac{9}{16\pi} 
   \Bigl\{F_{s \bar s} \Bigr\} 
   m_{\tilde \nu_{\tau}}
   \times
   \frac{\lambda_{322}'^4}
   {3\lambda_{311}'^2 + 3\lambda_{322}'^2
   + 4 \lambda_{312}^2 + 4 \lambda_{321}^2}. 
\label{Eq:cross_jet}   
\end{split}      
\end{equation}
The terms of $F_{u \bar d}$ and $F_{\bar u d}$ are the left-handed stau 
exchange contributions. Since the tau sneutrino and the stau are component 
of the $SU(2)_L$ doublet, we assumed their degeneracy in mass. 
In the case 1 (case 2), only the first line (third line) contributes to the dijet 
production.

\subsection{NSI}  \label{Sec:NSI} 

With the interaction Eq.~\eqref{Eq:RPV_L2}, there is  modification on neutrino
oscillation physics. It is called Non-Standard Interaction (NSI).
Particularly, there is a strong enhancement, called chiral enhancement.

Conventional beam experiments use neutrino emitted by $\pi$ decay.
In the presence of the interaction Eq.~\eqref{Eq:RPV_L2}, we have an effective
operator which causes a $\pi$ decay with LFV in the follwing way.

The effective Lagrangian is
\begin{eqnarray}
 {\mathcal L}=\frac{2 \lambda_{312}^*\lambda'_{311}}
{m^2_{\tilde{\tau}}}\bar\nu_e\mu_R\bar d_Ru_L 
+\frac{2 \lambda_{321}^*\lambda'_{311}}
{m^2_{\tilde{\tau}}}\bar\nu_\mu e_R\bar{d}_Ru_L 
+ {\rm h.c.}.
\end{eqnarray}
Amplitude for 
$\pi^+\rightarrow \mu^+\nu_e$ is proportional to
\begin{eqnarray}
 \mathcal{M}\propto <\nu_e\mu^+|\bar\nu_e\mu_R|0><0|\bar{d}_Ru_L|\pi^+>.
\end{eqnarray}
Since \cite{PionDecay}
\begin{eqnarray}
\bar{d}_Ru_L=\frac{i}{m}\partial_\mu(\bar{u}\gamma^\mu\gamma_5d)
\end{eqnarray}
using equation of motion and $m=m_u+m_d$, a sum of u- and d- quark masses.
Therefore the magnitude of the amplitude is enhanced by \cite{Ota:2002na}
\begin{eqnarray}
\frac{m_\pi^2}{m_\mu m}
\end{eqnarray}
comparing with usual current-current interaction. 
Here $m_\pi $ is $\pi$  mass.
This is the chiral
enhancement.
We can expect 30 times enhancement.
It interferes with the usual $\pi$ decay though it depends on the phase
 of $\lambda_{312}^*\lambda'_{311}$, and can affect the neutrino
 oscillation experiment with conventional beam.

The strength of the NSIs is parameterized by the relative strength with
the weak interaction. For the conventional beam experiment the effect of
$\pi^+\rightarrow\mu^+\nu_e$ is denoted by $\epsilon^S_{\mu e}$ and
\begin{equation}
 \epsilon^S_{\mu e}=
 \sqrt{2}\frac{m_\pi^2}{m_\mu m}\frac{2 \lambda_{312}^*\lambda'_{311}}
{G_Fm^2_{\tilde{\tau}}}. 
\end{equation}
With this interaction, the $\mu$ flavor eigenstate in the $\pi$ decay
, which is denoted by $(0,1,0)$
in the lepton flavor eigenstates, is deformed to be $(\epsilon^S_{\mu e},1,0)$.

Note that the operator $\bar e_R\nu_\mu\bar u_L d_R$ causes
$\pi^-\rightarrow e^-\bar\nu_\mu$. It has an electron final state.
Since there is $\mu$ in $\pi$ decay more than 99\% case 
it  cannot interfere with a usual $\pi$ decay and hence it has no effect on
neutrino oscillation experiment. Furthermore $\pi$ decay cannot be
caused by operators with $\lambda'_{322}$.
It means, in principle, with neutrino oscillation experiment 
operator with $\lambda^*_{312}\lambda'_{311}$ can be distinguished from
others.

In principle, there are other NSI processes in matter effect and
detection process. They are, however, absent or tiny.  Indeed
there is no matter effect as $\lambda_{311}$ is absent. 
The NSI effect detection process is suppressed by chirality
since the interaction is not (V-A)(V-A) type \cite{Ota:2001pw}.

\subsection{Muonium conversion}\label{Sec:muonium}

In the scenario, muonium ($M = \mu^+ e^-$) converts to autimuonium 
($\bar M = \mu^- e^+$) via the tau sneutrino exchange. The $M$-$\bar M$ 
conversion is described by $(V \pm A) \times (V \pm A)$ form 
interaction~\cite{Horikawa:1995ae}
\begin{equation}
\begin{split}
   \mathcal{L}(M \to \bar M) = \frac{G_{M \bar M}}{\sqrt{2}} 
   (\bar \mu \gamma_\mu P_L e) (\bar \mu \gamma^\mu P_R e) 
   + \text{h.c.}. 
\label{Eq:L_MMbar_1}   
\end{split}      
\end{equation}
Here $G_{M \bar M}$ is an effective coupling analogous to the Fermi 
coupling constant $G_F$. Latest experimental limit of $M$-$\bar M$ 
conversion is set on the $G_{M \bar M}$, $G_{M \bar M} \leq  3.0 
\times 10^{-3} G_F$~\cite{Willmann:1998gd}. 
We derive the interaction Lagrangian describing the $M$-$\bar M$ conversion 
by the Fierz transformation from the fundamental Lagrangian~\eqref{Eq:RPV_L2} 
as follows, 
\begin{equation}
\begin{split}
   \mathcal{L}(M \to \bar M) = 
   \frac{\lambda_{321} \lambda^*_{312}}{2 m_{\tilde \nu_\tau}^2} 
   (\bar \mu \gamma_\mu P_L e) (\bar \mu \gamma^\mu P_R e) 
   + \text{h.c.}. 
\label{Eq:L_MMbar_1}   
\end{split}      
\end{equation}
Thus the upper bound from $M$-$\bar M$ conversion search experiment is 
\begin{equation}
\begin{split}
   |\lambda_{321} \lambda^*_{312}| 
   \left( \frac{\text{1TeV}}{m_{\tilde \nu_\tau}} \right)^2 \leq 
   4.948 \times 10^{-2}. 
\label{Eq:bound_MMbar}   
\end{split}      
\end{equation}

\section{Numerical result}  \label{Sec:result} 

\begin{table}[t]
\caption{Current and future experimental limits on the $\mu$-$e$ 
conversion branching ratio 
and the upper limits on $\lambda' \lambda$ corresponding to each 
experimental limit. }
\vspace{2mm}
\hspace{-3mm}
\small{
{\renewcommand\arraystretch{1.7}
\begin{tabular}{llllll}
\hline
Experiment 
& BR limit
& Limit on $\lambda'_{311} \lambda$ (case-I)
& Limit on $\lambda'_{322} \lambda$ (case-I\hspace{-1pt}I)
& Limit on $\lambda' \lambda$ (case-I\hspace{-1pt}I\hspace{-1pt}I)
\\ \hline \hline
SINDRUM
& $7 \times 10^{-13}$~\cite{Bertl:2006up}
& $1.633 \times 10^{-7} \Bigl( \dfrac{m_{\tilde \nu_\tau}}{1\text{TeV}} \Bigr)^2$
& $3.170 \times 10^{-7} \Bigl( \dfrac{m_{\tilde \nu_\tau}}{1\text{TeV}} \Bigr)^2$
& $1.072 \times 10^{-7} \Bigl( \dfrac{m_{\tilde \nu_\tau}}{1\text{TeV}} \Bigr)^2$
\\[0.5mm]
DeeMe
& $5 \times 10^{-15}$~\cite{Natori:2014yba}
& $ 1.550 \times 10^{-8} \Bigl( \dfrac{m_{\tilde \nu_\tau}}{1\text{TeV}} \Bigr)^2$
& $ 2.915 \times 10^{-8} \Bigl( \dfrac{m_{\tilde \nu_\tau}}{1\text{TeV}} \Bigr)^2$
& $ 1.012 \times 10^{-8} \Bigl( \dfrac{m_{\tilde \nu_\tau}}{1\text{TeV}} \Bigr)^2$
\\[0.5mm]
COMET-I 
& $7 \times 10^{-15}$~\cite{Kuno:2013mha}
& $1.830 \times 10^{-8} \Bigl( \dfrac{m_{\tilde \nu_\tau}}{1\text{TeV}} \Bigr)^2$ 
& $3.504 \times 10^{-8} \Bigl( \dfrac{m_{\tilde \nu_\tau}}{1\text{TeV}} \Bigr)^2$
& $1.196 \times 10^{-8} \Bigl( \dfrac{m_{\tilde \nu_\tau}}{1\text{TeV}} \Bigr)^2$
\\[0.5mm]
COMET-I\hspace{-1pt}I 
&$3 \times 10^{-17}$~\cite{Kuno:2013mha}
& $1.198 \times 10^{-9} \Bigl( \dfrac{m_{\tilde \nu_\tau}}{1\text{TeV}} \Bigr)^2$
& $2.294 \times 10^{-9} \Bigl( \dfrac{m_{\tilde \nu_\tau}}{1\text{TeV}} \Bigr)^2$
& $7.827 \times 10^{-10} \Bigl( \dfrac{m_{\tilde \nu_\tau}}{1\text{TeV}} \Bigr)^2$
\\[0.5mm]
PRISM 
& $7 \times 10^{-19}$~\cite{Kuno:2013mha}
& $1.830 \times 10^{-10} \Bigl( \dfrac{m_{\tilde \nu_\tau}}{1\text{TeV}} \Bigr)^2$
& $3.504 \times 10^{-10} \Bigl( \dfrac{m_{\tilde \nu_\tau}}{1\text{TeV}} \Bigr)^2$
& $1.196 \times 10^{-10} \Bigl( \dfrac{m_{\tilde \nu_\tau}}{1\text{TeV}} \Bigr)^2$
\\ \hline
\end{tabular} 
}
}
\label{Tab:mue_conv}
\end{table}

We are now in a position to show numerical results. 
Table~\ref{Tab:mue_conv} shows the current experimental limit and 
the future single event sensitivity for $\mu$-$e$ conversion process, 
and shows the upper limits on the combination of the RPV couplings, 
$\lambda' \lambda$, corresponding to the limit and the sensitivities in 
each experiment. 
In the calculation of the upper limits, we take Au, Si, and Al for target 
nucleus of SINDRUM-I\hspace{-1pt}I, DeeMe, and other experiments, 
respectively.

$\mu$-$e$ conversion search is a reliable probe to both the RPV 
couplings and tau sneutrino mass. 
The current experimental limit puts strict limit on the RPV couplings, 
$\lambda' \lambda \lesssim 10^{-7}$ for $m_{\tilde \nu_\tau} = 
1\text{TeV}$ and $\lambda' \lambda \lesssim 10^{-5}$ for 
$m_{\tilde \nu_\tau} = 3\text{TeV}$, respectively. 
In near future, the accessible RPV couplings will be extended by more 
than 3 orders of current limits, $\lambda' \lambda \simeq 10^{-10}$ 
for $m_{\tilde \nu_\tau} = 1\text{TeV}$ and $\lambda' \lambda 
\simeq 10^{-8}$ for $m_{\tilde \nu_\tau} = 3\text{TeV}$, 
respectively.

The $\mu$-$e$ conversion process is one of the clear signatures for 
the RPV scenario, but it is not the sufficient evidence of the scenario. 
We must check the correlations among the reaction rates of 
$\mu$-$e$ conversion process, the cross sections of $pp \to 
\mu^- e^+$ and $pp \to jj$, and so on in order to discriminate 
the case-I, -I\hspace{-1pt}I, and -I\hspace{-1pt}I\hspace{-1pt}I 
each other and to confirm the RPV scenario. 
In the following subsections, in each case, we show the correlations, 
and discuss the parameter determination.

\subsection{Case-I ($\lambda'_{311} \neq 0$, $\lambda'_{322} = 0$)}  
\label{Sec:311} 

\begin{figure}[t!]
\hspace{-7mm}
\begin{tabular}{cc}
\subfigure[$m_{\tilde \nu_\tau} = 1$TeV. $\sqrt{s}= 14$TeV. ]{
\includegraphics[scale=0.68]{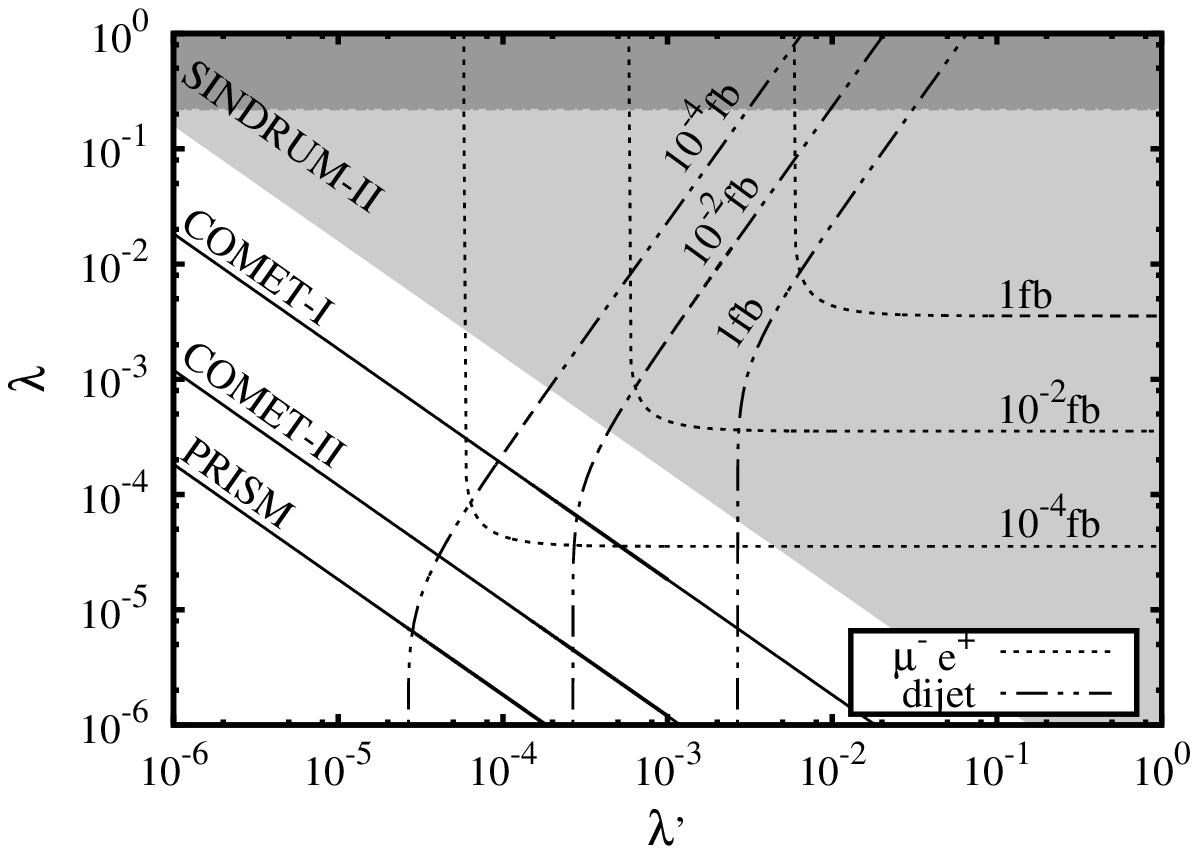}
\label{left1}
} & \hspace{-12mm}
\subfigure[$m_{\tilde \nu_\tau} = 1$TeV. $\sqrt{s}= 100$TeV. ]{
\includegraphics[scale=0.68]{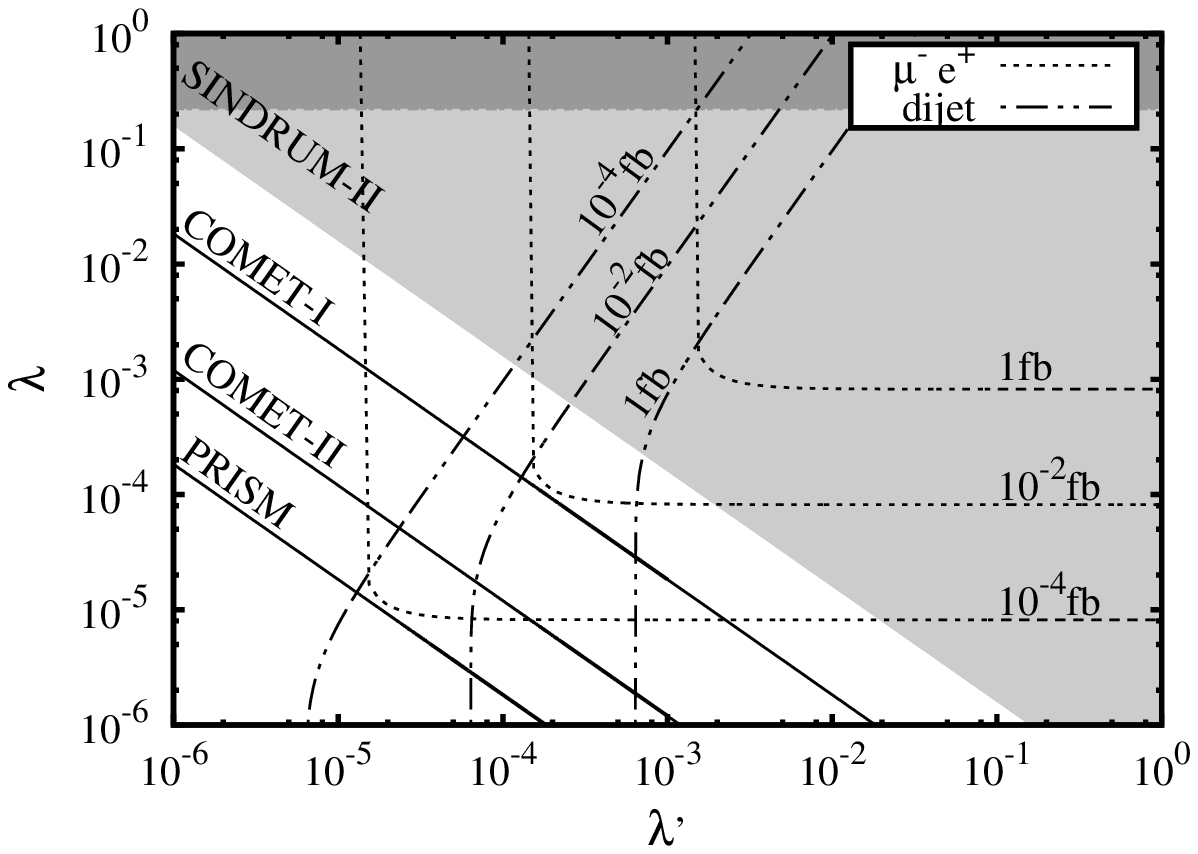}
\label{right1}
} \\
\subfigure[$m_{\tilde \nu_\tau} = 3$TeV. $\sqrt{s}= 14$TeV. ]{
\includegraphics[scale=0.68]{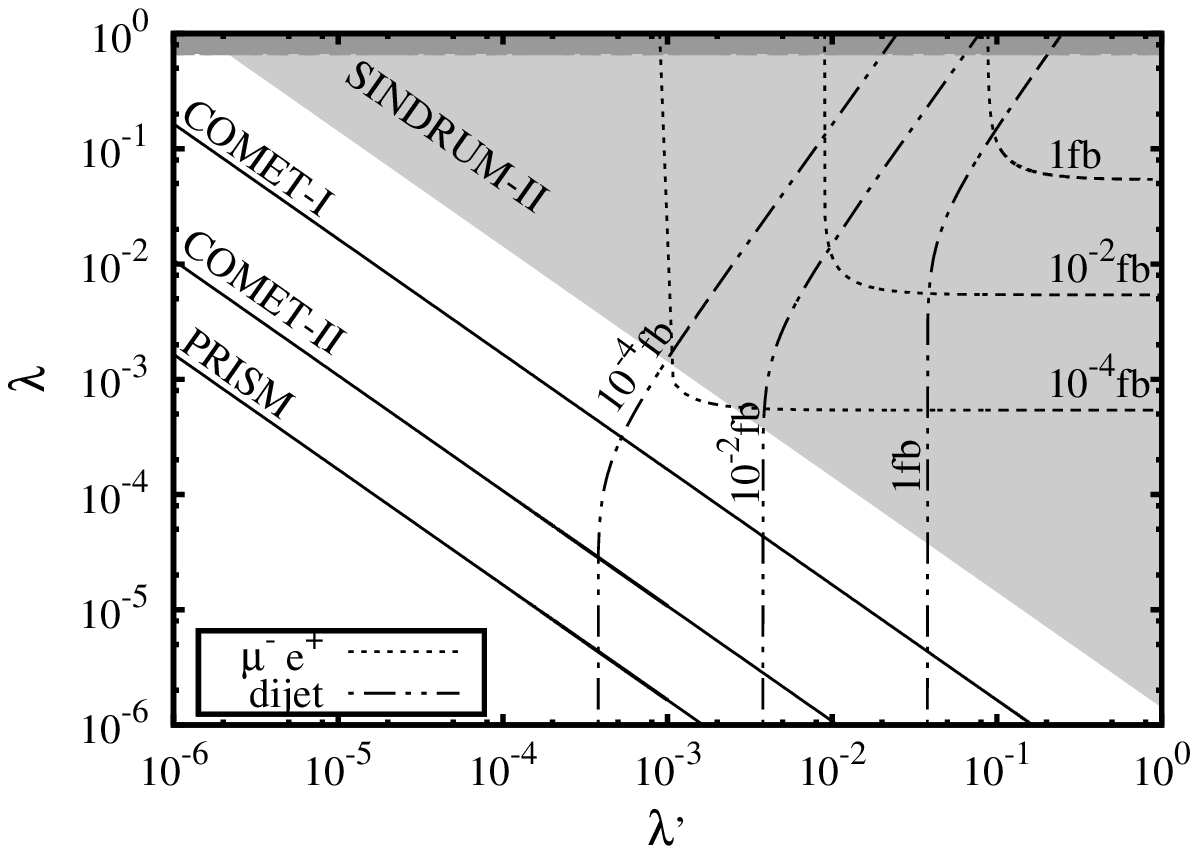}
\label{left2}
} & \hspace{-12mm}
\subfigure[$m_{\tilde \nu_\tau} = 3$TeV. $\sqrt{s}= 100$TeV. ]{
\includegraphics[scale=0.68]{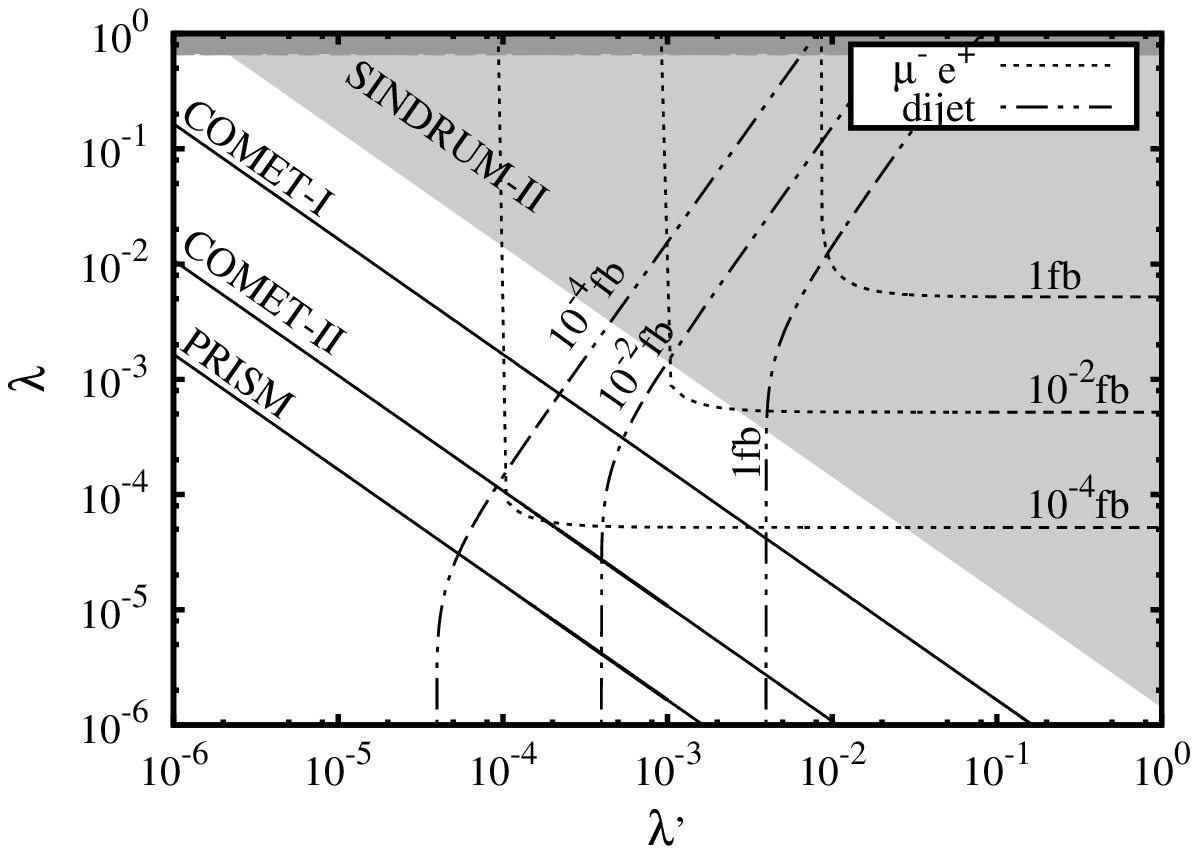}
\label{right2}
} \\
\end{tabular}
\caption{Contour plot of $\sigma(pp \to \mu^- e^+)$, $\sigma(pp 
\to dijet)$, and BR($\mu^- N \to e^- N$) in the case-I for 
(a) $m_{\tilde \nu_\tau} = 1$TeV and $\sqrt{s}=14$TeV 
(b) $m_{\tilde \nu_\tau} = 1$TeV and $\sqrt{s}=100$TeV 
(c) $m_{\tilde \nu_\tau} = 3$TeV and $\sqrt{s}=14$TeV 
(d) $m_{\tilde \nu_\tau} = 3$TeV and $\sqrt{s}=100$TeV. 
For simplicity, we take universal RPV coupling, $\lambda \equiv 
\lambda_{312} = \lambda_{321} = -\lambda_{132} = 
-\lambda_{231} $. Light shaded region is excluded by the 
$\mu$-$e$ conversion search~\cite{Bertl:2006up}, 
and dark shaded band is excluded region by the $M$-$\bar M$ 
conversion search~\cite{Willmann:1998gd}.}
\label{Fig:cont_I}
\end{figure}

\begin{figure}[t!]
\hspace{-6mm}
\begin{tabular}{cc}
\subfigure[$\text{N}=\text{C}$ and $\sqrt{s}= 14$TeV.]{
\includegraphics[scale=0.63]{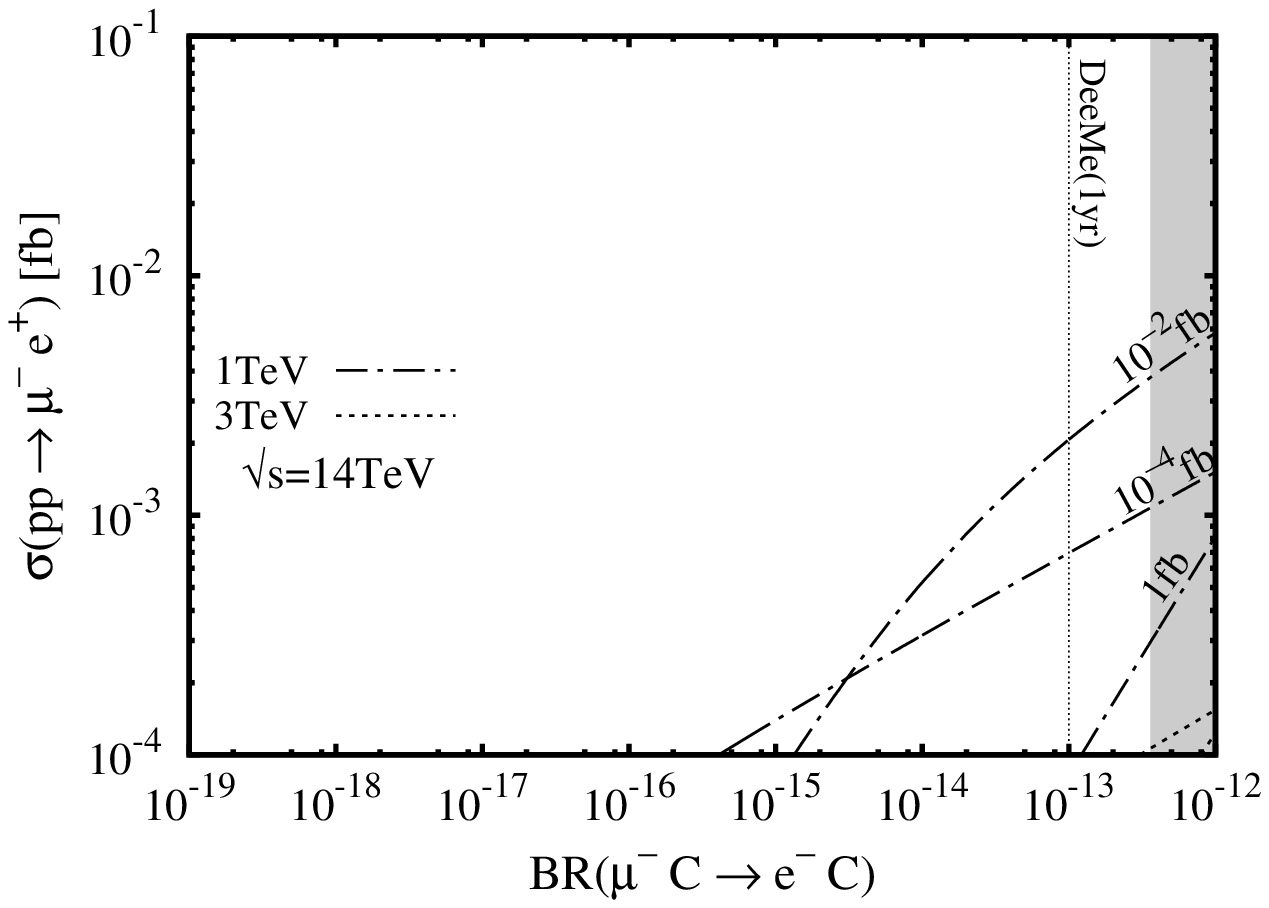}
\label{case1_a}
} & \hspace{-14mm}
\subfigure[$\text{N}=\text{C}$ and $\sqrt{s}= 100$TeV.]{
\includegraphics[scale=0.63]{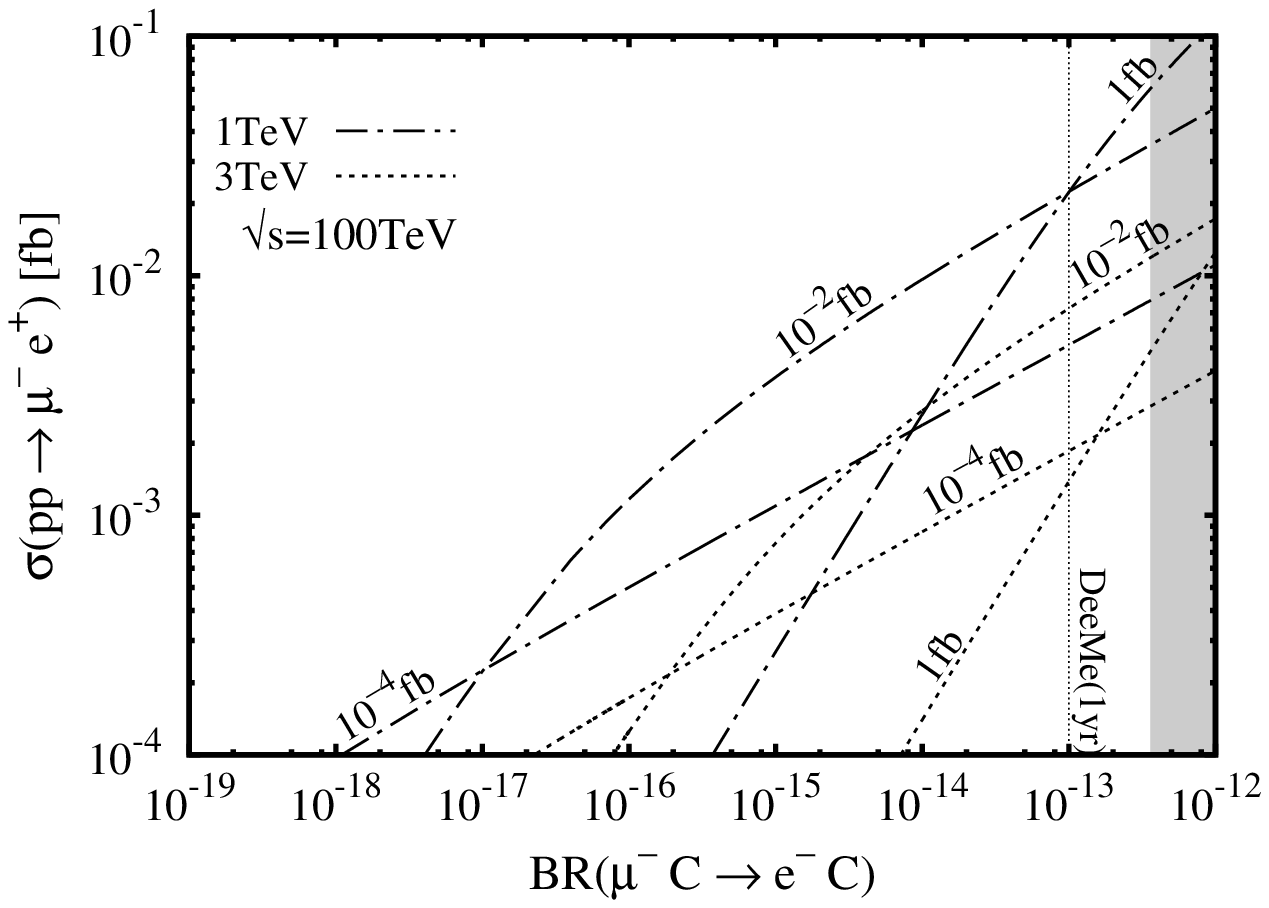}
\label{case1_b}
} \\[-4.5mm]
\subfigure[$\text{N}=\text{Si}$ and $\sqrt{s}= 14$TeV.]{
\includegraphics[scale=0.63]{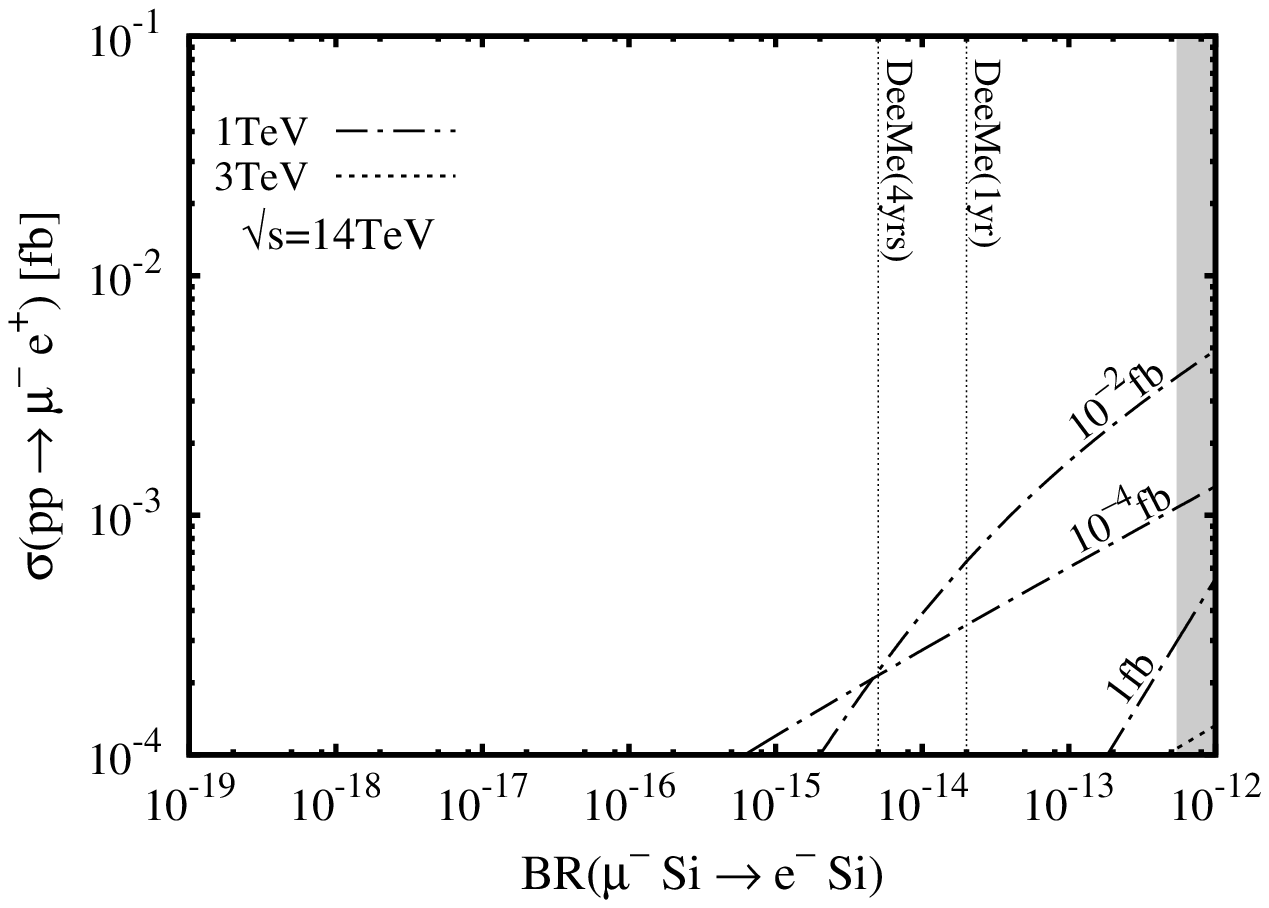}
\label{case1_c}
} & \hspace{-14mm}
\subfigure[$\text{N}=\text{Si}$ and $\sqrt{s}= 100$TeV.]{
\includegraphics[scale=0.63]{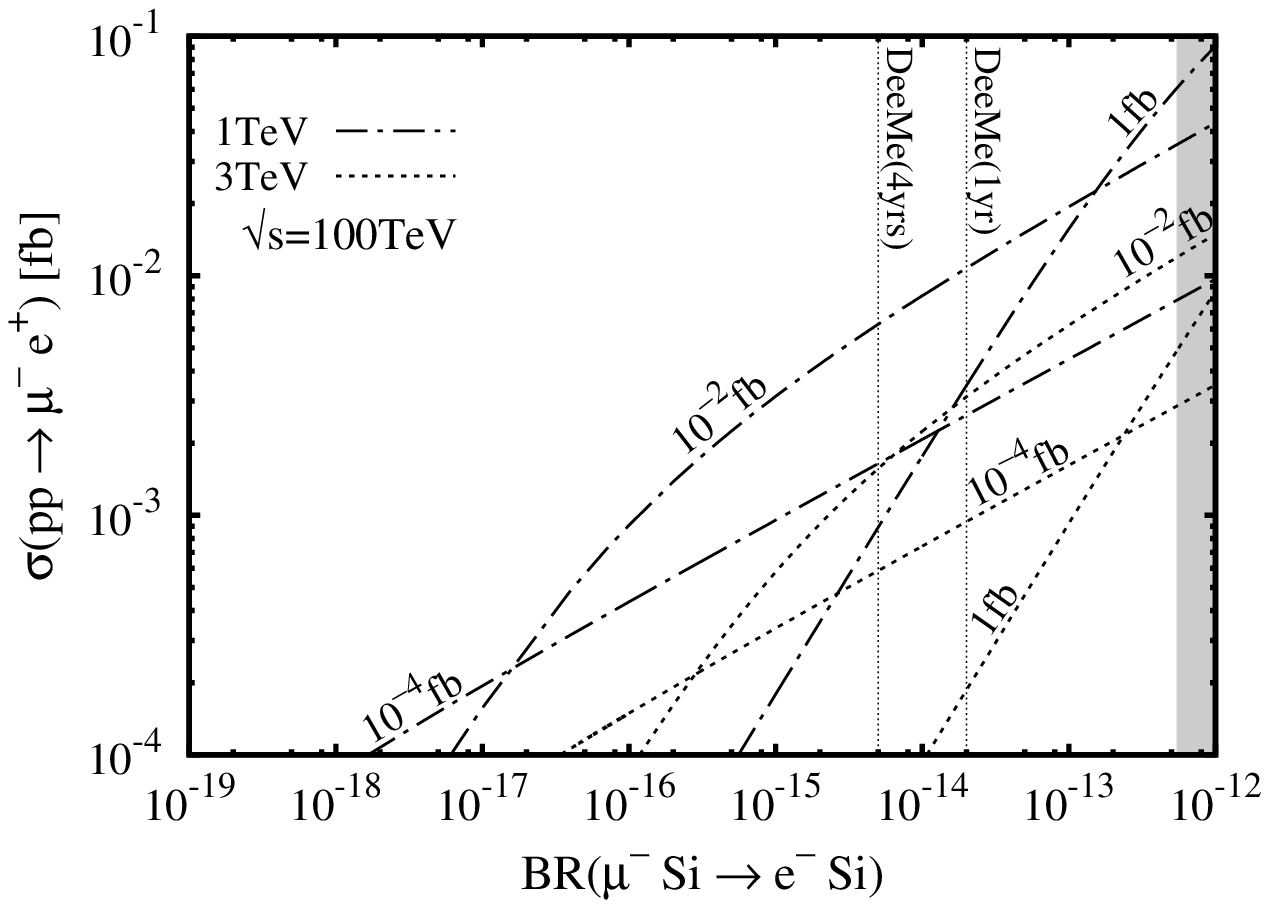}
\label{case1_d}
} \\
\end{tabular}
\caption{$\sigma (pp \to \mu^- e^+)$ as a function of $\text{BR} 
(\mu^- N \to e^- N)$ for each $\sigma (pp \to jj)$ 
in the case-I. $\sigma (pp \to jj)$ are attached on each line. 
Results for $m_{\tilde \nu_\tau} = 1\text{TeV}$ 
($m_{\tilde \nu_\tau} = 3\text{TeV}$) are given by dot-dashed 
line (dotted line). Shaded region in each panel is the excluded 
region by the SINDRUM-I\hspace{-1pt}I experiment. Left panels 
show the results 
for the collision energy $\sqrt{s} = 14\text{TeV}$, and right panels 
show the results for $\sqrt{s} = 100\text{TeV}$. We take C [(a) 
and (b)],  and Si [(c) and (d)] for the target nucleus of 
$\mu$-$e$ conversion process. }
\label{Fig:sigma_vs_BR_I_1}
\end{figure}

\begin{figure}[t!]
\hspace{-6mm}
\begin{tabular}{cc}
\subfigure[$\text{N}=\text{Al}$ and $\sqrt{s}= 14$TeV.]{
\includegraphics[scale=0.63]{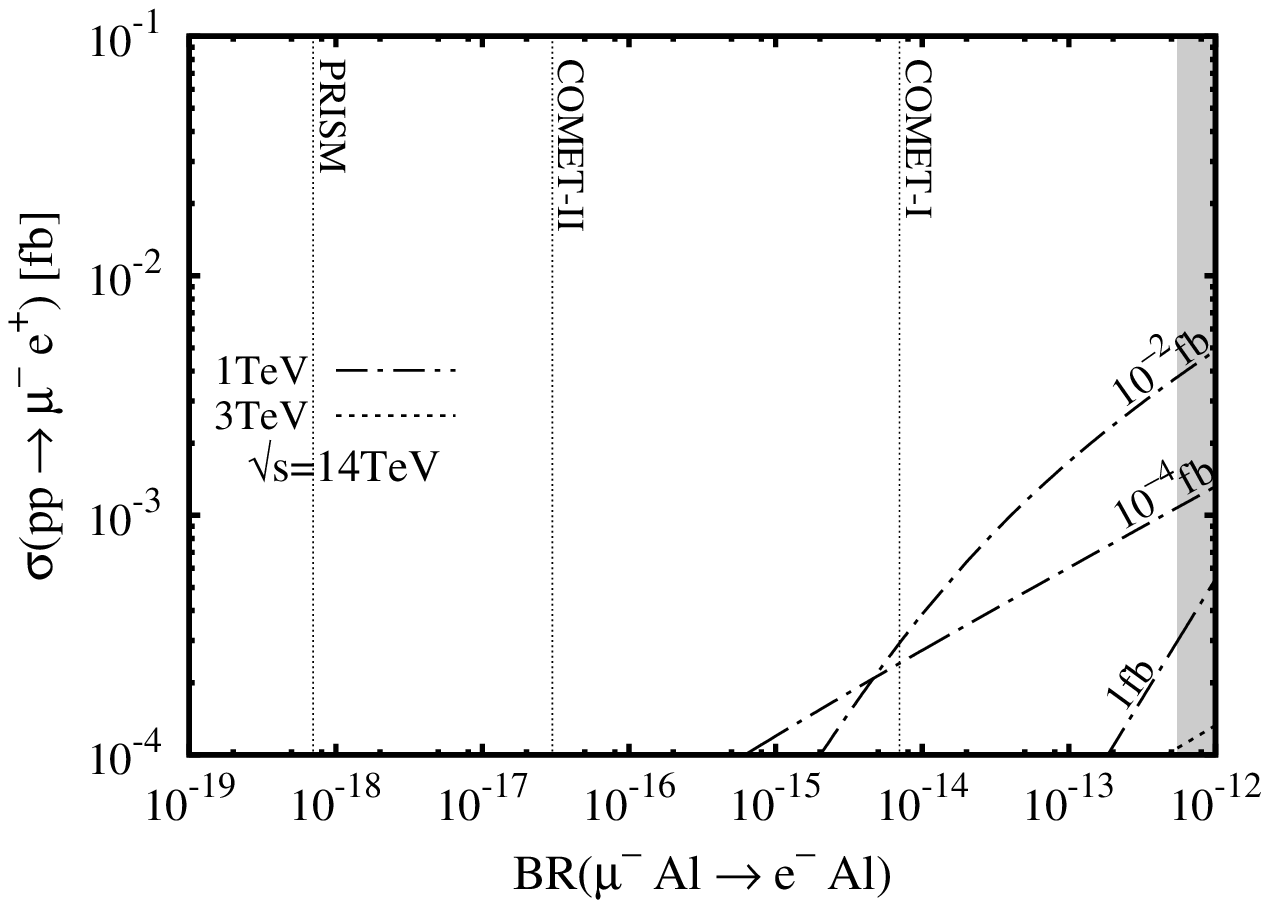}
\label{case1_e}
} & \hspace{-14mm}
\subfigure[$\text{N}=\text{Al}$ and $\sqrt{s}= 100$TeV.]{
\includegraphics[scale=0.63]{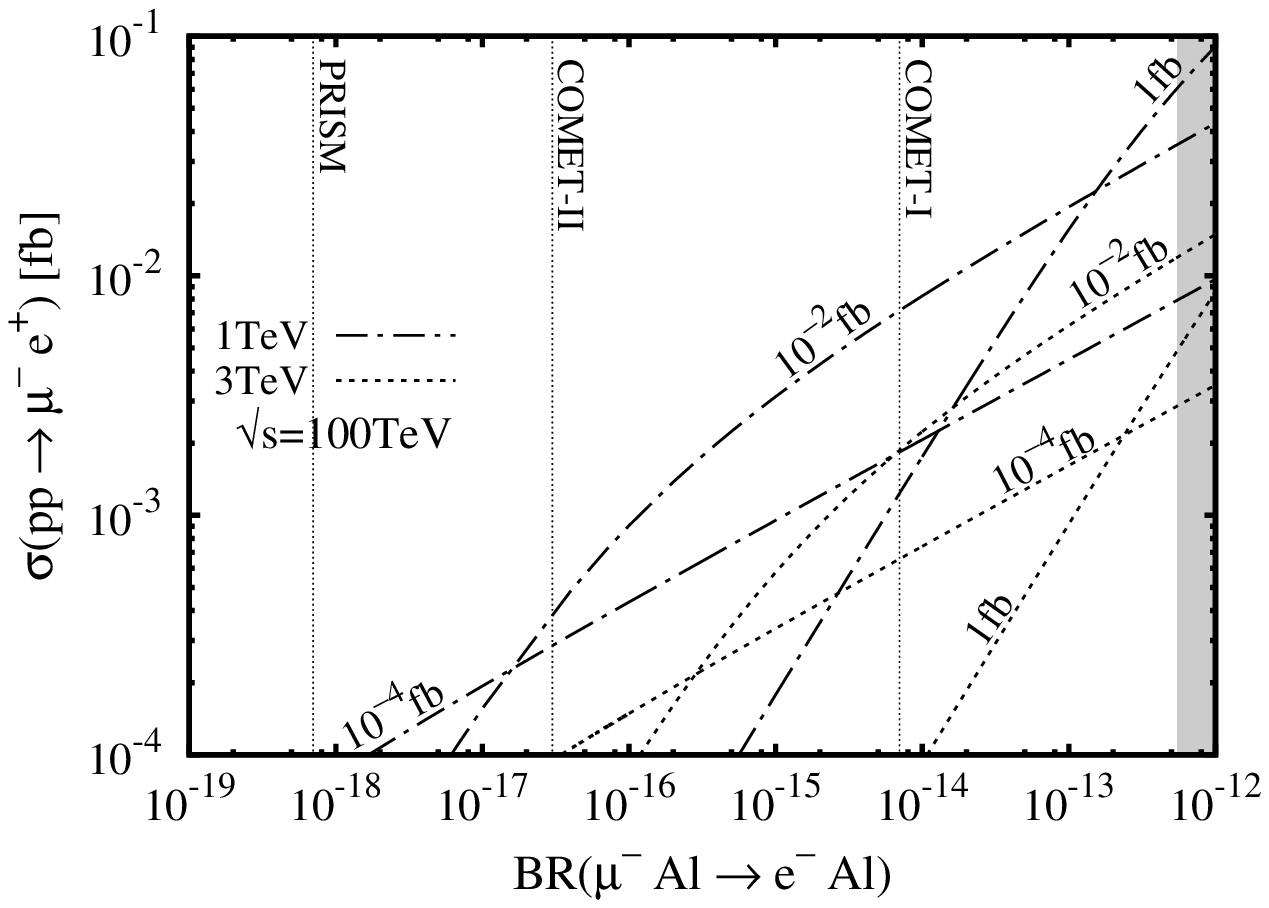}
\label{case1_f}
} \\[-4.5mm]
\subfigure[$\text{N}=\text{Ti}$ and $\sqrt{s}= 14$TeV.]{
\includegraphics[scale=0.63]{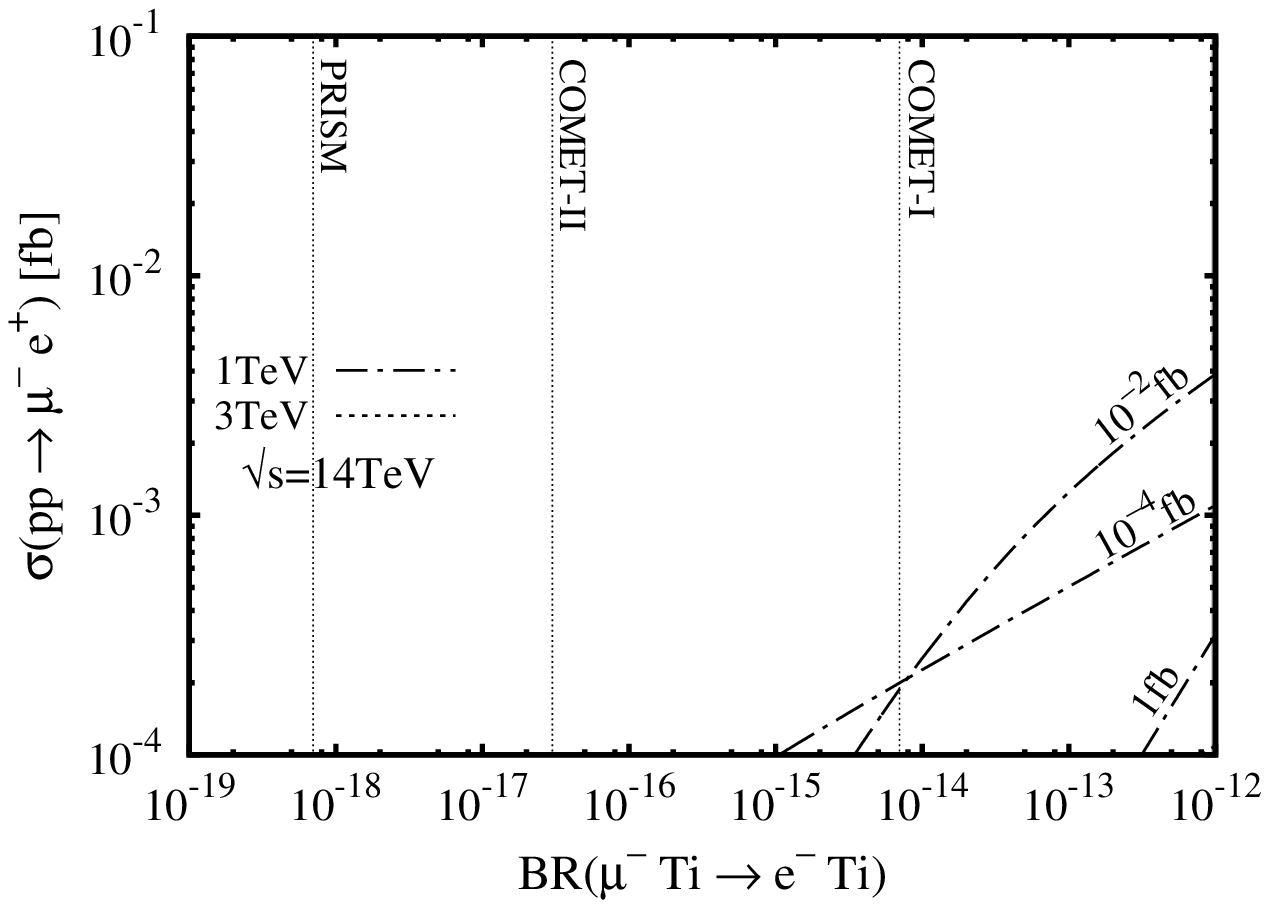}
\label{case1_g}
} & \hspace{-14mm}
\subfigure[$\text{N}=\text{Ti}$ and $\sqrt{s}= 100$TeV.]{
\includegraphics[scale=0.63]{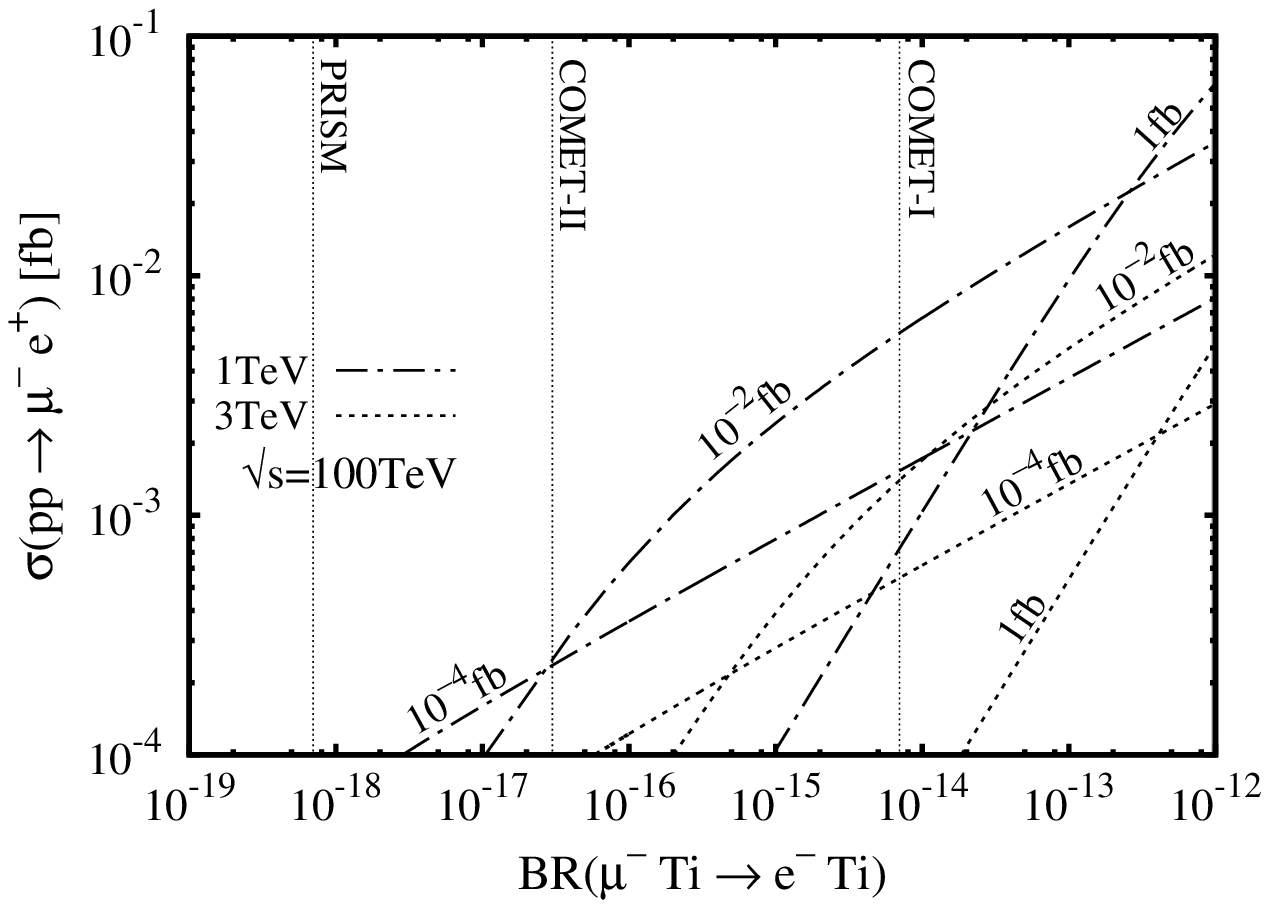}
\label{case1_h}
} \\
\end{tabular}
\caption{Same as Fig.~\ref{Fig:sigma_vs_BR_I_1} except for 
target nucleus. We take Al [(a) and (b)], and Ti [(c) and (d)] for 
the target nucleus of $\mu$-$e$ conversion process. }
\label{Fig:sigma_vs_BR_I_2}
\end{figure}

The parameter dependence of $\sigma(pp \to \mu^- e^+)$, 
$\sigma(pp \to jj)$, and $\text{BR}(\mu^- N \to e^- N)$ are depicted 
in Fig.~\ref{Fig:cont_I}. 
Dashed and dot-dashed lines are contours of $\sigma(pp \to \mu^- e^+)$ 
and $\sigma(pp \to jj)$ at $\sqrt{s}=14$TeV (left panels) and $\sqrt{s} 
=100$TeV (right panels), respectively. Solid lines are contours of 
$\text{BR}(\mu^- \text{Al} \to e^- \text{Al})$, which are translated 
from the single event sensitivities of each experiments (see 
Table~\ref{Tab:mue_conv}).  
Light shaded region is excluded by the $\mu$-$e$ conversion search at 
the SINDRUM-I\hspace{-1pt}I experiment~\cite{Bertl:2006up}, 
and dark shaded 
band is excluded region by the $M$-$\bar M$ conversion search experiment 
at the Paul Scherrer Institute (PSI)~\cite{Willmann:1998gd}. 
We take $m_{\tilde \nu_\tau} = 1$TeV for panels (a) and (b), and 
$m_{\tilde \nu_\tau} = 3$TeV for panels (c) and (d). 
For simplicity, we take the couplings universally in leptonic RPV sector: 
$\lambda \equiv \lambda_{312} = \lambda_{321} = -\lambda_{132} 
= -\lambda_{231}$.

Figure~\ref{Fig:cont_I} displays the strong potential of $\mu$-$e$ 
conversion search to explore the RPV scenarios. The PRISM experiment will 
cover almost parameter space wherein the LHC experiment can survey. 
In the parameter range between the SINDRUM-I\hspace{-1pt}I limit 
and the PRISM reach, 
combining the measurement results of $\sigma(pp \to \mu^- e^+)$, 
$\sigma(pp \to jj)$, and $\text{BR}(\mu^- \text{Al} \to e^- \text{Al})$, 
the RPV couplings and the tau sneutrino mass will be precisely determined.

Figures~\ref{Fig:sigma_vs_BR_I_1} and \ref{Fig:sigma_vs_BR_I_2} show 
$\sigma(pp \to \mu \bar e)$ as a function of $\text{BR} (\mu + N \to e 
+ N)$ in the case-I. 
Candidate materials for the target of $\mu$-$e$ conversion search are 
carbon (C) and silicon (Si) at the DeeMe experiment, and are aluminum (Al) 
or titanium (Ti) at the COMET, Mu2e, and PRISM experiment. 
Vertical dotted lines show the experimental reach of DeeMe 1-year running 
(DeeMe(1yr)), DeeMe 4-years running (DeeMe(4yrs)), COMET phase-I 
(COMET-I), COMET phase-I\hspace{-1pt}I (COMET-I\hspace{-1pt}I), 
and PRISM (PRISM). Shaded regions are the excluded region by the 
SINDRUM-I\hspace{-1pt}I experiment~\cite{Bertl:2006up}, which 
are translated into the limit for each nucleus from that for Au. 
The experimental reach of Mu2e experiment is planned to be similar of 
the COMET phase-I\hspace{-1pt}I~\cite{Mu2e}. 
Left and right panels show the results of $\sqrt{s}= 14\text{TeV}$ 
and $\sqrt{s}=100\text{TeV}$, respectively. Results for 
$m_{\tilde \nu_\tau} = 1\text{TeV}$ and $m_{\tilde \nu_\tau} = 
3\text{TeV}$ are given by dot-dashed line and dotted line, respectively. 
Each line corresponds to the dijet production cross section at the LHC, 
$\sigma(pp \to jj)$, at $\sqrt{s}=14\text{TeV}$ (left panels) and at 
$\sqrt{s}=100\text{TeV}$ (right panels), respectively. 
For simplicity, we take universal RPV coupling, $\lambda \equiv 
\lambda_{312} = \lambda_{321} = -\lambda_{132} = 
-\lambda_{231}$.

Figures~\ref{Fig:sigma_vs_BR_I_1} and \ref{Fig:sigma_vs_BR_I_2} show 
the clear correlations among $\sigma (pp \to \mu^- e^+)$, $\sigma(pp \to jj)$, 
and $\text{BR} (\mu^- N \to e^- N)$. Checking the correlations makes 
possible to distinguish the RPV scenario and other new physics scenarios.

\begin{table}[t]
\begin{center}
\caption{Numerical value of $k_N$ for a target nucleus $N$ in each case. }
\vspace{2mm}
\hspace{-3mm}
{\renewcommand\arraystretch{1.8}
\begin{tabular}{llllll}
\hline 

& C
& Al
& Si
& Ti
& Au
\\[-0.3mm] \hline \hline
case-I
& $13.83 \Bigl( \dfrac{1\text{TeV}}{m_{\tilde \nu_\tau}} \Bigr)^4$
& $20.92 \Bigl( \dfrac{1\text{TeV}}{m_{\tilde \nu_\tau}} \Bigr)^4$
& $20.80 \Bigl( \dfrac{1\text{TeV}}{m_{\tilde \nu_\tau}} \Bigr)^4$
& $35.71 \Bigl( \dfrac{1\text{TeV}}{m_{\tilde \nu_\tau}} \Bigr)^4$
& $26.26 \Bigl( \dfrac{1\text{TeV}}{m_{\tilde \nu_\tau}} \Bigr)^4$
\\[0.5mm]
case-I\hspace{-1pt}I
& $3.913 \Bigl( \dfrac{1\text{TeV}}{m_{\tilde \nu_\tau}} \Bigr)^4$
& $5.881 \Bigl( \dfrac{1\text{TeV}}{m_{\tilde \nu_\tau}} \Bigr)^4$
& $5.886 \Bigl( \dfrac{1\text{TeV}}{m_{\tilde \nu_\tau}} \Bigr)^4$
& $9.962 \Bigl( \dfrac{1\text{TeV}}{m_{\tilde \nu_\tau}} \Bigr)^4$
& $7.185 \Bigl( \dfrac{1\text{TeV}}{m_{\tilde \nu_\tau}} \Bigr)^4$
\\[0.5mm]
case-I\hspace{-1pt}I\hspace{-1pt}I
& $32.46 \Bigl( \dfrac{1\text{TeV}}{m_{\tilde \nu_\tau}} \Bigr)^4$
& $48.97 \Bigl( \dfrac{1\text{TeV}}{m_{\tilde \nu_\tau}} \Bigr)^4$
& $48.83 \Bigl( \dfrac{1\text{TeV}}{m_{\tilde \nu_\tau}} \Bigr)^4$
& $83.38 \Bigl( \dfrac{1\text{TeV}}{m_{\tilde \nu_\tau}} \Bigr)^4$
& $60.91 \Bigl( \dfrac{1\text{TeV}}{m_{\tilde \nu_\tau}} \Bigr)^4$
\\ \hline
\end{tabular} 
}
\end{center}
\label{Tab:kN}
\end{table}

In Figs~\ref{Fig:sigma_vs_BR_I_1} and \ref{Fig:sigma_vs_BR_I_2}, 
behavior of the correlations are not so intuitive. We quantitatively analyze 
the behavior. We infer the $\sigma(pp \to \mu^- e^+)$ from the 
$\sigma(pp \to jj)$
and $\text{BR} (\mu^- N \to e^- N)$.

As we formulated in Secs.~\ref{Sec:mue_conv} 
and \ref{Sec:collider}, $\text{BR} \equiv \text{BR} (\mu^- N \to e^- 
N)$ and $\sigma_{jet} \equiv \sigma(pp \to jj)$ are divided into the 
kinematics part and RPV coupling dependent part as follows, 
\begin{equation}
\begin{split}
   \text{BR} = k_N (\lambda'_{311} \lambda)^2, 
   \label{Eq:BR_k}   
\end{split}      
\end{equation}
\begin{equation}
\begin{split}
   \sigma_{jet} \equiv \sigma(pp \to jj) = 
   F_{jet} \frac{\displaystyle \lambda'^4_{311}}
   {\displaystyle 3\lambda'^2_{311} + 8 \lambda^2}. 
   \label{Eq:sigma_jet_Fjet}   
\end{split}      
\end{equation}
Here $k_N$ is a coefficient depending on a target nucleus $N$ and the 
sneutrino mass, which values are calculated by 
Eqs.~\eqref{Eq:BR_C}-\eqref{Eq:BR_Ti} and are listed in Table 4.
$F_{jet}$ includes the numerical factor and kinematical factor in 
$\sigma_{jet}$, and is calculated from Eq.~\eqref{Eq:cross_jet}, $F_{jet} = 
\frac{\displaystyle 9}{\displaystyle 16\pi} \left\{ F_{d \bar d} + F_{u \bar d} 
+ F_{\bar u d} \right\} m_{\tilde \nu_\tau}$. 
We have a cubic equation of $\lambda'^2_{311}$ from 
Eqs.~\eqref{Eq:BR_k} and \eqref{Eq:sigma_jet_Fjet}, 
\begin{equation}
\begin{split}
   k_{N} F_{jet} (\lambda'^2_{311})^3 
   -3 k_{N} (\lambda'^2_{311})^2 
   -8 \sigma_{jet} \text{BR} = 0. 
\end{split}      
\end{equation}
By solving the cubic equation, we obtain an analytic expression of 
$\lambda'^2_{311}$ as a function of $\text{BR}$, 
\begin{equation}
\begin{split}
   \lambda'^2_{311} 
   &= 
   \Biggl\{ 
   \biggl( \frac{2 \sigma_{jet} \text{BR}}{k_N F_{jet}} \biggr) 
   + \left( \frac{\sigma_{jet}}{F_{jet}} \right)^3 
   + \sqrt{
   \biggl( 
   \frac{2 \sigma_{jet} \text{BR}}{k_N F_{jet}} 
   + \biggl( \frac{\sigma_{jet}}{F_{jet}} \biggr)^3 
   \biggr)^2 
   - \biggl( \frac{\sigma_{jet}}{F_{jet}} \biggr)^6 
   }
   \hspace{1mm} 
   \Biggr\}^{1/3} 
   \\& \ + 
   \Biggl\{ 
   \biggl( \frac{2 \sigma_{jet} \text{BR}}{k_N F_{jet}} \biggr) 
   + \left( \frac{\sigma_{jet}}{F_{jet}} \right)^3 
   - \sqrt{
   \biggl( 
   \frac{2 \sigma_{jet} \text{BR}}{k_N F_{jet}} 
   + \biggl( \frac{\sigma_{jet}}{F_{jet}} \biggr)^3 
   \biggr)^2 
   - \biggl( \frac{\sigma_{jet}}{F_{jet}} \biggr)^6 
   }
   \hspace{1mm} 
   \Biggr\}^{1/3} 
   \\& \ + 
   \frac{\sigma_{jet}}{F_{jet}}. 
\label{Eq:lam_prime_BRfunc}   
\end{split}      
\end{equation}
$\lambda^2$ is easily obtained from Eqs.~\eqref{Eq:BR_k} and 
\eqref{Eq:lam_prime_BRfunc}, 
\begin{equation}
\begin{split}
   \lambda^2 
   &= 
   \frac{\text{BR}}{k_N \lambda'^2_{311}} 
   \\& = 
   \frac{\text{BR}}{k_N} 
   \Biggl\{ 
   \frac{\sigma_{jet}}{F_{jet}} 
   + \biggl[  
   \biggl( \frac{2 \sigma_{jet} \text{BR}}{k_N F_{jet}} \biggr) 
   + \left( \frac{\sigma_{jet}}{F_{jet}} \right)^3 
   + \sqrt{
   \biggl( 
   \frac{2 \sigma_{jet} \text{BR}}{k_N F_{jet}} 
   + \biggl( \frac{\sigma_{jet}}{F_{jet}} \biggr)^3 
   \biggr)^2 
   - \biggl( \frac{\sigma_{jet}}{F_{jet}} \biggr)^6 
   }
   \hspace{1mm} 
   \biggr]^{1/3} 
   \\& \hspace{10mm} + 
   \biggl[  
   \biggl( \frac{2 \sigma_{jet} \text{BR}}{k_N F_{jet}} \biggr) 
   + \left( \frac{\sigma_{jet}}{F_{jet}} \right)^3 
   - \sqrt{
   \biggl( 
   \frac{2 \sigma_{jet} \text{BR}}{k_N F_{jet}} 
   + \biggl( \frac{\sigma_{jet}}{F_{jet}} \biggr)^3 
   \biggr)^2 
   - \biggl( \frac{\sigma_{jet}}{F_{jet}} \biggr)^6 
   }
   \hspace{1mm} 
   \biggr]^{1/3}
   \Biggr\}^{-1}. 
   \label{Eq:lam_BRfunc}
\end{split}      
\end{equation}

As a result, by substituting $\lambda'^2_{311}$ and $\lambda^2$ into the 
expression of $\sigma(pp \to \mu^- e^+)$ [Eq.~\eqref{Eq:sigma_muebar}], 
we obtain the prediction of  $\sigma(pp \to \mu^- e^+)$ as a function of 
$\text{BR}$ and $\sigma_{jet}$, 
\begin{equation}
\begin{split}
   \sigma(pp \to \mu^- e^+) = \frac{12}{16 \pi} 
   F_{d \bar d} m_{\tilde \nu_\tau} 
   \frac{(\text{BR}/k_N)}{3 \lambda'^2_{311} + 8 \lambda^2}. 
   \label{Eq:muebar_BRfunc}
\end{split}      
\end{equation}
Once $\sigma_{jet}$ is measured, we can evaluate $\sigma(pp \to \mu^- 
e^+)$ as a function of $\text{BR}$ with the Eq.~\eqref{Eq:muebar_BRfunc}. 
Note that the solution Eqs.~\eqref{Eq:lam_prime_BRfunc} and 
\eqref{Eq:lam_BRfunc} is uniquely determined as read in
Fig.~\ref{Fig:cont_I},
and hence $\sigma(pp \to \mu^- e^+)$ is also uniquely inferred.
We cannot, however, determine $\sigma_{jet}$ uniquely from BR and
$\sigma(pp \to \mu^- e^+)$ since as a function of BR the latter is two-valued
function as is n in Fig.~\ref{Fig:cont_I}. Therefore there are crosses of 
two lines in Figs.~\ref{Fig:sigma_vs_BR_I_1} and \ref{Fig:sigma_vs_BR_I_2}.

We quantitatively analyze the behavior for 2 reference points. As a first 
reference point, we take $N=\text{Al}$, $m_{\tilde \nu_\tau}=1\text{TeV}$, 
$\sqrt{s}=100\text{TeV}$, and $\sigma_{jet} = 1\text{fb}$. 
In this point, when $\text{BR} \lesssim 10^{-13}$, $\lambda'^2_{311}$ 
and $\lambda^2$ are approximately calculated from 
Eqs.~\eqref{Eq:lam_prime_BRfunc} and \eqref{Eq:lam_BRfunc} as 
follows, 
\begin{equation}
\begin{split}
   \lambda'^2_{311} \simeq  
   3 \left( \frac{\sigma_{jet}}{F_{jet}} \right), \ \ 
   \lambda^2 
   = \frac{\text{BR}}{k_\text{Al} \lambda'^2_{311}} 
   = \frac{\text{BR}}{3 k_\text{Al}} 
       \left( \frac{F_{jet}}{\sigma_{jet}} \right). 
\end{split}      
\end{equation}
By substituting $\lambda'^2_{311}$ and $\lambda^2$ into 
Eq.~\eqref{Eq:sigma_muebar}, we obtain the approximate expression of 
$\sigma (pp \to \mu^- e^+)$, and find the $\text{BR}$ dependence 
on $\sigma (pp \to \mu^- e^+)$ as follows, 
\begin{equation}
\begin{split}
   \sigma(pp \to \mu^- e^+) 
   &\simeq 
   \frac{12}{16\pi} F_{d \bar d} m_{\tilde \nu_\tau} 
   \frac{\text{BR}/k_\text{Al}}
   {3 \cdot 3\left( \dfrac{\sigma_{jet}}{F_{jet}} \right) 
   +8 \cdot \dfrac{\text{BR}}{3k_{Al}} 
   \left( \dfrac{F_{jet}}{\sigma_{jet}} \right)} 
   \\ &\simeq 
   \frac{1}{12\pi} F_{d \bar d} m_{\tilde \nu_\tau}  
   \left( \frac{F_{jet}}{\sigma_{jet} k_\text{Al}} \right) 
   \text{BR} . 
\end{split}      
\end{equation}
The $\text{BR}$ dependence is consistent with the numerical result 
in Fig.~\ref{Fig:sigma_vs_BR_I_2}. 
As a second reference point, we take $N=\text{Al}$, $m_{\tilde \nu_\tau} 
= 1\text{TeV}$, $\sqrt{s}=100\text{TeV}$, and $\sigma_{jet} = 
10^{-4}\text{fb}$. In this point, when $\text{BR} \gtrsim 10^{-21}$, 
$\lambda'^2_{311}$ and $\lambda^2$ are approximately calculated 
from Eqs.~\eqref{Eq:lam_prime_BRfunc} and \eqref{Eq:lam_BRfunc} 
as follows, 
\begin{equation}
\begin{split}
   \lambda'^2_{311} \simeq 4 
   \left( \frac{\sigma_{jet} \text{BR}} 
   {k_\text{Al} F_{jet}} \right)^{1/3}, \ \ 
   \lambda^2 
   = \frac{\text{BR}}{k_\text{Al} \lambda'^2_{311}} 
   = \frac{1}{4} \left( \frac{(\text{BR})^2 F_{jet}}
   {k_\text{Al}^2 \sigma_{jet}} \right)^{1/3}. 
\end{split}      
\end{equation}
By substituting $\lambda'^2_{311}$ and $\lambda^2$ into 
Eq.~\eqref{Eq:sigma_muebar}, we obtain the approximate expression of 
$\sigma (pp \to \mu^- e^+)$, and find the $\text{BR}$ dependence 
on $\sigma (pp \to \mu^- e^+)$ as follows, 
\begin{equation}
\begin{split}
   \sigma(pp \to \mu^- e^+) 
   &\simeq 
   \frac{12}{16\pi} F_{d \bar d} m_{\tilde \nu_\tau} 
   \frac{\text{BR}/k_\text{Al}}
   {3 \cdot 4\left( \dfrac{\sigma_{jet} \text{BR}} 
   {k_\text{Al} F_{jet}} \right)^{1/3} 
   +8 \cdot \dfrac{1}{4} \left( \dfrac{(\text{BR})^2 F_{jet}}
   {k_\text{Al}^2 \sigma_{jet}} \right)^{1/3}} 
   \\ &\simeq 
   \frac{3}{8\pi} F_{d \bar d} m_{\tilde \nu_\tau}  
   \left( \frac{\sigma_{jet}}{k_\text{Al} F_{jet}} \right)^{1/3} 
   \left( \text{BR} \right)^{1/3}. 
\end{split}      
\end{equation}
The $\text{BR}$ dependence is consistent with the numerical result 
in Fig.~\ref{Fig:sigma_vs_BR_I_2}. Also in other points, we can 
similarly check the $\text{BR}$ dependence, and find its consistency.

In Figs.~\ref{Fig:sigma_vs_BR_I_1} and \ref{Fig:sigma_vs_BR_I_2}, 
in some regions of $\text{BR}(\mu^- N \to e^- N)$, larger 
$\sigma(pp \to jj)$ suggests smaller $\sigma(pp \to \mu^- e^+)$. 
This strange relation is simply understood as follows. 
Large $\sigma(pp \to jj)$ for a fixed $\text{BR}(\mu^- N \to e^- N)$ 
leads large $\lambda'_{311}$ and small $\lambda$ (see 
Eqs.~\eqref{Eq:omega_conv_2} and \eqref{Eq:cross_jet}). In this 
case, as is shown in Eq.~\eqref{Eq:muebar_BRfunc}, $\sigma(pp \to 
\mu^- e^+) \propto 1/\lambda'^2_{311}$. 
Thus, in some regions, we find the strange relation. 
This is one of the unique relation in the RPV scenario. In other models, 
if mediator universally couples to both quarks and leptons, we will not 
find the difference between $\sigma(pp \to jj)$ and $\sigma(pp \to 
\mu^- e^+)$ (except for color factor). We can distinguish such models 
from the RPV scenarios by checking the unique relation.

\subsection{Case-I\hspace{-1pt}I 
($\lambda'_{311} = 0$ and $\lambda'_{322} \neq 0$)} \label{Sec:322} 

\begin{figure}[t!]
\hspace{-7mm}
\begin{tabular}{cc}
\subfigure[$m_{\tilde \nu_\tau} = 1$TeV. $\sqrt{s}= 14$TeV. ]{
\includegraphics[scale=0.68]{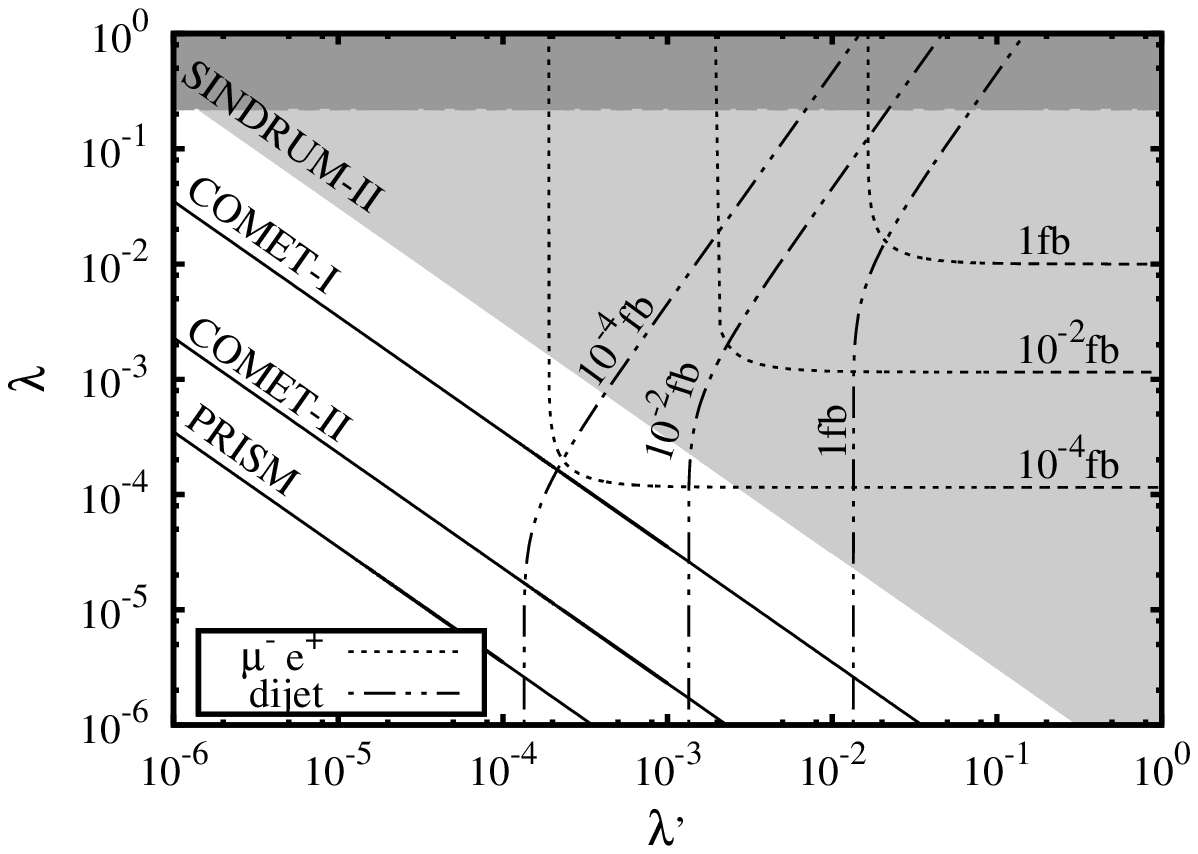}
\label{left1}
} & \hspace{-12mm}
\subfigure[$m_{\tilde \nu_\tau} = 1$TeV. $\sqrt{s}= 100$TeV. ]{
\includegraphics[scale=0.68]{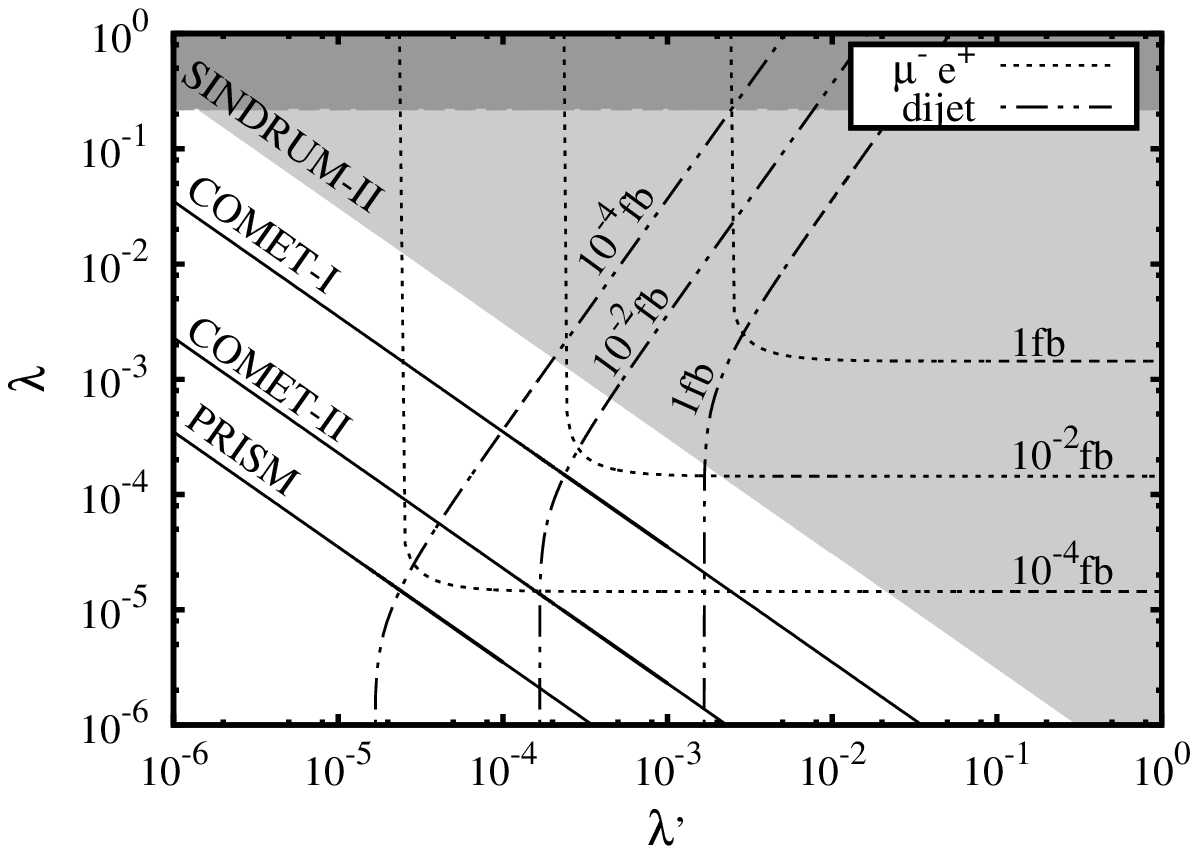}
\label{right1}
} \\
\subfigure[$m_{\tilde \nu_\tau} = 3$TeV. $\sqrt{s}= 14$TeV. ]{
\includegraphics[scale=0.68]{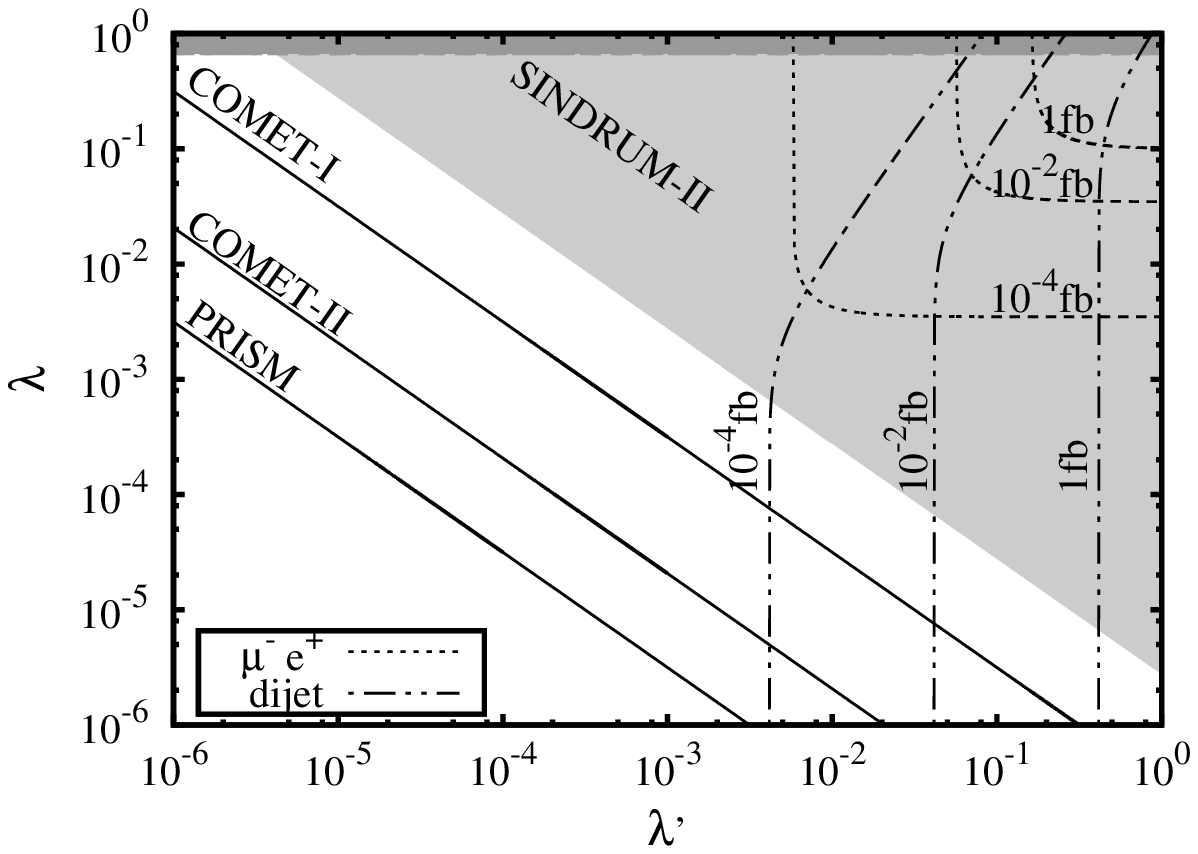}
\label{left2}
} & \hspace{-12mm}
\subfigure[$m_{\tilde \nu_\tau} = 3$TeV. $\sqrt{s}= 100$TeV. ]{
\includegraphics[scale=0.68]{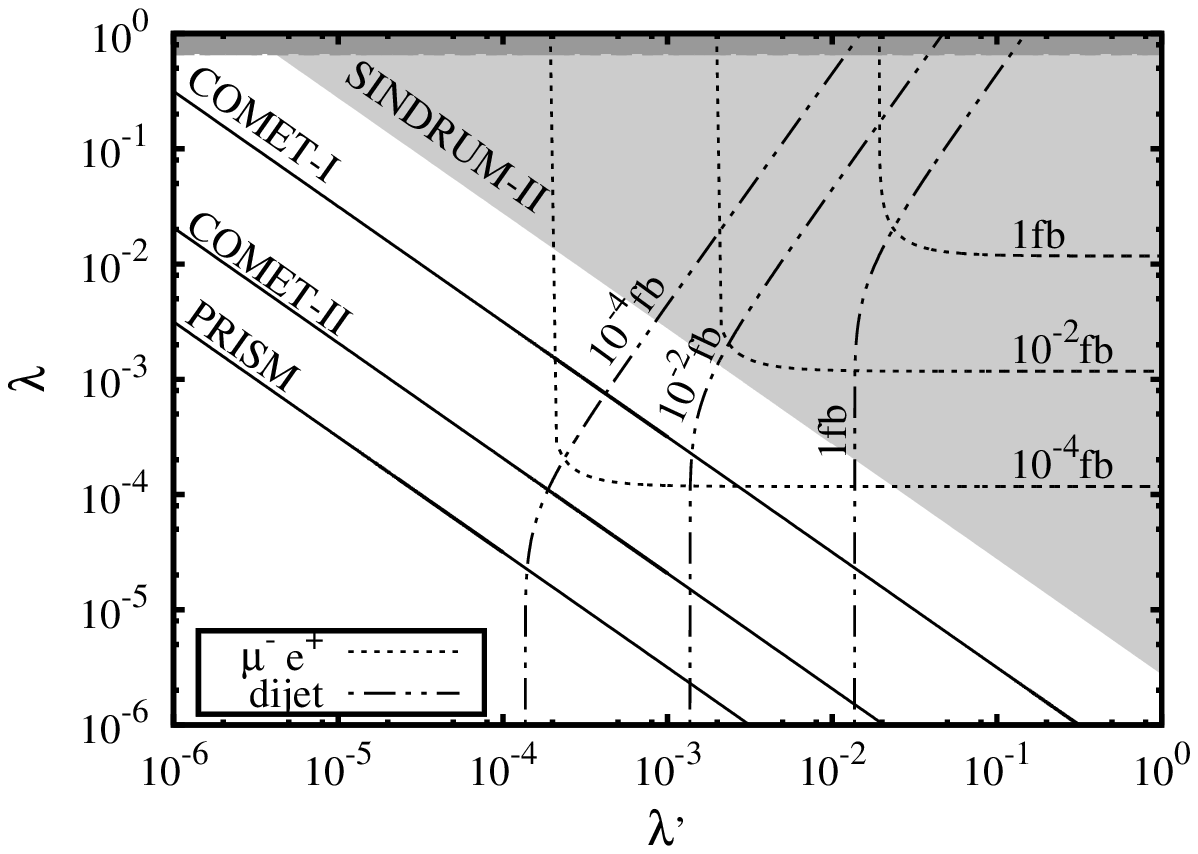}
\label{right2}
} \\
\end{tabular}
\caption{Contour plot of $\sigma(pp \to \mu^- e^+)$, 
$\sigma(pp \to jj)$, and BR($\mu^- N \to e^- N$) in 
the case-I\hspace{-1pt}I for 
(a) $m_{\tilde \nu_\tau} = 1$TeV and $\sqrt{s}=14$TeV 
(b) $m_{\tilde \nu_\tau} = 1$TeV and $\sqrt{s}=100$TeV 
(c) $m_{\tilde \nu_\tau} = 3$TeV and $\sqrt{s}=14$TeV 
(d) $m_{\tilde \nu_\tau} = 3$TeV and $\sqrt{s}=100$TeV. 
For simplicity, we take universal RPV coupling, $\lambda \equiv 
\lambda_{312} = \lambda_{321} = -\lambda_{132} = 
-\lambda_{231} $. Light shaded region is excluded by the 
$\mu$-$e$ conversion search~\cite{Bertl:2006up}, 
and dark shaded band is excluded region by the $M$-$\bar M$ 
conversion search~\cite{Willmann:1998gd}.}
\label{Fig:cont_II}
\end{figure}

\begin{figure}[t!]
\hspace{-6mm}
\begin{tabular}{cc}
\subfigure[$N=\text{C}$ and $\sqrt{s}= 14$TeV.]{
\includegraphics[scale=0.63]{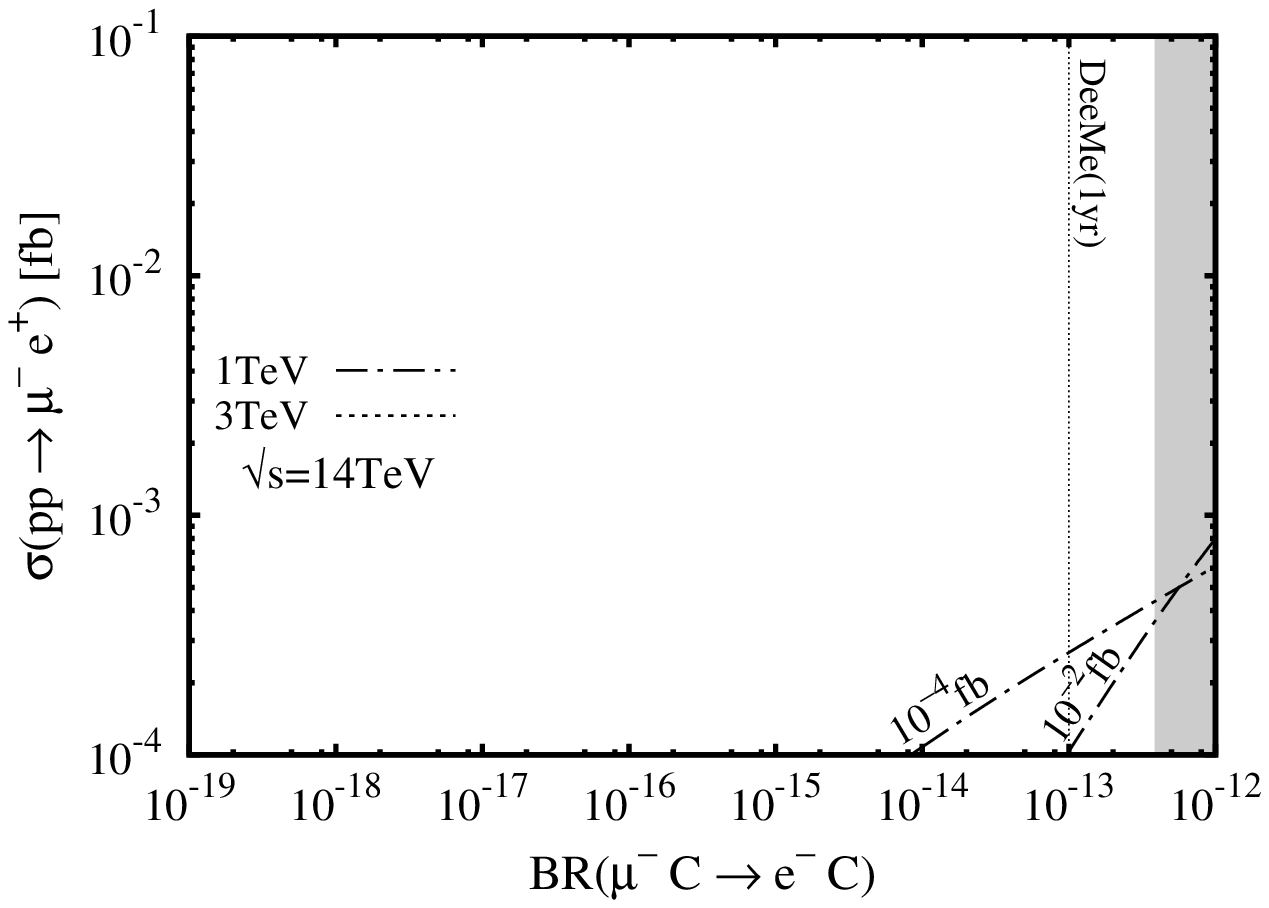}
\label{case1_a}
} & \hspace{-14mm}
\subfigure[$N=\text{C}$ and $\sqrt{s}= 100$TeV.]{
\includegraphics[scale=0.63]{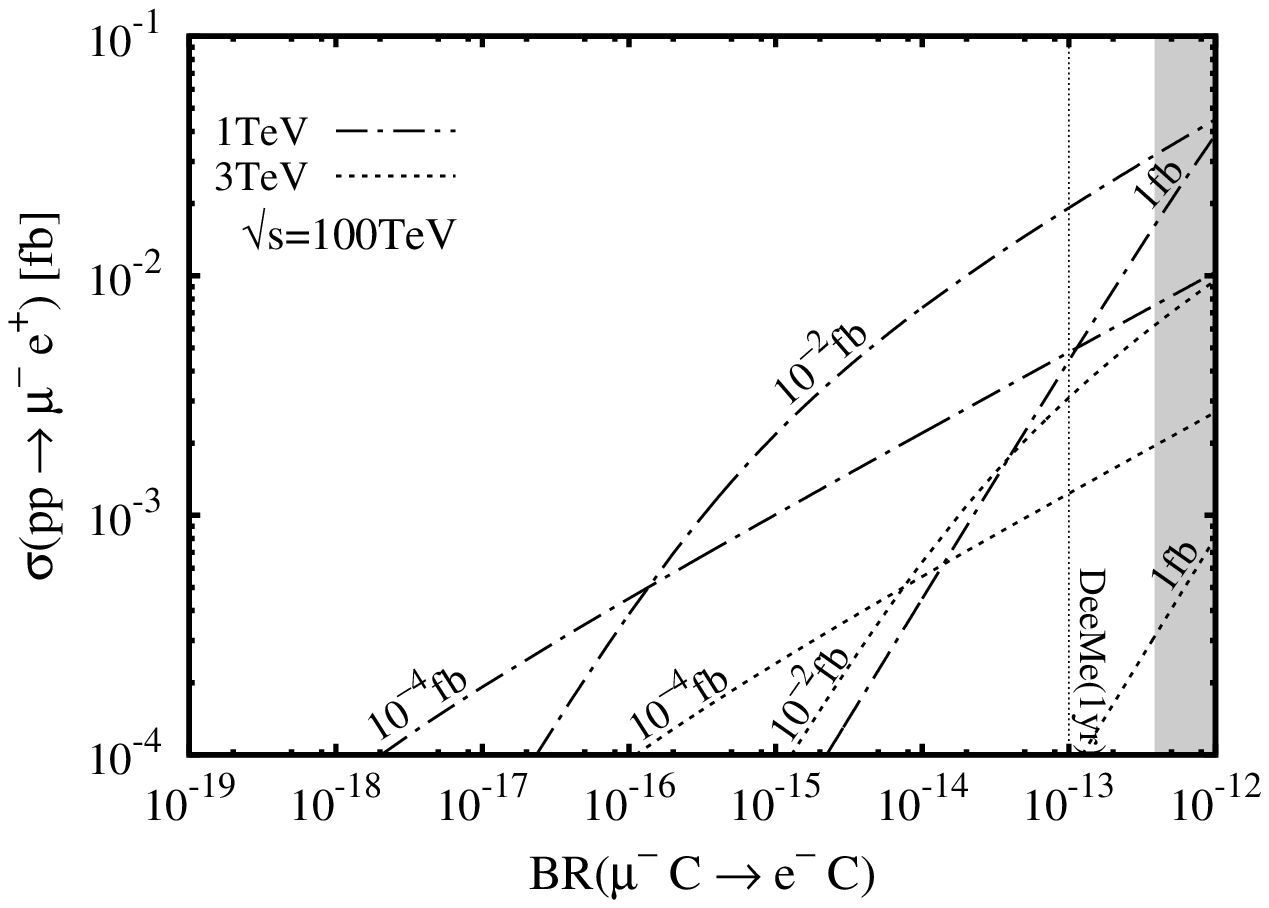}
\label{case1_b}
} \\[-4.5mm]
\subfigure[$N=\text{Si}$ and $\sqrt{s}= 14$TeV.]{
\includegraphics[scale=0.63]{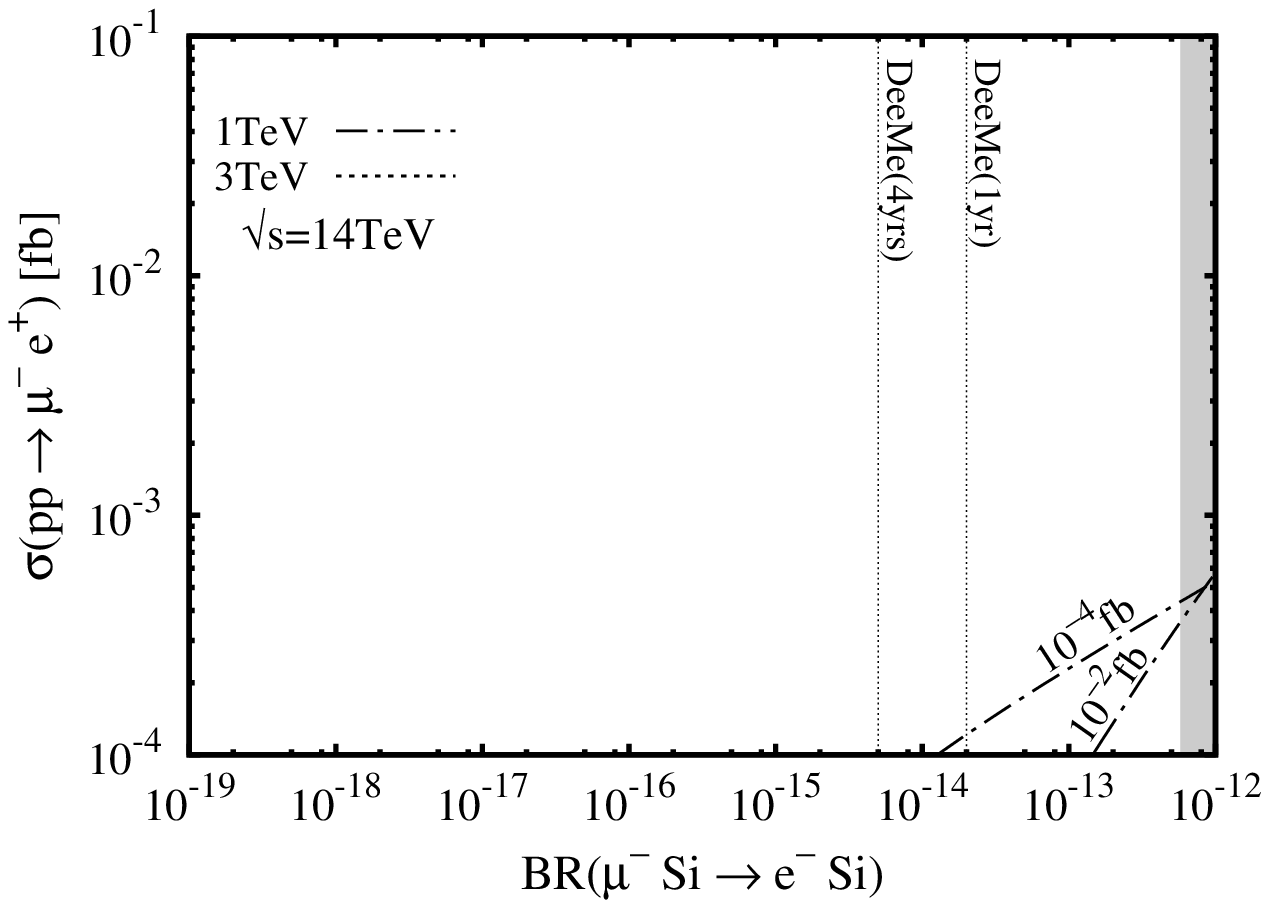}
\label{case1_c}
} & \hspace{-14mm}
\subfigure[$N=\text{Si}$ and $\sqrt{s}= 100$TeV.]{
\includegraphics[scale=0.63]{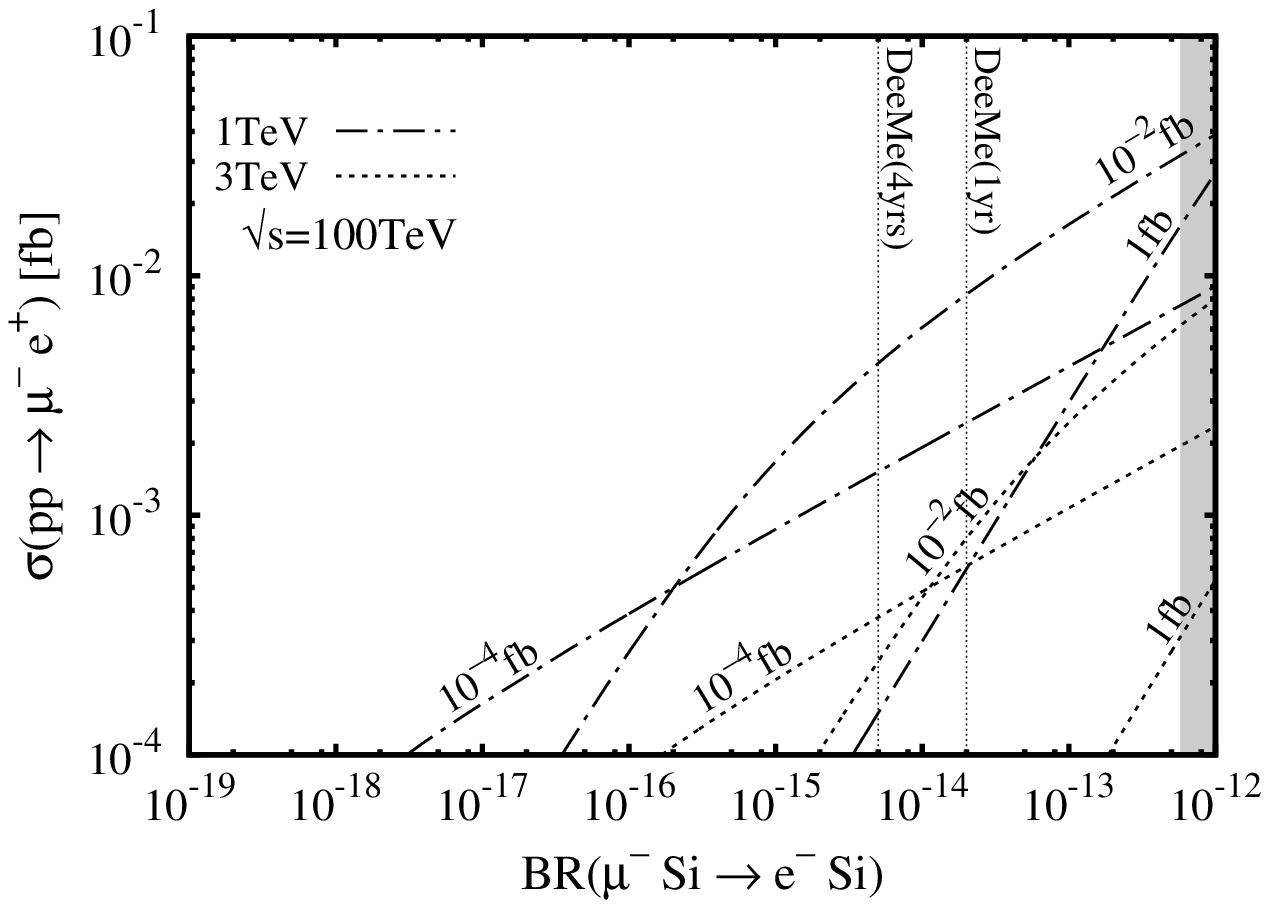}
\label{case1_d}
} \\
\end{tabular}
\caption{$\sigma (pp \to \mu^- e^+)$ as a function of $\text{BR} 
(\mu^- N \to e^- N)$ for each $\sigma (pp \to jj)$ in the 
case-I\hspace{-1.5pt}I. $\sigma (pp \to jj)$ are attached on each line. 
Results for $m_{\tilde \nu_\tau} = 1\text{TeV}$ 
($m_{\tilde \nu_\tau} = 3\text{TeV}$) are given by dot-dashed 
line (dotted line). Shaded regions are the excluded region by the 
SINDRUM-I\hspace{-1pt}I experiment. Left panels show the results 
for the collision energy $\sqrt{s} = 14\text{TeV}$, and right panels 
show the results for $\sqrt{s} = 100\text{TeV}$. We take C [(a) 
and (b)],  and Si [(c) and (d)] for the target nucleus of $\mu$-$e$ 
conversion process. }
\label{Fig:sigma_vs_BR_II_1}
\end{figure}

\begin{figure}[t!]
\hspace{-6mm}
\begin{tabular}{cc}
\subfigure[$\text{N}=\text{Al}$ and $\sqrt{s}= 14$TeV.]{
\includegraphics[scale=0.63]{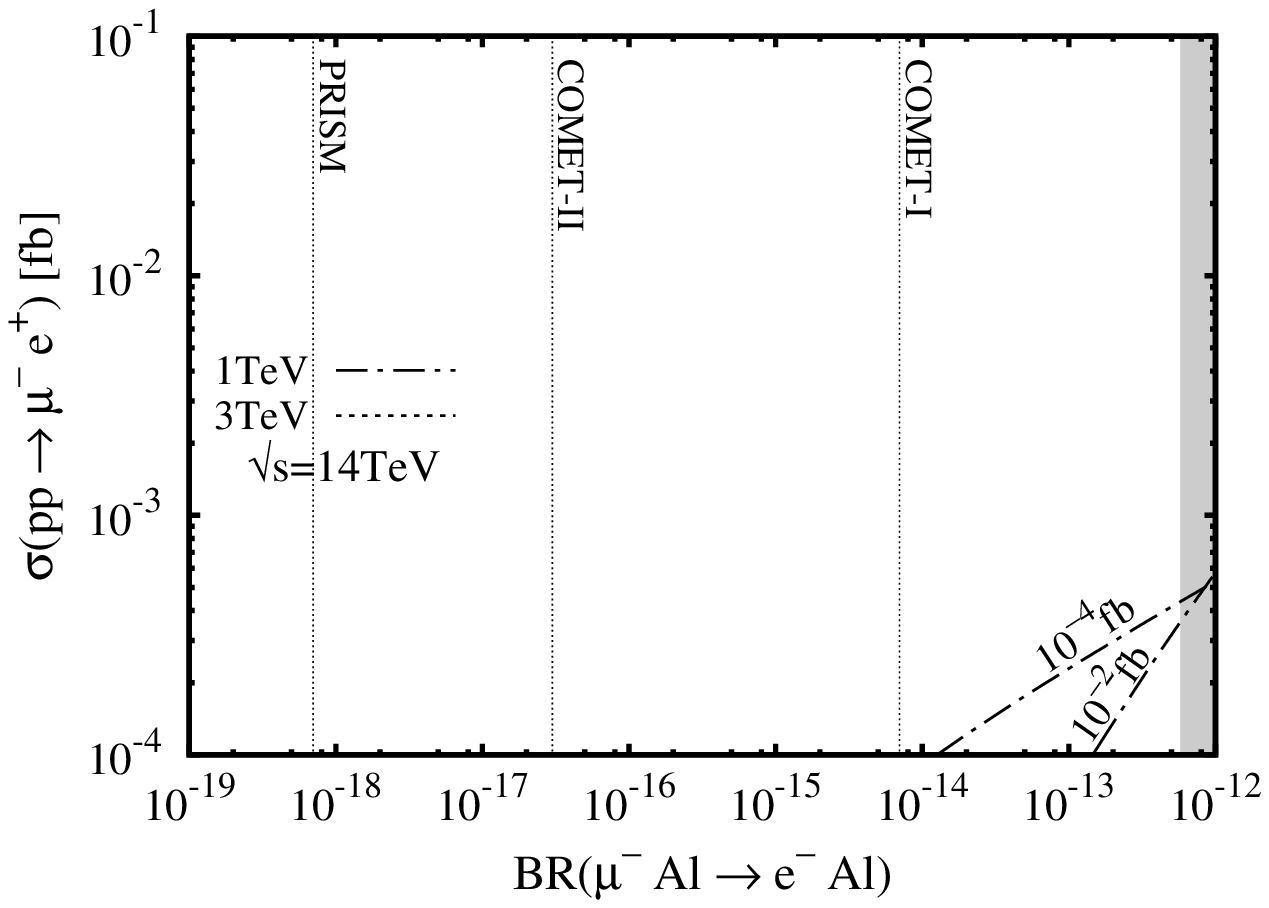}
\label{case1_e}
} & \hspace{-14mm}
\subfigure[$\text{N}=\text{Al}$ and $\sqrt{s}= 100$TeV.]{
\includegraphics[scale=0.63]{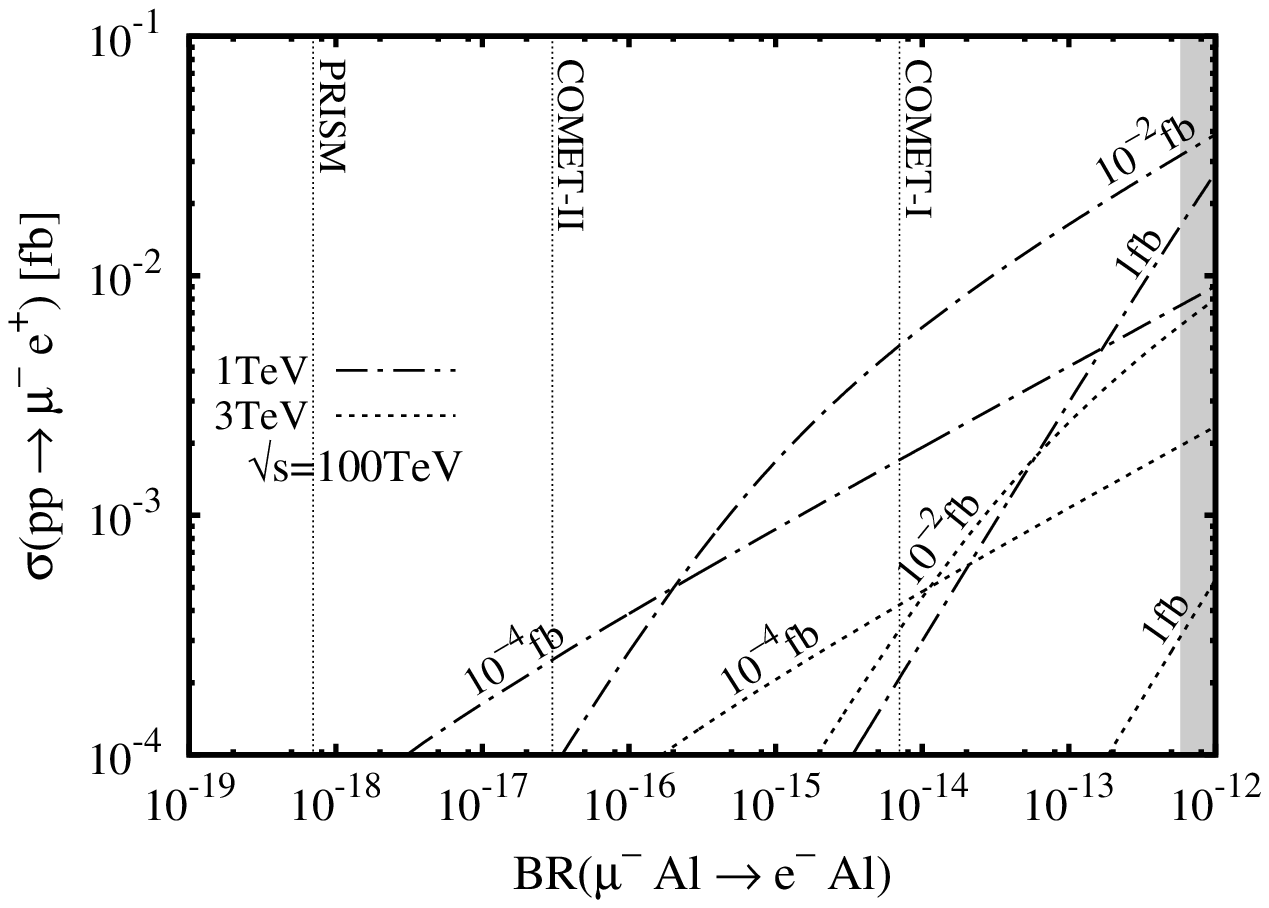}
\label{case1_f}
} \\[-4.5mm]
\subfigure[$\text{N}=\text{Ti}$ and $\sqrt{s}= 14$TeV.]{
\includegraphics[scale=0.63]{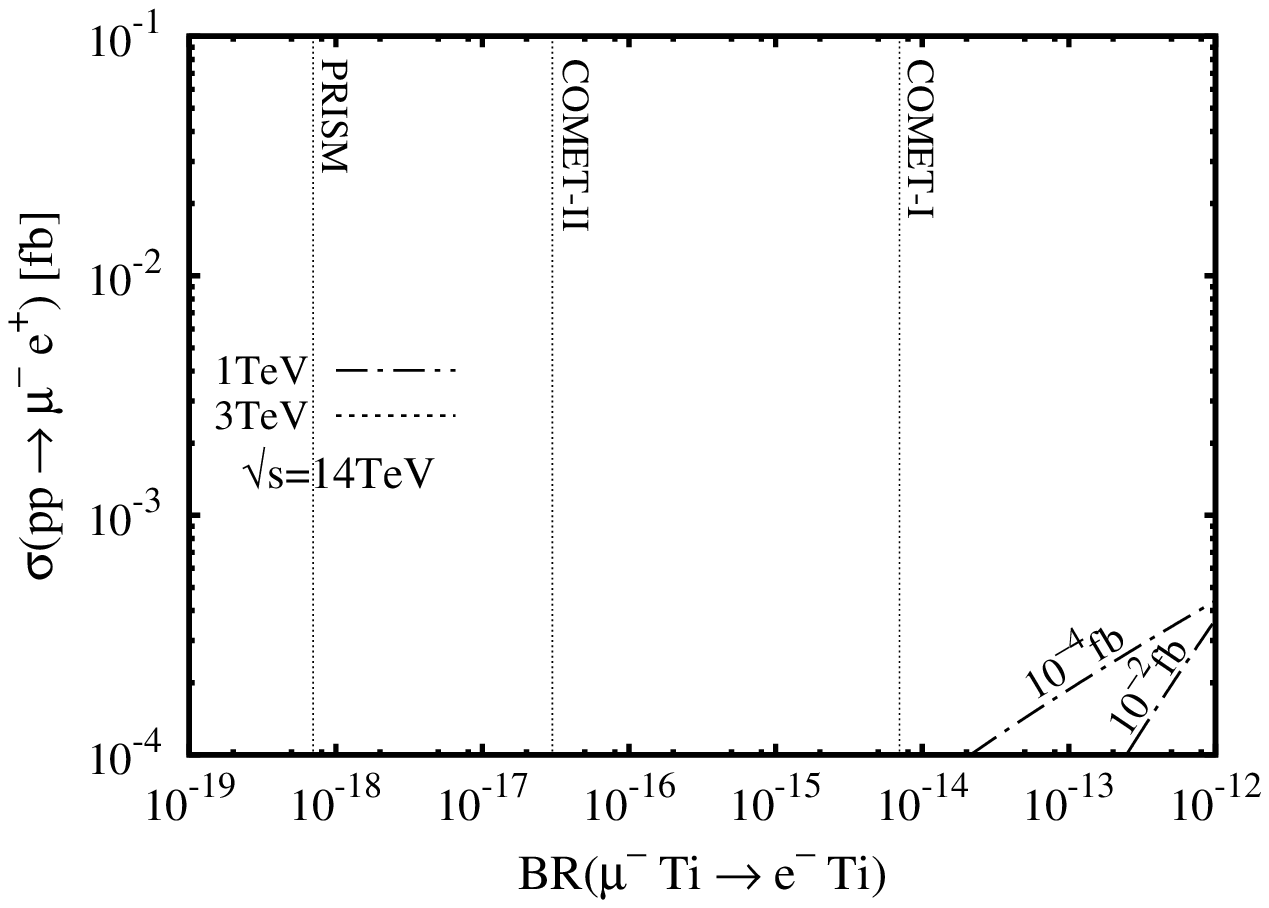}
\label{case1_g}
} & \hspace{-14mm}
\subfigure[$\text{N}=\text{Ti}$ and $\sqrt{s}= 100$TeV.]{
\includegraphics[scale=0.63]{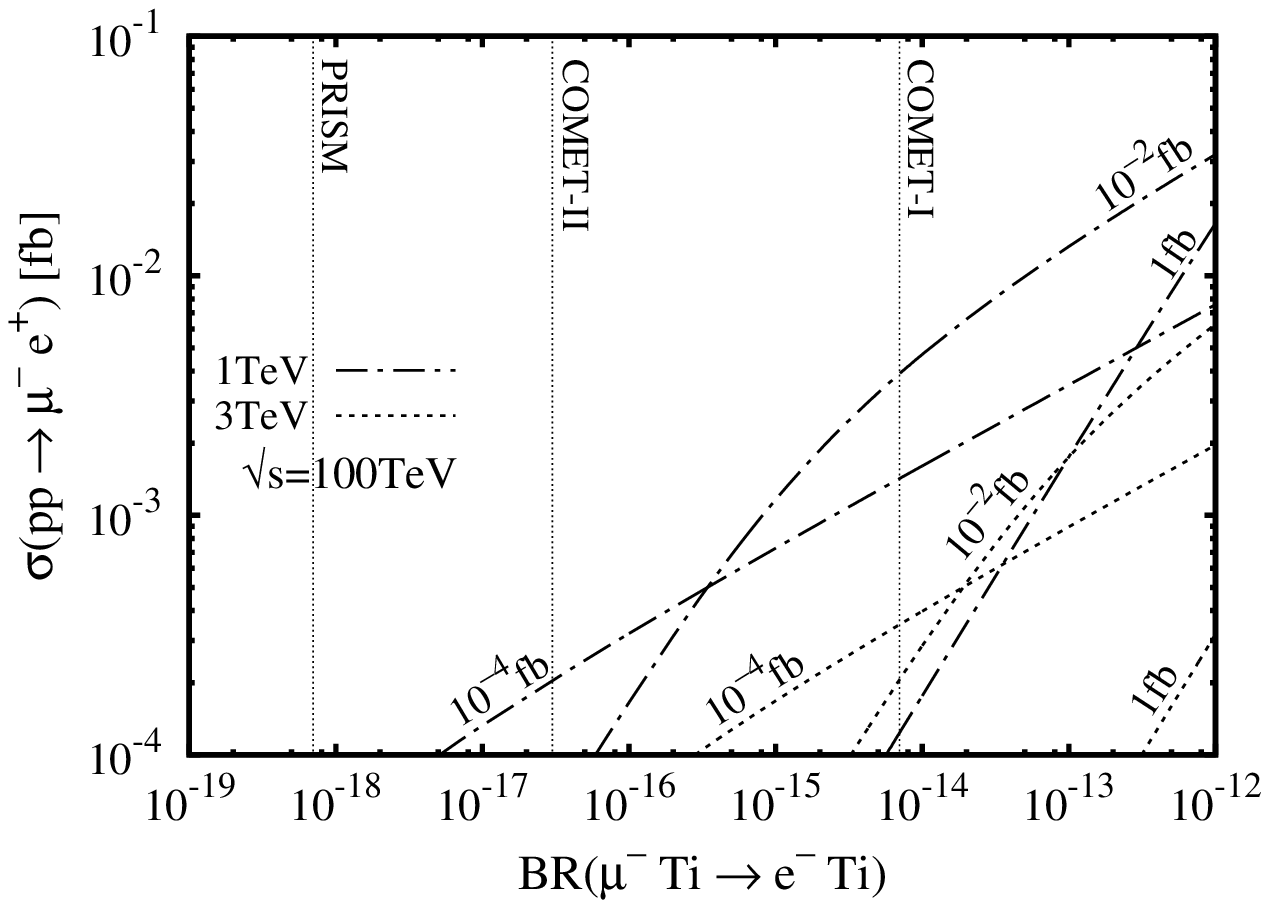}
\label{case1_h}
} \\
\end{tabular}
\caption{Same as Fig.~\ref{Fig:sigma_vs_BR_II_1} except for 
target nucleus. We take Al [(a) and (b)], and Ti [(c) and (d)] for 
the target nucleus of $\mu$-$e$ conversion process. }
\label{Fig:sigma_vs_BR_II_2}
\end{figure}

Figure~\ref{Fig:cont_II} displays the parameter dependence of 
$\sigma (pp \to \mu^- e^+)$, $\sigma(pp \to jj)$, and $\text{BR} 
(\mu^- \text{Al} \to e^- \text{Al})$ in the case-I\hspace{-1pt}I. 
The description of Fig.~\ref{Fig:cont_II} is same as that of 
Fig.~\ref{Fig:cont_I}. 
Figures~\ref{Fig:sigma_vs_BR_II_1} and \ref{Fig:sigma_vs_BR_II_2} 
show $\sigma(pp \to \mu^- e^+)$ as a function of $\text{BR} 
(\mu^- \text{Al} \to e^- \text{Al})$ in the case-I\hspace{-1pt}I. 
The descriptions of the figures are same as those of 
Figs.~\ref{Fig:sigma_vs_BR_I_1} and \ref{Fig:sigma_vs_BR_I_2}.

The RPV parameters are determined by measuring $\sigma (pp \to 
\mu^- e^+)$, $\sigma(pp \to jj)$, and $\text{BR}(\mu^- \text{Al} 
\to e^- \text{Al})$, and plot the point on Fig.~\ref{Fig:cont_II}. 
Since $\sigma (pp \to \mu^- e^+)$ at 14TeV LHC is too small for 
the parameter determination, we must focus on the invariant mass 
from dijet. Precise measurements both of the tau sneutrino mass and 
$\sigma(pp \to jj)$ specify a contour of $\sigma(pp \to jj)$ in 
$\lambda'_{322}$-$\lambda$ plane. Then precise measurement 
of $\text{BR} (\mu^- \text{Al} \to e^- \text{Al})$ can pin  
down the right parameter set on the contour. 
The accuracy of the pin-down strongly depends on the accuracy 
both of the invariant mass reconstruction and measurement 
of $\text{BR} (\mu^- \text{Al} \to e^- \text{Al})$. We will 
discuss the issue in detail in a separate publication~\cite{future}.

After the discovery of $\mu$-$e$ conversion signal, if the constructed 
invariant mass is heavier than 1TeV in measuring $pp \to \mu^- e^+$ 
and $pp \to jj$ at $\sqrt{s}=14\text{TeV}$, the case-I\hspace{-1pt}I 
is ruled out. 
In the case-I\hspace{-1pt}I, accessible parameter space at the LHC 
with $\sqrt{s} = 14\text{TeV}$ collision is limited to within the space 
for lighter tau sneutrino, $m_{\tilde \nu_\tau} \lesssim 1\text{TeV}$. 
This is because both $\sigma (pp \to \mu^- e^+)$ and $\sigma (pp 
\to jj)$ are too small due to the low density of strange quark component 
in a proton (see Fig.~\ref{fig:Fqqbar}). 
We need the 100TeV hadron collider to explore the parameter space for
heavier sneutrino, $m_{\tilde \nu_\tau} \gtrsim 1\text{TeV}$, in the 
case-I\hspace{-1pt}I.

Because of the low density of strange quark component in a proton, the 
reaction rate of $\mu$-$e$ conversion in case-I\hspace{-1pt}I is clearly 
different from that in case-I and -I\hspace{-1pt}I\hspace{-1pt}I. 
For a fixed combination of $\sigma (pp \to \mu^- e^+)$ and $\sigma 
(pp \to jj)$, the expected $\text{BR} (\mu^- N \to e^- N)$ is small 
compared with that in case-I and -I\hspace{-1pt}I\hspace{-1pt}I 
(Eqs.~\eqref{Eq:BR_C} - \eqref{Eq:BR_Ti}), and hence it is easy to 
discriminate case-I\hspace{-1pt}I scenario and case-I and 
-I\hspace{-1pt}I\hspace{-1pt}I by checking the correlations in 
Figs.~\ref{Fig:sigma_vs_BR_II_1} and \ref{Fig:sigma_vs_BR_II_2}. 
It is important to emphasize that we have to exhibit the correlations 
in order for verification of RPV scenarios wherein cLFV processes will 
never be found except for $\mu$-$e$ conversion. It is first time that the 
correlations are graphically shown in RPV SUSY models.

\subsection{Case-I\hspace{-1pt}I\hspace{-1pt}I ($\lambda'_{311} \neq 0$ 
and $\lambda'_{322} \neq 0$)}  \label{Sec:both} 

\begin{figure}[t!]
\hspace{-7mm}
\begin{tabular}{cc}
\subfigure[$m_{\tilde \nu_\tau} = 1$TeV. $\sqrt{s}= 14$TeV. ]{
\includegraphics[scale=0.68]{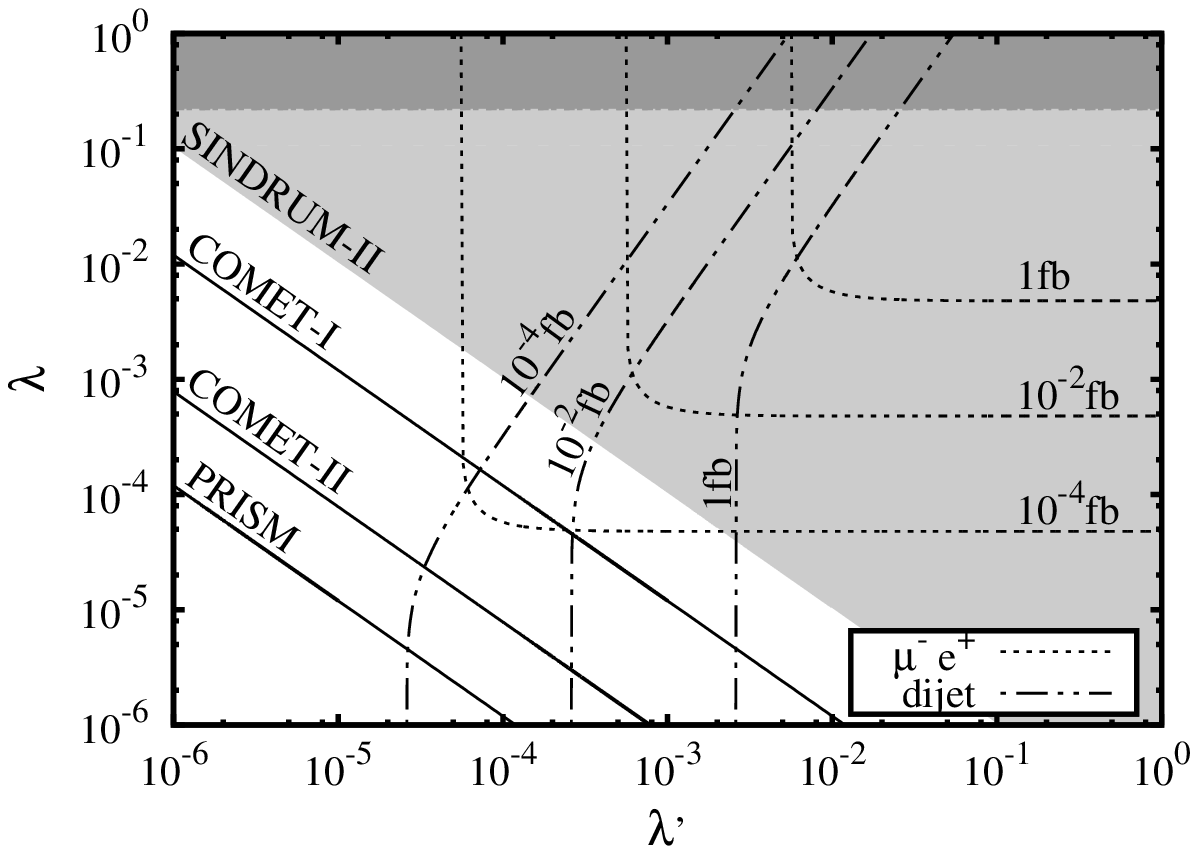}
\label{left1}
} & \hspace{-12mm}
\subfigure[$m_{\tilde \nu_\tau} = 1$TeV. $\sqrt{s}= 100$TeV. ]{
\includegraphics[scale=0.68]{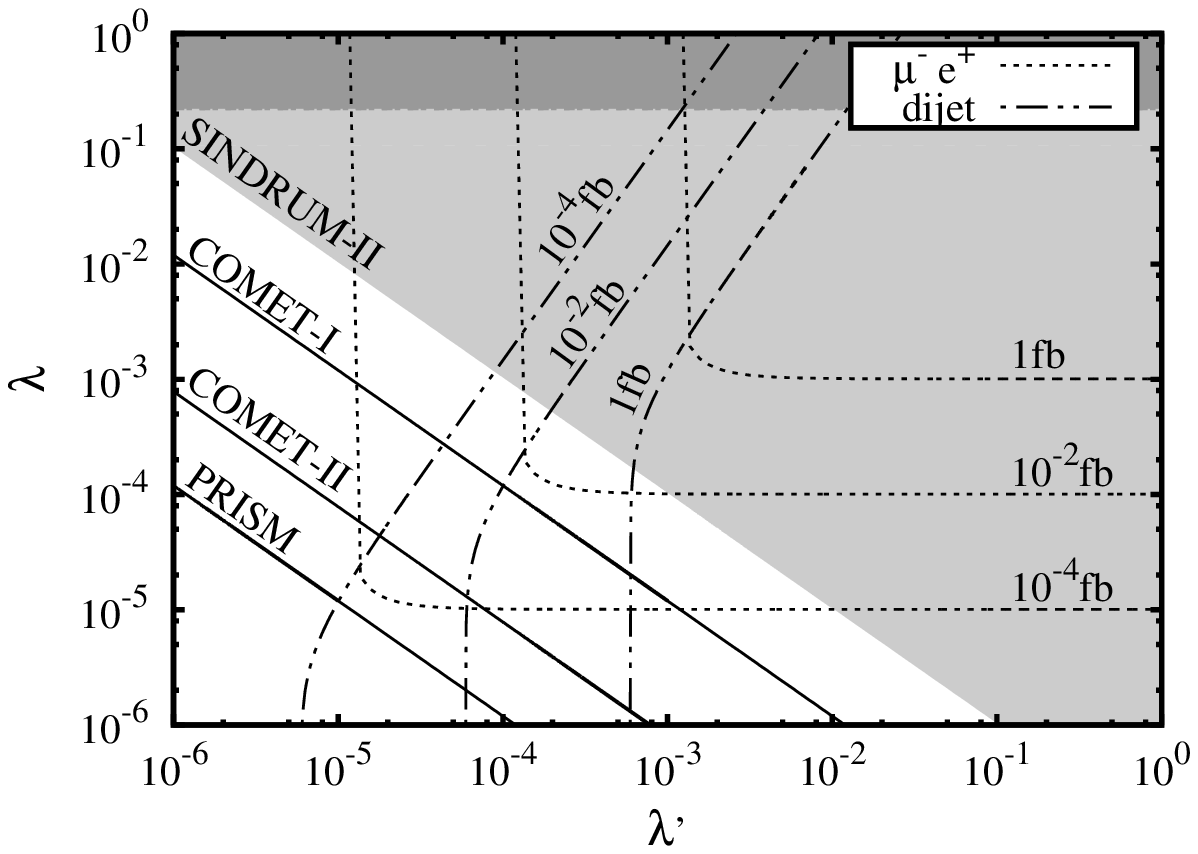}
\label{right1}
} \\
\subfigure[$m_{\tilde \nu_\tau} = 3$TeV. $\sqrt{s}= 14$TeV. ]{
\includegraphics[scale=0.68]{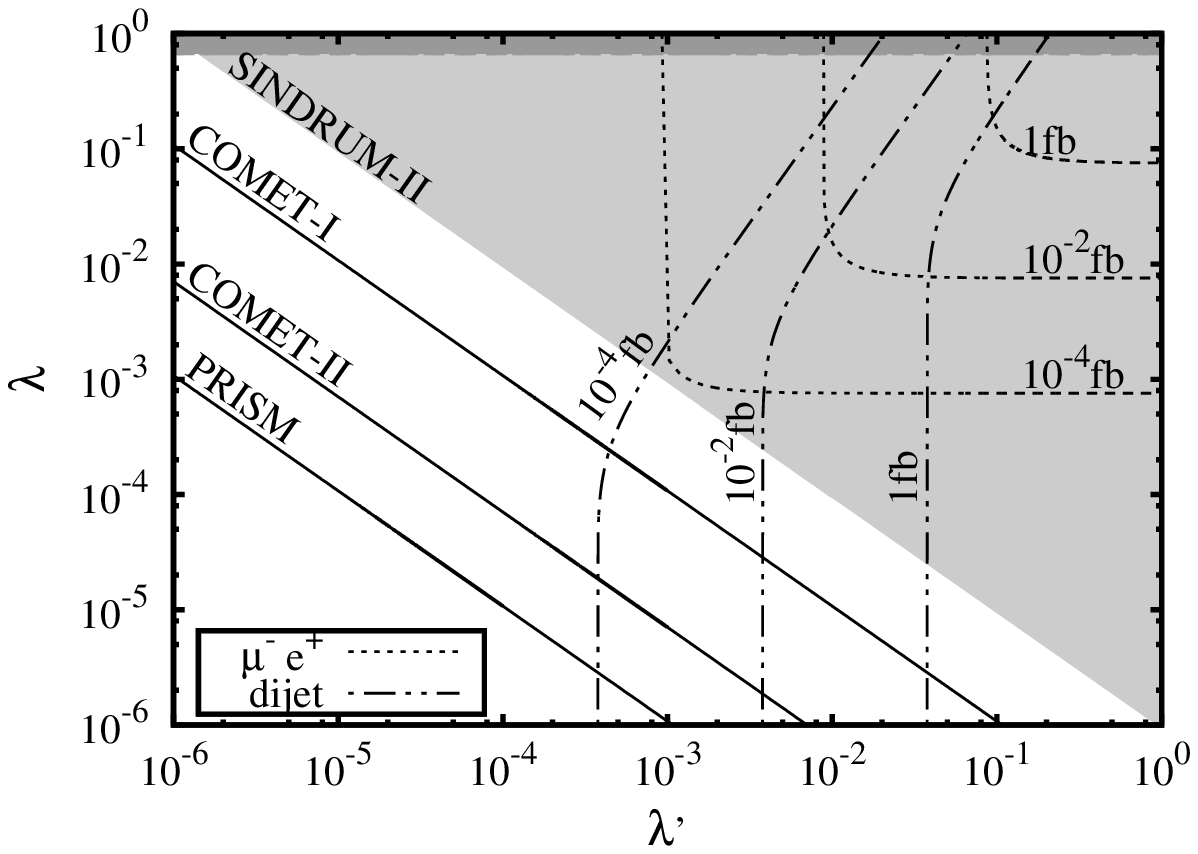}
\label{left2}
} & \hspace{-12mm}
\subfigure[$m_{\tilde \nu_\tau} = 3$TeV. $\sqrt{s}= 100$TeV. ]{
\includegraphics[scale=0.68]{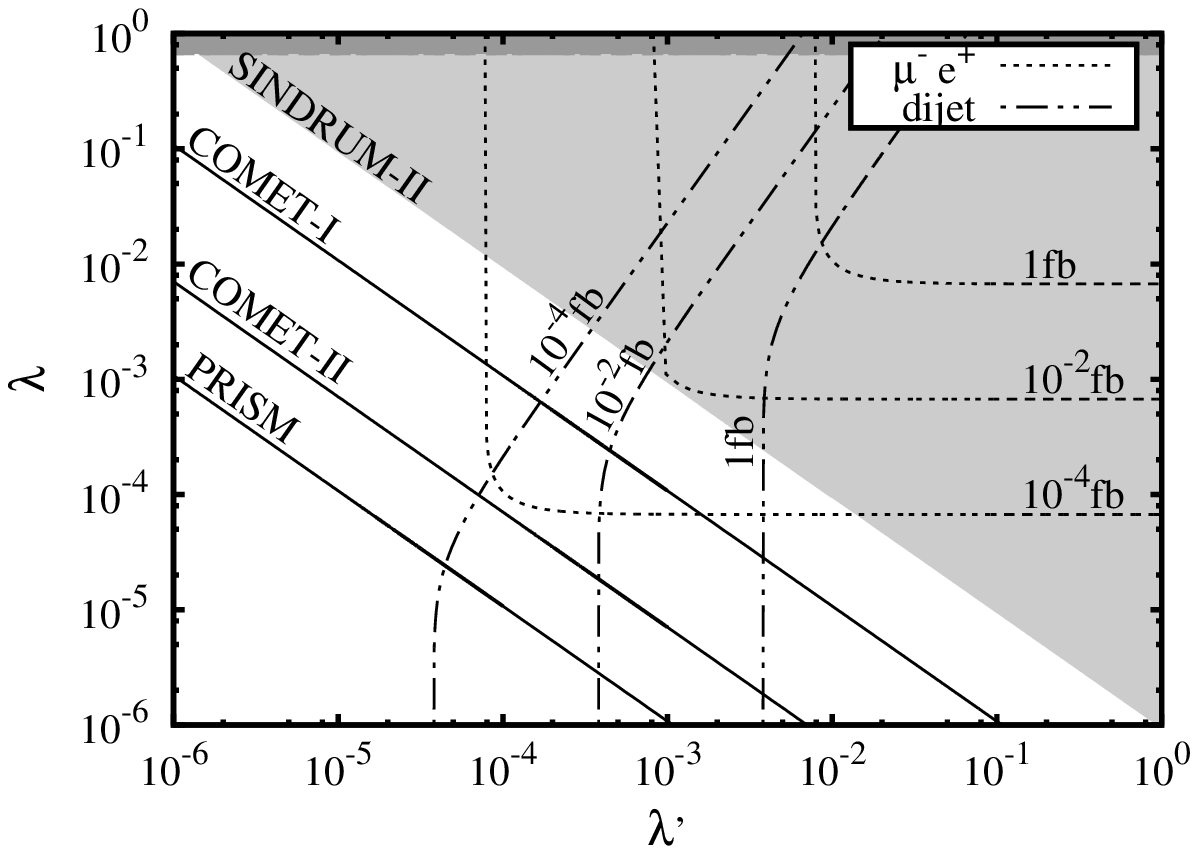}
\label{right2}
} \\
\end{tabular}
\caption{Contour plot of $\sigma(pp \to \mu^- e^+)$, 
$\sigma(pp \to jj)$, and BR($\mu^- N \to e^- N$) in the 
case-I\hspace{-1pt}I\hspace{-1pt}I for 
(a) $m_{\tilde \nu_\tau} = 1$TeV and $\sqrt{s}=14$TeV 
(b) $m_{\tilde \nu_\tau} = 1$TeV and $\sqrt{s}=100$TeV 
(c) $m_{\tilde \nu_\tau} = 3$TeV and $\sqrt{s}=14$TeV 
(d) $m_{\tilde \nu_\tau} = 3$TeV and $\sqrt{s}=100$TeV. 
For simplicity, we take universal RPV coupling, $\lambda \equiv 
\lambda_{312} = \lambda_{321} = -\lambda_{132} = 
-\lambda_{231} $. Light shaded region is excluded by the 
$\mu$-$e$ conversion search~\cite{Bertl:2006up}, 
and dark shaded band is excluded region by the $M$-$\bar M$ 
conversion search~\cite{Willmann:1998gd}.}
\label{Fig:cont_III}
\end{figure}

\begin{figure}[t!]
\hspace{-6mm}
\begin{tabular}{cc}
\subfigure[$\text{N}=\text{C}$ and $\sqrt{s}= 14$TeV.]{
\includegraphics[scale=0.63]{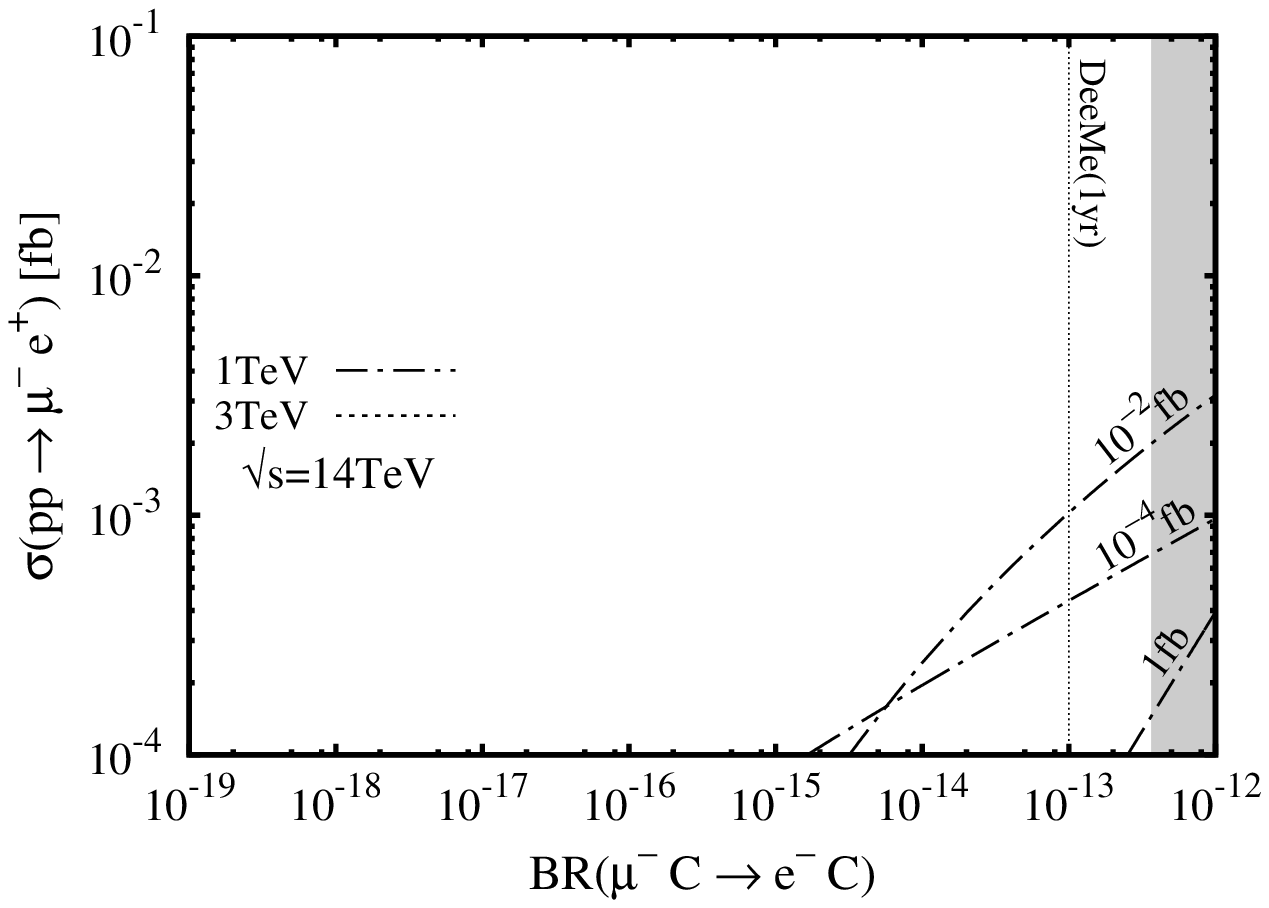}
\label{case3_C_14}
} & \hspace{-14mm}
\subfigure[$\text{N}=\text{C}$ and $\sqrt{s}= 100$TeV.]{
\includegraphics[scale=0.63]{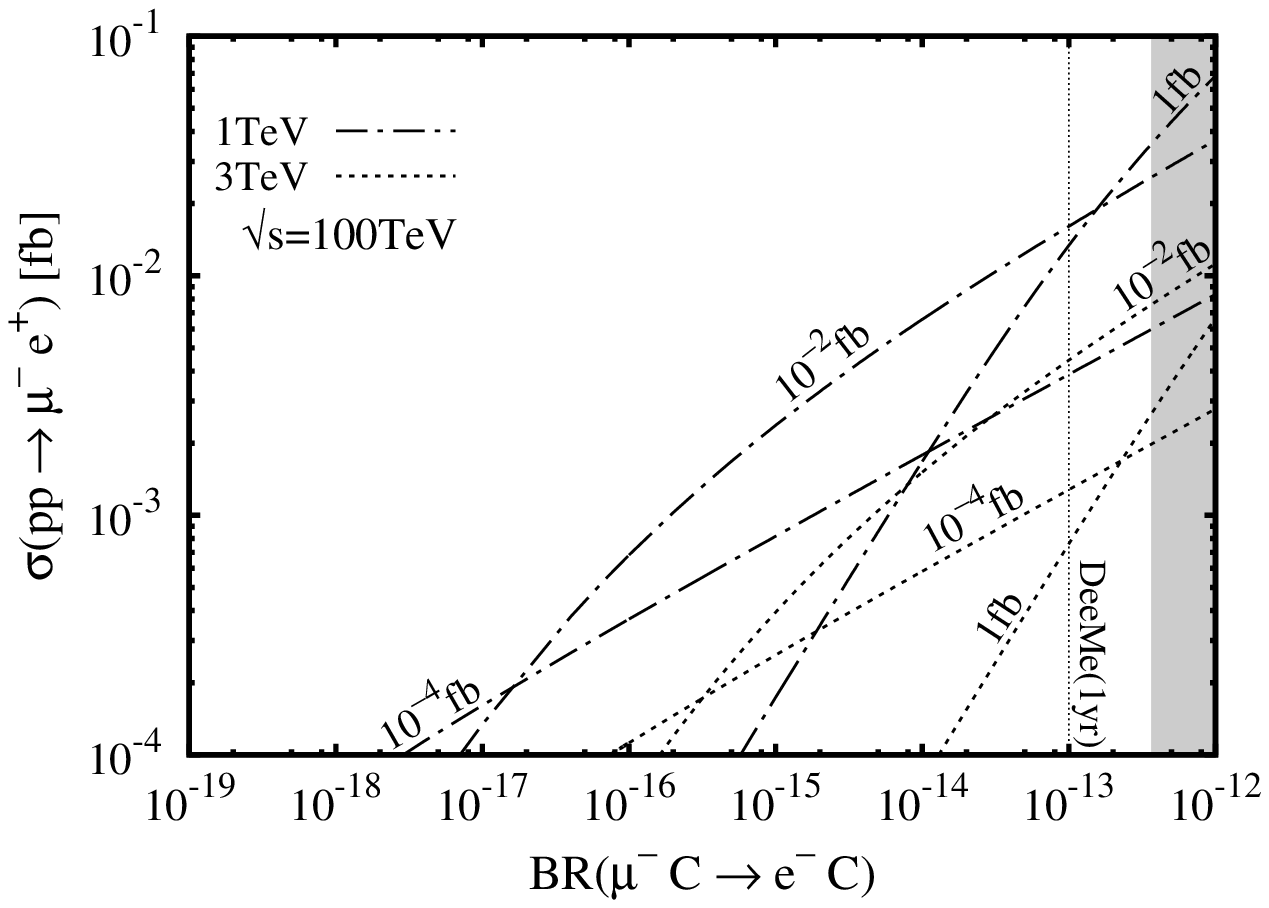}
\label{case3_C_100}
} \\[-4.5mm]
\subfigure[$\text{N}=\text{Si}$ and $\sqrt{s}= 14$TeV.]{
\includegraphics[scale=0.63]{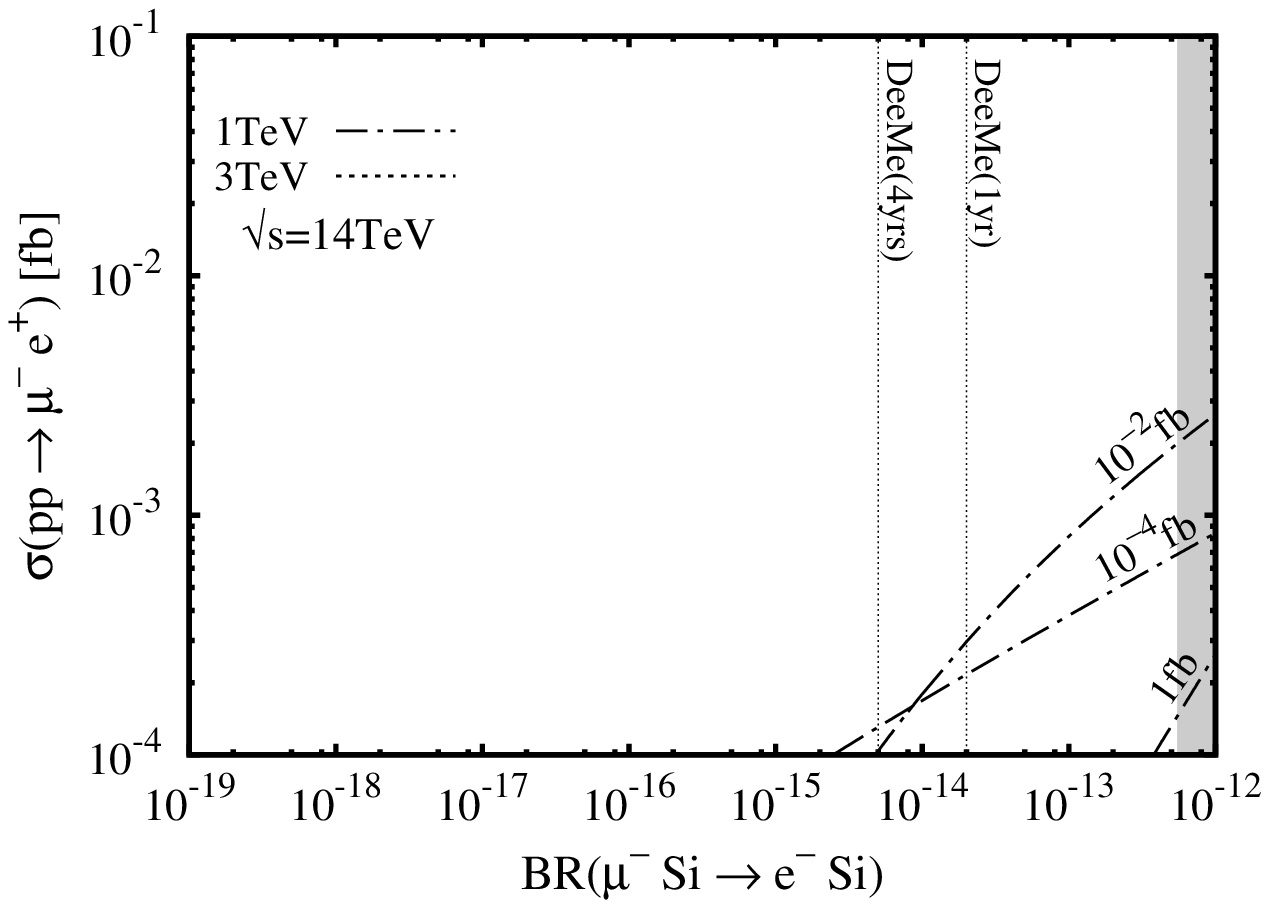}
\label{case3_Si_14}
} & \hspace{-14mm}
\subfigure[$\text{N}=\text{Si}$ and $\sqrt{s}= 100$TeV.]{
\includegraphics[scale=0.63]{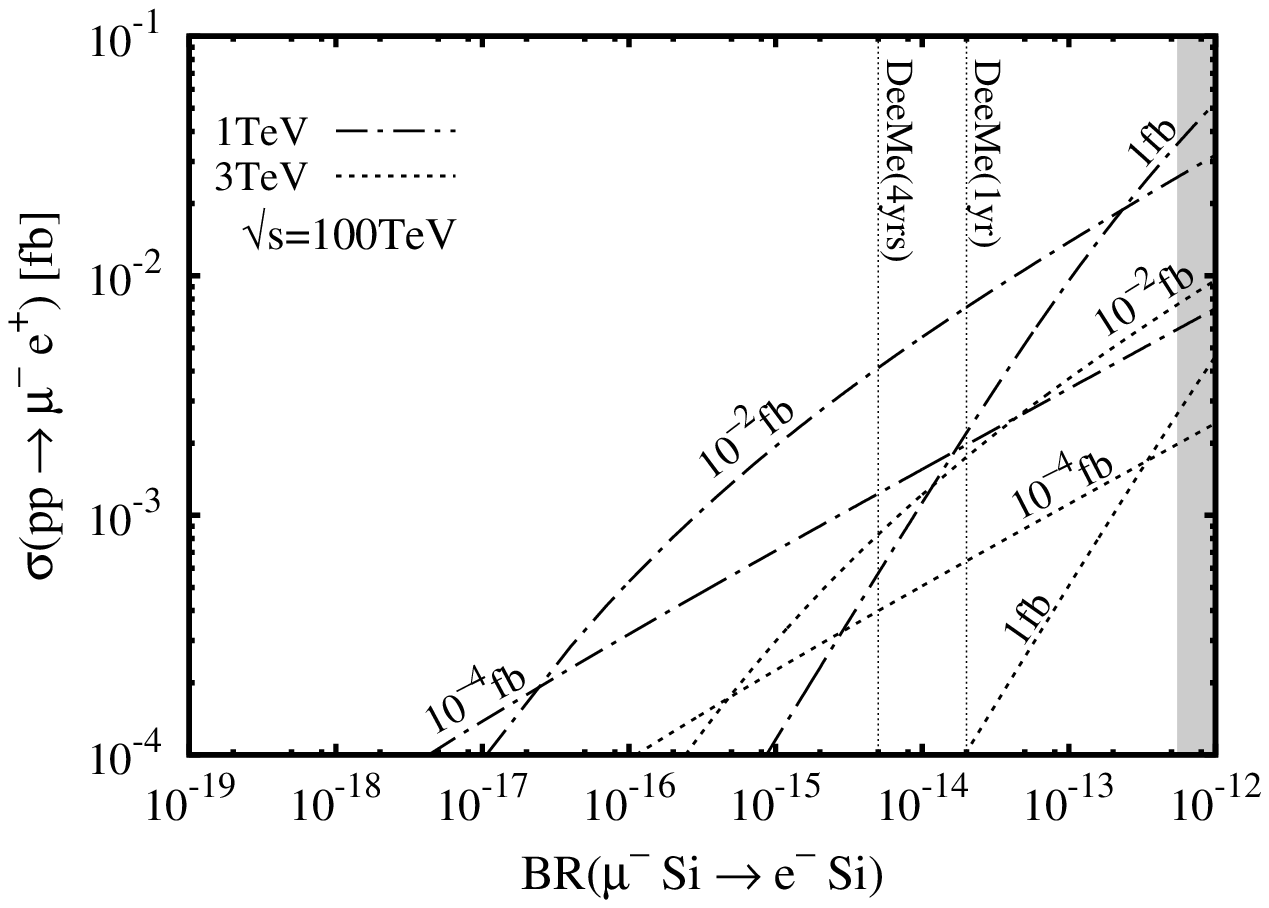}
\label{case3_Si_100}
} \\
\end{tabular}
\caption{$\sigma (pp \to \mu^- e^+)$ as a function of 
$\text{BR} (\mu^- N \to e^- N)$ for each $\sigma (pp 
\to jj)$ in the case-I\hspace{-1pt}I\hspace{-1pt}I. 
$\sigma (pp \to jj)$ are attached on each line. 
Results for $m_{\tilde \nu_\tau} = 1\text{TeV}$ 
($m_{\tilde \nu_\tau} = 3\text{TeV}$) are given by 
dot-dashed line (dotted line). Shaded region in each panel 
is the excluded region by the SINDRUM-I\hspace{-1pt}I 
experiment. Left panels show the results for the collision 
energy $\sqrt{s} = 14\text{TeV}$, and right panels show 
the results for $\sqrt{s} = 100\text{TeV}$. We take C [(a) 
and (b)],  and Si [(c) and (d)] for the target nucleus of 
$\mu$-$e$ conversion process. }
\label{Fig:sigma_vs_BR_III_1}
\end{figure}

\begin{figure}[h!]
\hspace{-6mm}
\begin{tabular}{cc}
\subfigure[$\text{N}=\text{Al}$ and $\sqrt{s}= 14$TeV.]{
\includegraphics[scale=0.63]{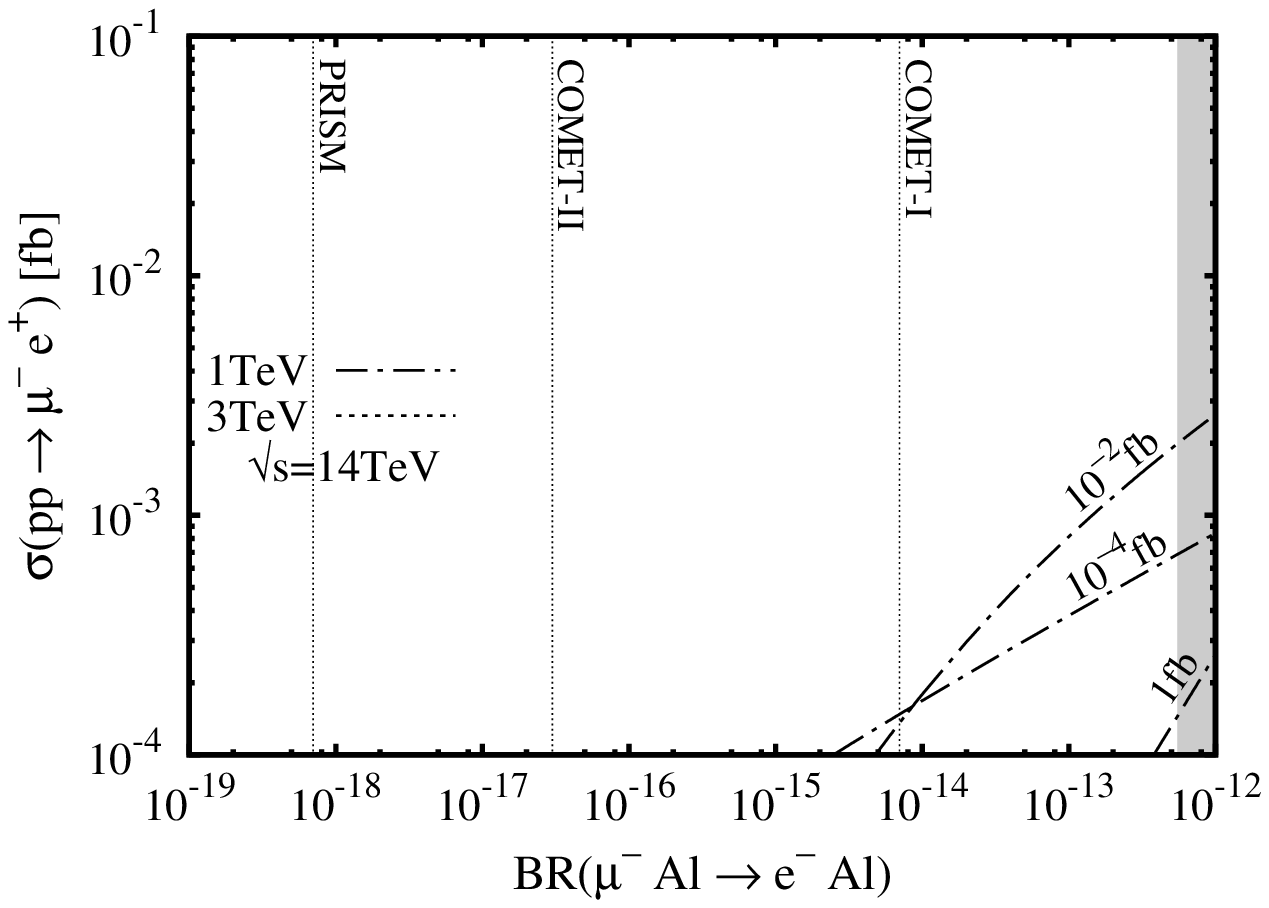}
\label{case3_Al_14}
} & \hspace{-14mm}
\subfigure[$\text{N}=\text{Al}$ and $\sqrt{s}= 100$TeV.]{
\includegraphics[scale=0.63]{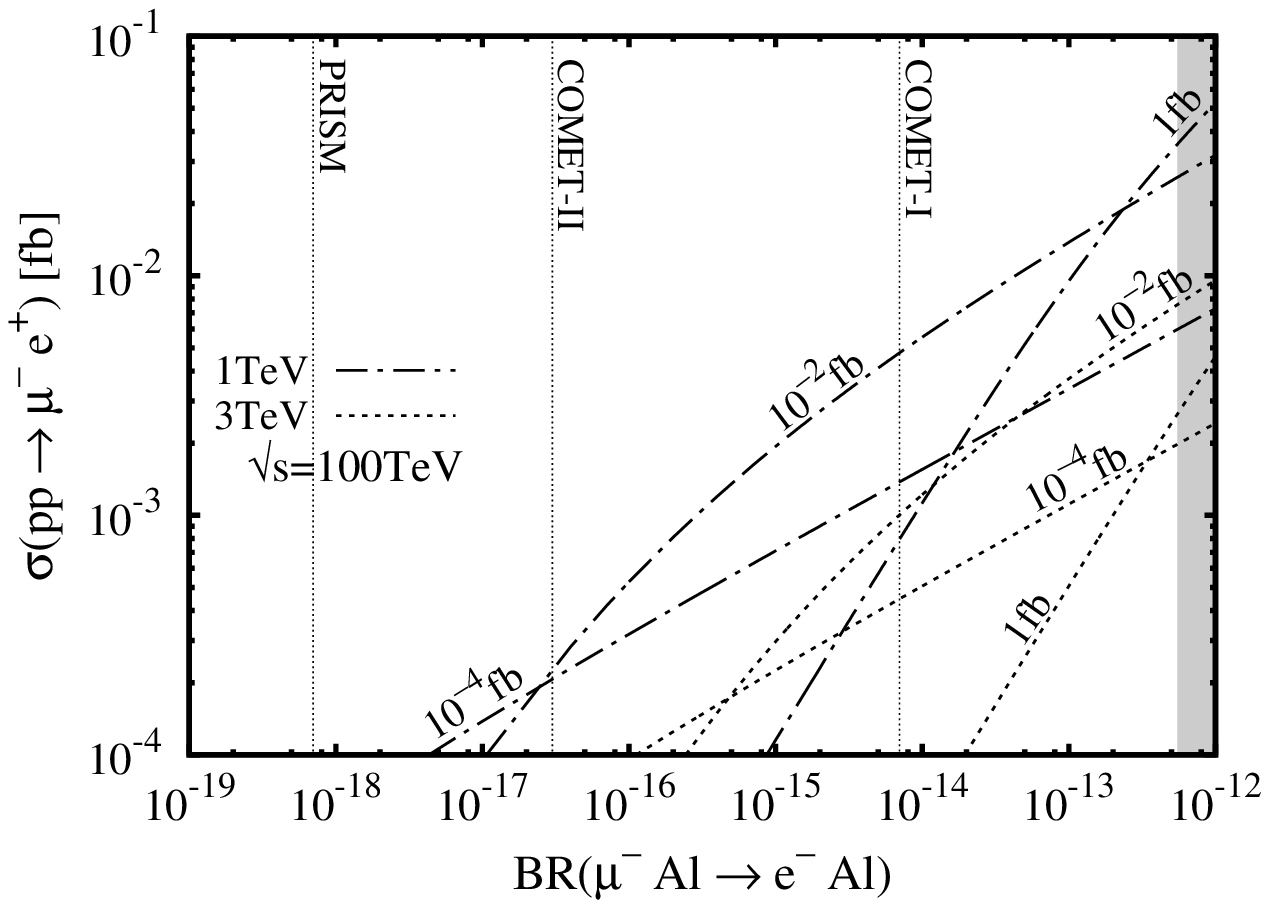}
\label{case3_Al_100}
} \\[-4.5mm]
\subfigure[$\text{N}=\text{Ti}$ and $\sqrt{s}= 14$TeV.]{
\includegraphics[scale=0.63]{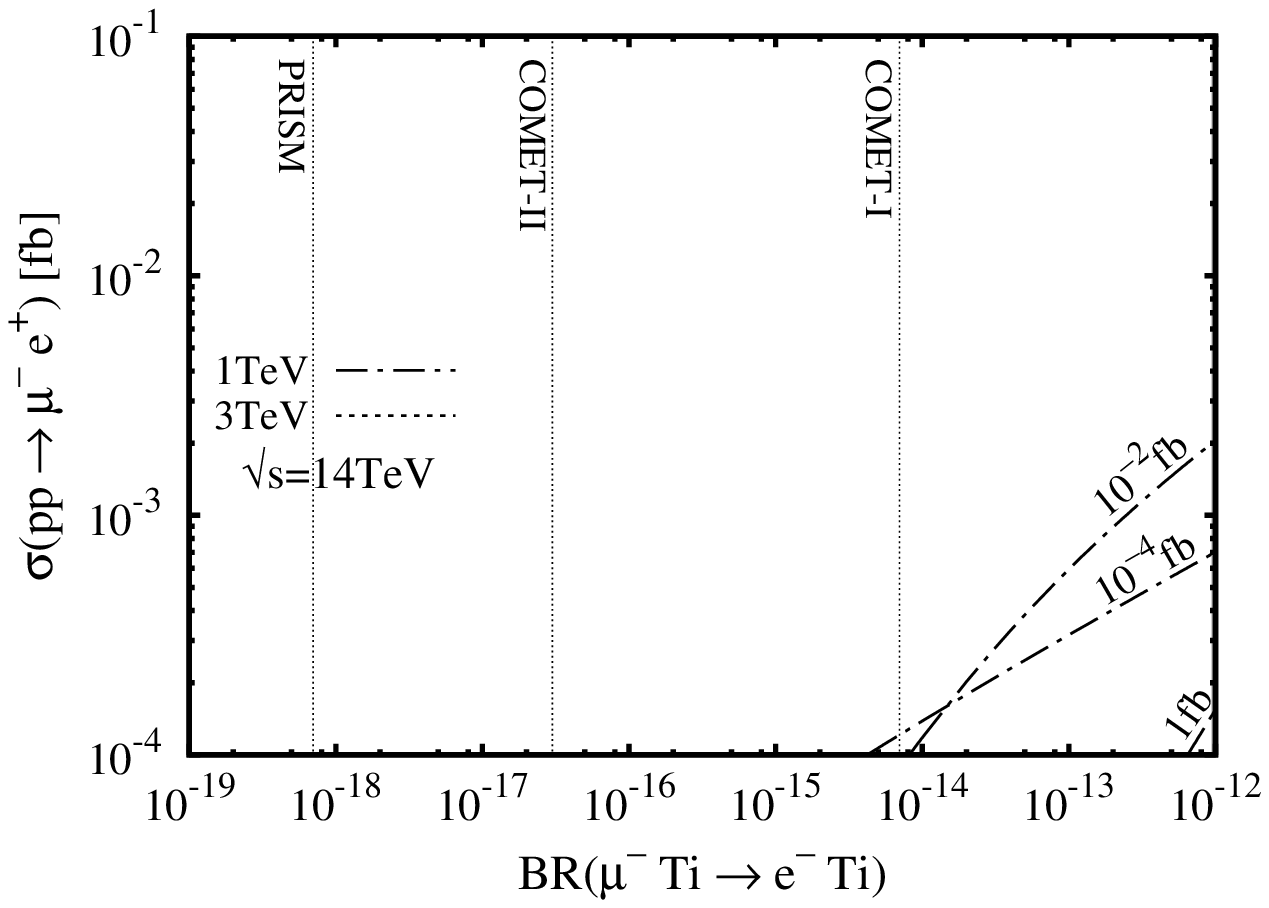}
\label{case3_Ti_14}
} & \hspace{-14mm}
\subfigure[$\text{N}=\text{Ti}$ and $\sqrt{s}= 100$TeV.]{
\includegraphics[scale=0.63]{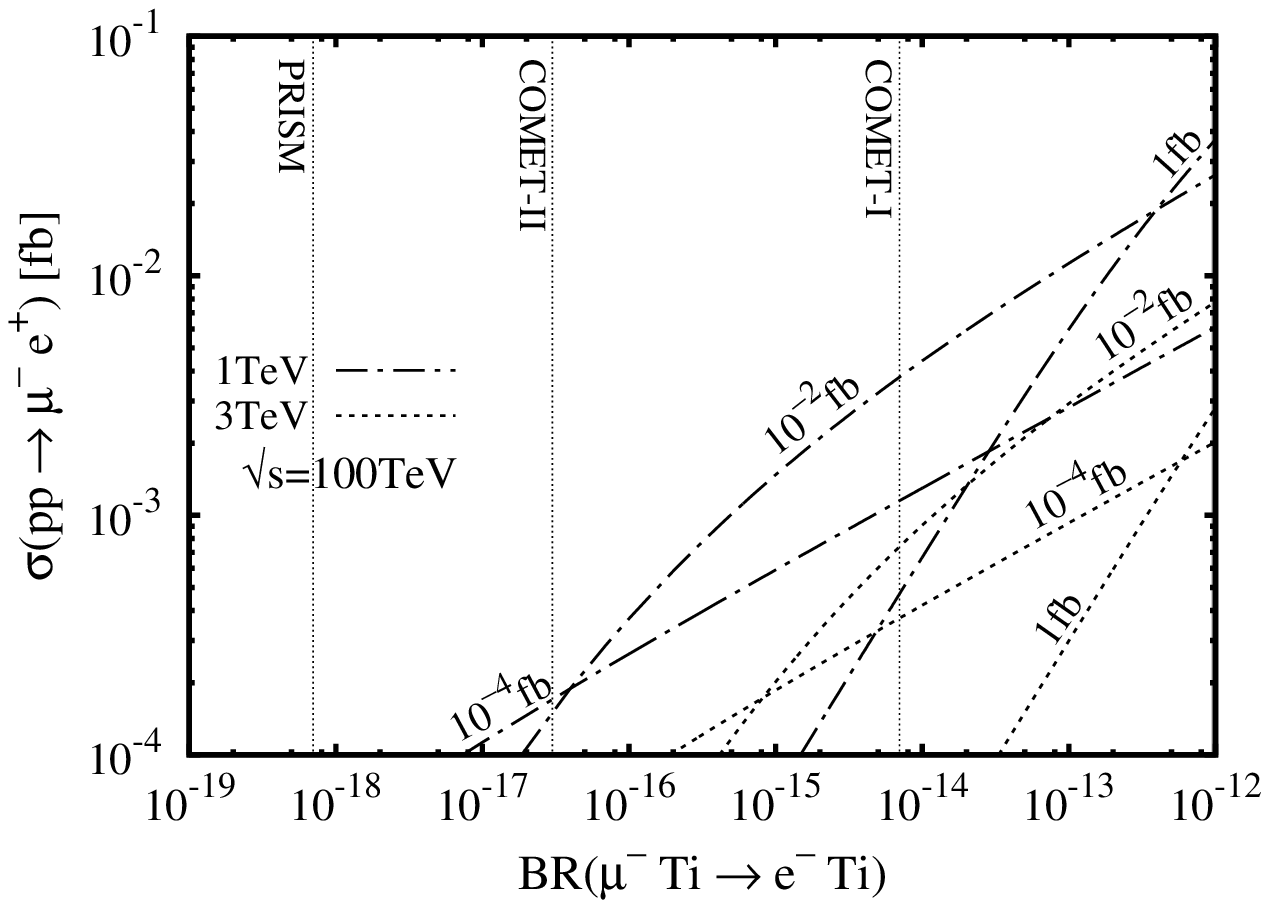}
\label{case3_Ti_100}
} \\
\end{tabular}
\caption{Same as Fig.~\ref{Fig:sigma_vs_BR_III_1} except for 
target nucleus. We take Al [(a) and (b)], and Ti [(c) and (d)] for 
the target nucleus of $\mu$-$e$ conversion process. }
\label{Fig:sigma_vs_BR_III_2}
\end{figure}

Figure~\ref{Fig:cont_III} displays the parameter dependence of 
$\sigma (pp \to \mu^- e^+)$, $\sigma(pp \to jj)$, and $\text{BR} 
(\mu^- \text{Al} \to e^- \text{Al})$ in the 
case-I\hspace{-1pt}I\hspace{-1pt}I. The description of 
Fig.~\ref{Fig:cont_III} is same as that of Fig.~\ref{Fig:cont_I}. 
Figures~\ref{Fig:sigma_vs_BR_III_1} and \ref{Fig:sigma_vs_BR_III_2} 
show $\sigma(pp \to \mu^- e^+)$ as a function of $\text{BR}(\mu^- 
\text{Al} \to e^- \text{Al})$ in the case-I\hspace{-1pt}I\hspace{-1pt}I. 
The description of the figure is same as those of 
Figs.~\ref{Fig:sigma_vs_BR_I_1} and \ref{Fig:sigma_vs_BR_I_2}.

We can check the nice consistency between theoretical calculations and 
the behavior of plots in Figs.~\ref{Fig:sigma_vs_BR_III_1} and 
\ref{Fig:sigma_vs_BR_III_2} by repeating the same quantitative 
analyze in Sec.~\ref{Sec:311} with $F_{jet}$ and $k_N$ for the 
case-I\hspace{-1pt}I\hspace{-1pt}I (see Table 4).

We can discriminate case-I, -I\hspace{-1pt}I, and 
-I\hspace{-1pt}I\hspace{-1pt}I by checking the correlations of $\sigma 
(pp \to \mu^- e^+)$, $\sigma(pp \to jj)$, and $\text{BR} (\mu^- 
\text{Al} \to e^- \text{Al})$ with Figs.~\ref{Fig:sigma_vs_BR_I_1}, 
\ref{Fig:sigma_vs_BR_I_2}, \ref{Fig:sigma_vs_BR_II_1}, 
\ref{Fig:sigma_vs_BR_II_2}, \ref{Fig:sigma_vs_BR_III_1}, and 
\ref{Fig:sigma_vs_BR_III_2}. 
And, as we discussed in case-I and -I\hspace{-1pt}I, the RPV couplings 
are precisely determined via the measurement $\sigma (pp \to \mu^- e^+)$, 
$\sigma(pp \to jj)$, and $\text{BR} (\mu^- \text{Al} \to e^- \text{Al})$ 
by using Fig.~\ref{Fig:cont_III}.


\subsection{comment for NSI}  \label{Sec:res_NSI} 

In Figs~\ref{Fig:cont_III}, \ref{Fig:sigma_vs_BR_III_1}, and 
\ref{Fig:sigma_vs_BR_III_2}, for simplicity, we take $\lambda'_{311} 
= \lambda'_{322}$. 
When we take $\lambda'_{311} \neq \lambda'_{322}$, as is studied 
in Sec.~\ref{Sec:mue_conv} and Sec.~\ref{Sec:collider}, behavior of 
the plots are basically same with Figs~\ref{Fig:cont_III}, 
\ref{Fig:sigma_vs_BR_III_1}, and \ref{Fig:sigma_vs_BR_III_2}. 
In such a case, in order to determine $\lambda'_{311}$ and
$\lambda'_{322}$ separately, we need another measurement, say that of
the NSI at next-generation neutrino experiments.

It is said that $\epsilon^S_{\mu e}$ of ~($10^{-4}$) can be searched
in near future~\cite{Kopp:2007ne}. However, from the current limit of the branching ratio
of $\mu\rightarrow e$ conversion it must be less than $10^{-6}$ which is
far below than the expected sensitivity.

 We leave the detailed study 
for future work~\cite{future}.

\clearpage
\section{Summary and discussion}  \label{Sec:summary} 

We have studied a supersymmetric standard model without R parity
as a benchmark case that COMET/DeeMe observe $\mu - e$ conversion 
prior to all the other experiments observing new physics. 

In this case with the assumption that only the third generation sleptons
contribute to such a process, we need to assume that $\{\lambda'_{311} 
{\rm \hspace{1.2mm} and/or} \hspace{1.2mm} \lambda'_{322}\} \times
\{\lambda_{312}{\rm{ \hspace{1.2mm} and/or \hspace{1.2mm} }}
\lambda_{321}\}$ must be sufficiently large. 
Though other combinations of coupling constants can lead a significant 
$\mu-e$ conversion process, only those are considered here. 
This is because in most of scenarios in the supersymmetric theory, the 
third generation of the scalar lepton has the lightest mass.

With these assumptions, we calculated the effects on future experiments.
First we considered the sensitivity of the future $\mu - e$ conversion
experiments on the couplings and the masses. To do this we considered
the three cases; I) $\lambda'_{311}$ is dominant, I\hspace{-1pt}I) 
$\lambda'_{322}$ is dominant, I\hspace{-1pt}I\hspace{-1pt}I) both 
are dominant. Since the matrix element of $\bar q q$
in nucleus is different for down quark and strange quark, we got a 
different sensitivity on them.

Then with the sensitivity kept into mind
we estimated the reach to the couplings by calculating the cross section
of $pp \rightarrow \mu^- e^+$ and $pp \rightarrow jj$  as a function
of the slepton masses and the couplings. To have a signal of $\mu^- e^+$
both the coupling $\lambda'$ and $\lambda$ must be large and hence
there are  lower bounds for them
while to observe dijet event via the slepton only the coupling
$\lambda'$ must be large and hence there is a lower bound on it
(Figs.~\ref{Fig:cont_I}, \ref{Fig:cont_II}, and \ref{Fig:cont_III}). 
In all cases we have a chance to get confirmation of $\mu - e $
conversion in LHC indirectly. In addition, we put a bound
on the couplings by comparing both modes.

On the contrary to the hope on LHC, unfortunately the current bound
by $\mu -e $ conversion gives the much smaller Non-Standard Interaction
on neutrino physics than the sensitivity in near future experiment.
Instead of this fact, with this we can distinguish $\lambda_{312}$ 
and $\lambda_{321}$ and it is worth searching it.

Finally we considered muonium conversion. If $\lambda'$ is very small
we cannot expect a signal from LHC. In this case at least one of $\lambda_{312}$
and $\lambda_{321}$ must be very large and if it is lucky, that is both
of them are very large we can expect muonium conversion.

There are other opportunities to check the result on $\mu -e$ conversion.
For example we can distinguish $\lambda_{312}$ and $\lambda_{321}$
in linear collider with polarized beam. We can also expect the
signal $p e^- \to p \mu^-$ in LHeC. It is however beyond the scope of this
paper to estimate their sensitivities and we leave them in future work~\cite{future}.

\section*{Acknowledgments}   

This  work was supported in part by the Grant-in-Aid for the Ministry of Education,
Culture, Sports, Science, and Technology, Government of Japan,
No. 24340044 and No. 25105009 (J.S.). and No. 25003345 (M.Y.).


\end{document}